\pdfoutput=1
\documentclass[11pt]{article}

\usepackage{arxiv}
\hypersetup{hidelinks,hypertexnames=false}

\usepackage{amsmath}
\usepackage{amsthm}

\usepackage[T1]{fontenc}
\usepackage[utf8]{inputenc}
\usepackage{textcomp}
\usepackage{lmodern}
\usepackage{microtype}

\usepackage{amssymb,amsfonts,latexsym,mathrsfs,stmaryrd,bm}
\usepackage{graphicx}
\usepackage{subcaption}
\usepackage{float}
\usepackage{placeins}
\usepackage{booktabs,multirow}
\usepackage{enumerate}
\usepackage{url}
\usepackage{xcolor}
\usepackage{tikz}
\usetikzlibrary{graphs}
\usetikzlibrary{cd}
\usepackage{scalerel}
\usepackage{cleveref}

\makeatletter
\def\maxwidth{\ifdim\Gin@nat@width>\linewidth\linewidth\else\Gin@nat@width\fi}
\def\maxheight{\ifdim\Gin@nat@height>\textheight\textheight\else\Gin@nat@height\fi}
\def\fps@figure{htbp}
\makeatother
\setkeys{Gin}{width=\maxwidth,height=\maxheight,keepaspectratio}
\newtheorem{teo}{Theorem}
\newtheorem{prop}{Proposition}
\newtheorem{defi}{Definition}
\newtheorem{rmk}{Remark}
\newtheorem{lem}{Lemma}
\newtheorem{cor}{Corollary}
\newtheorem{assumption}{Assumption}

\crefname{teo}{Theorem}{Theorems}
\crefname{prop}{Proposition}{Propositions}
\crefname{defi}{Definition}{Definitions}
\crefname{rmk}{Remark}{Remarks}
\crefname{lem}{Lemma}{Lemmas}
\crefname{cor}{Corollary}{Corollaries}
\crefname{assumption}{Assumption}{Assumptions}
\crefname{claim}{Claim}{Claims}
\crefname{app}{Appendix}{Appendices}
\Crefname{app}{Appendix}{Appendices}

\DeclareMathOperator*{\LCA}{LCA}

\DeclareMathOperator{\BV}{BV}
\DeclareMathOperator{\V}{V}

\newcommand{\R}{\mathbb{R}}

\newcommand{\virgolette}[1]{``#1''}

\newcommand{\Xvar}{\mathcal{X}}
\newcommand{\Yvar}{\mathcal{Y}}

\firstpageno{1}

\begin{document}

\title{Functional Data Representation with Merge Trees}

\author{Matteo Pegoraro\thanks{Faculty of Informatics, USI, Lugano, Switzerland.\protect\newline
       E-mail: \texttt{pegorm@usi.ch}}
  \and
  Piercesare Secchi\thanks{MOX -- Department of Mathematics, Politecnico di Milano, Milano, Italy.}}

\maketitle

\begin{abstract}
In this paper, we address the problem of representing functional data using tools from algebraic topology. We model functions through merge trees, which, like the more widely used persistence diagrams, are invariant under homeomorphic reparameterizations of their domain. This invariance allows statistical analyses that are robust to functional misalignment. We study a recently introduced metric for merge trees, establish its stability properties, and prove the consistency and convergence rates of a corresponding class of estimators.
To illustrate the advantages of our topological approach to functional data analysis, we apply it to the AneuRisk65 dataset, replicating a well-known supervised classification task that made this dataset a benchmark for misaligned functional data. We also compare our results with those obtained using the Fisher–Rao metric, which further supports the effectiveness of the proposed method.
The supplementary material includes an extensive comparison between merge trees and persistence diagrams, outlining their similarities and differences to help analysts choose the most suitable representation for a given application.
\end{abstract}


\section{Introduction}

Since the foundational works of \citet{book_fda} and \citet{ferraty_fda}, Functional Data Analysis (FDA) has become essential for studying data where each unit is best represented as a real-valued function over a continuous, often one-dimensional, domain $D$.
These functional observations arise from smoothing discrete measurements, embedding them in a suitable function space, and addressing potential misalignments. A well-established example of this FDA pipeline is provided by \cite{aneurisk_jasa}, which analyzes the AneuRisk65 dataset.

 Common approaches to smoothing include kernel smoothing \citep{nadaraya1964estimating, schuster1979contributions, muller1984smooth, mack1989derivative} or projection onto basis functions such as splines or trigonometric polynomials. The choice of smoother depends on signal-to-noise ratio, required differentiability, and the structure of the embedding space. For example, when derivatives are of interest or when non-negligible noise affects data points, interpolation is avoided in favor of smoother representations, often within Sobolev spaces \citep{aneurisk_splines}.

\section*{Amplitude and Phase Variation}

Functional data often exhibit multiple sources of variability \citep{vantini}, notably amplitude and phase variation, the latter referring to domain-wise misalignment. The AneuRisk65 dataset has become a benchmark for methods that seek to disentangle these sources (see the Special Section on Time Warpings and Phase Variation, Electronic Journal of Statistics, Vol. 8(2)). If not properly accounted for, phase variation can obscure meaningful patterns in the data \citep{chemo, marron_registration}.

A common strategy to deal with phase variation problems is to embed the data in a Hilbert space and define equivalence classes via alignment or registration procedures \citep{sanga_warp_review}, akin to approaches in shape analysis \citep{stat_shape} and pattern theory \citep{pattern_theory}. Functions defined on compact intervals are aligned using warping functions of different kinds, ranging from affine functions \citep{kmean_aling} to homeomorphisms and diffeomorphisms \citep{sriva_0, sriva_1, bharath2020distribution}. Once optimal representatives are selected (typically by minimizing an appropriate loss function; \citealp{aneurisk_kmeans}), subsequent analysis is conducted on the aligned data.

However, this framework has limitations. The embedding space’s metric may not respect the equivalence classes \citep{sfere_marron}, and complications arise when $D$ is not a compact interval. Alternatives include metrics like the Fisher–Rao metric \citep{sriva_0,sriva_1, fisher_rao_densities}, which naturally incorporate diffeomorphic warpings and directly compare equivalence classes. Yet, even these approaches face challenges outside the compact interval setting.

An alternative approach to phase variation is to represent functional data using statistics that are invariant under reparametrization, yet still informative for analysis. Algebraic topology offers a natural framework for this, providing tools to identify features that remain unchanged under continuous deformations of the domain \citep{hatcher}.

Topological Data Analysis (TDA) builds on these ideas to develop methods for extracting shape-related features from data \citep{PH_survey}.
A key tool in TDA is persistent homology \citep{PD_1}, which tracks the evolution of homological features—like path connected components (0-dimensional holes) and loops (1-dimensional holes)—across a filtration, that is, a nested family of topological spaces.

Of particular interest for our work are filtrations generated by real-valued functions. Given a topological space $X$ and a function $f : X \rightarrow \mathbb{R}$, one considers the sublevel sets $X_t = f^{-1}((-\infty, t])$. The changing topology of the sets $\{X_t\}_{t \in \mathbb{R}}$ encodes information about the structure of $f$, and importantly, it is invariant under homeomorphic warpings of the domain. This makes TDA particularly well-suited for handling phase variation in functional data.

Our objective is therefore to explore how topological representations of functions can complement classical functional data analysis. Studying these methods in a context where alignment and registration techniques are already well established, rather than in general shape analysis settings where alignment is substantially more complex or even infeasible, allows for a more direct and meaningful comparison.

\section*{Merge Trees}

To enable statistical analysis, persistent homology outputs are typically encoded through topological summaries endowed with a metric structure. Persistence diagrams (PDs) \citep{cohen_PD} are arguably the most widely used among these and have found applications across diverse fields, including biology, robotics, signal processing, and materials science. Alternative summaries include persistence landscapes, images, and silhouettes \citep{landscapes, pers_img, silhouettes}.

In this work, we focus on a different type of summary: merge trees. These trees parametrize the evolution of path-connected components in the sublevel sets of $f$, and can be viewed as a special case of Reeb graphs \citep{reeb, reeb_2}. Merge trees have a well-established role in both topology, data analysis, and computer vision \citep{merge_parall_2}, largely due to the fact that they represent a richer set of information compared to PDs.
Numerous frameworks have been proposed to measure distances between merge trees.
These can be broadly divided into two main categories:
(a) approaches based on the interleaving distance and other metrics with strong theoretical guarantees
\citep{merge, merge_interl, interl_approx, merge_intrins, merge_frechet, curry2021decorated, cardona2021universal}, and
(b) approaches defining edit distances aimed primarily at computational efficiency
\citep{wetzels2022deformation, bollen2022computing, pont2022principal, wetzels2023merge, merge_farlocca, merge_farlocca_2, merge_wass, sridharamurthy2021comparative}.

However, methods in category (a) often entail prohibitive computational costs, even when approximation algorithms are available, while those in category (b) generally lack rigorously established stability properties required for reliable data analysis, and thus often rely on ad hoc workarounds.
The only exception is \cite{wetzels2022deformation}, whose relationship with the metric employed in this work is discussed in \cite{pegoraro2025finitely}.
For a more detailed comparison, see also \cite{wetzels2025categorizing}.

We also mention very recent work applying the Gromov--Wasserstein distance to merge trees \citep{li2025flexible}; see also \Cref{rmk:GH} for further discussion.

\section*{Main Contributions}

To explore the potential of merge trees as robust representations for functional data under phase variation, we adopt the recently introduced metric from \cite{pegoraro2025finitely}.

Our main theoretical contributions are as follows:
\begin{enumerate}

    \item We provide a proof that persistence diagrams and merge trees are invariant under homeomorphic reparameterizations of the domain. While this fact is well known in the TDA community as a consequence of standard homological invariance, we expect this explicit derivation to be of independent interest for the FDA researchers, and we are not aware of earlier references, prior to preliminary versions of this manuscript, where such proofs are provided.

    \item We provide a theoretical analysis of the stability of the operator that maps a function to its merge tree representation.
    In \cite{pegoraro2025finitely}, it is shown that the stability properties established here are consistent with those characterizing commonly used distances for PDs, such as the \(1\)-Wasserstein distance.
    Furthermore, \cite{wetzels2025categorizing} present a related result; in particular, \Cref{teo:stability}, which also appeared in an earlier preprint of this work, generalizes Theorem~6 therein.

    \item Building on this analysis, we establish additional key properties that enable the construction of consistent estimators of merge trees from functions observed on a grid and affected by noise.
\end{enumerate}

To illustrate the practical utility of our approach, we revisit the AneuRisk65 dataset \citep{aneurisk_jasa}, a benchmark in the analysis of misaligned functional data, replicating and extending a well-known supervised classification task. Our methods achieve comparable or superior performance to standard FDA pipelines, including those based on Fisher–Rao alignment, while inherently avoiding the need for explicit alignment through the invariance properties of merge trees.

\section*{Outline of the Paper}

In \Cref{sec:merge_trees}, we present the merge tree representation of a function and define the metric structure for the space of merge trees. In \Cref{sec:vs_PD}, we provide a brief overview of PDs to facilitate comparisons with merge trees.
\Cref{sec:stability} contains the primary theoretical contributions of the paper, addressing the stability of the merge tree representation implied by the properties of continuity of the operator that maps a function to its merge tree, computational approaches for approximating merge trees, and the problem of consistently estimating merge trees from sampled observations of a function.
In \Cref{sec:case_study}, we explore a functional data classification problem inspired by \cite{aneurisk_jasa}, comparing their findings with those obtained using the TDA approach advocated in this work. Finally, \Cref{sec:discussion_FDA} concludes the paper with a discussion highlighting ideas related to our topological approach to functional data analysis.

In \Cref{sec:supplementary-material} we describe briefly the content of the supplementary material.

\begin{figure}
    \begin{subfigure}[c]{0.47\textwidth}
    	\centering
    	\includegraphics[width = \textwidth]{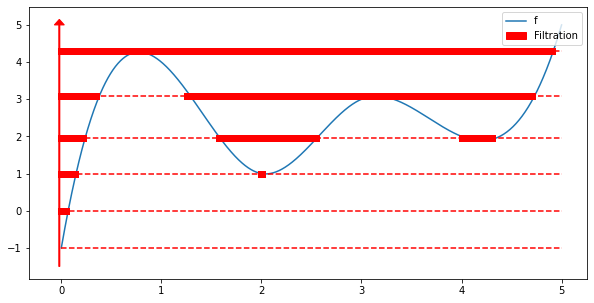}
    	\caption{ }
    	\label{fig:sublvl}
	\end{subfigure}
    \begin{subfigure}[c]{0.47\textwidth}
    	\centering
	    \includegraphics[width = \textwidth]{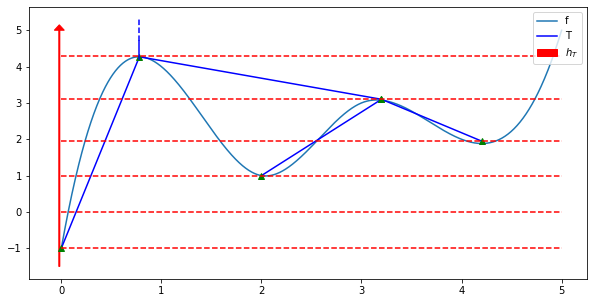}
    	\caption{}
    	\label{fig:func_tree}
		\end{subfigure}
\caption{Sublevel sets of a function (a);  the same function with its associated merge tree (b).}
\end{figure}

\section{Merge Trees of Functions}
\label{sec:merge_trees}

\subsection{Sublevel Sets}
In the upcoming sections we define the merge tree representation of a function.
Merge trees are an already established tool in topology and, to some extent, also in statistics since dendrograms can be regarded as merge trees.
However, we will dedicate a few lines to defining them within the framework proposed by \cite{pegoraro2025finitely}, which differs from the traditional approach described, for example, in \cite{merge_parall_2}.
Roughly speaking, the pipeline to obtain a merge tree is the following: we transform the given function into a sequence of nested subsets and then we track the topological changes along this sequence. Such information is then turned into a tree.

Consider a function $f:X\rightarrow \mathbb{R}$, with $X$ being any topological space.
We call sublevel set at height $t \in \mathbb{R}$, the set $X_t:= f^{-1}((-\infty,t])\subset X$ . The key property of the family $\{X_t\}_{t\in \mathbb{R}}$ is that such subsets are nested: if $t\leq t'$ then $X_t\subset X_{t'}$.
Note that the sequence $\{X_t\}_{t\in \mathbb{R}}$ is fully determined by the shape of the function $f$; see \Cref{fig:sublvl}. In fact, for \(x \in X,\) $f(x) = \text{inf}\{t\in\mathbb{R}: x\in X_t\}$, hence no information carried by \(f\) is lost by its representation $\{X_t\}_{t\in \mathbb{R}}$.

\subsection{Path-Connected Components}\label{sec:path_conn}
A topological space $X$ is path-connected if for every couple of points $x,y \in X$ there is a continuous curve $\alpha:[0,1]\rightarrow X$ such that $\alpha(0)=x$ and $\alpha(1)=y$.
The biggest path-connected subsets contained in a topological space are called path-connected components.

The covariant functor \citep{maclane} of path-connected components is usually referred to as $\pi_0$, meaning that, given a topological space $X$, $\pi_0(X)$ is the set of the path-connected components of $X$. If $q:X\rightarrow Y$ is a continuous function between two topological spaces $X$ and $Y$, one can define the function $\pi_0(q):\pi_0(X)\rightarrow \pi_0(Y)$ as follows: $U\mapsto V  \text{ if }q(U)\subset V.$
The functor $\pi_0$ satisfies a number of properties. Among them, we emphasize the following: for two continuous functions $p,q$ that can be composed into the function $p\circ q$, it is true that $\pi_0(p\circ q)=\pi_0(p)\circ \pi_0(q)$.

Path-connected components are the source of information we want to track along the family $\{X_t\}_{t\in \mathbb{R}}.$
For \(t\in \mathbb{R},\) let $\pi_0(X_t) = \{U_i^t\}_{i\in I}$ be the set of the path-connected components of \(X_t\)
and consider $f:X\rightarrow \mathbb{R}$ continuous. By definition the sets $X_t$ are closed subsets of $X$ and, for \(t \leq t'\), the inclusions $i_t^{t'}:X_t\hookrightarrow X_{t'}$ are continuous maps. Hence
$\pi_0(i_t^{t'}):\pi_0(X_t)\rightarrow \pi_0(X_{t'})$ is such that
$U_i^t\subset\pi_0(i_t^{t'})(U_i^t) $ for all $U_i^t \in \pi_0(X_t)$.

We now introduce a technical definition that will be needed to rigorously develop the next sections. Although somewhat formal, it essentially ensures that the path-connected components of the sublevel sets defined above behave properly.
See \Cref{sec:top_rmk} or \cite{chazal2016structure} for more details.

\begin{defi}[Tameness; \citealp{chazal2016structure}]
\label{defi:tame}
A family of topological spaces $\{X_t\}_{t\in \mathbb{R}}$ with $X_t\subset X_{t'}$, $t\leq t'$, is called tame if:
\begin{itemize}
    \item there exists a finite collection of real numbers $\{t_1<t_2<\ldots<t_n\}$, called critical set, such that, given $t<t',$
if $t,t'\in (t_i,t_{i+1})$ or $t,t'>t_n$, then $\pi_0(i_t^{t'})$ is bijective;
\item for every \(t\in \mathbb{R},\) $\pi_0(X_t)$ is finite.
\end{itemize}

Likewise, a function \(f:X\rightarrow \mathbb{R}\) is said to be tame if its sublevel set filtration $\{X_t\}_{t\in \mathbb{R}}$ is tame. The critical values of $f$ are defined as the critical values of the filtration.
\end{defi}

\subsection{Assumptions on $f:X\rightarrow \mathbb{R}$}

In this work, we address two main sets of assumptions: a more general set, which involves technical details that can disrupt the flow of the narrative, and a simplified set of assumptions that significantly reduce the complexity of introducing some of the upcoming constructions. We now present these sets of assumptions, further dividing the general set into two subcases, and provide commentary on each.

\begin{itemize}
\item[]{General assumptions}
    \begin{itemize}
        \item[(A0)] $f:X\rightarrow \mathbb{R}$ is a tame function  on a path-connected topological space $X$;
        \item[(A1)] $f:X\rightarrow \mathbb{R}$ is a tame  continuous function on a path-connected topological space $X$.
    \end{itemize}
\item[]{Simplified assumptions}
    \begin{itemize}
        \item[(B)] $f:X\rightarrow \mathbb{R}$, with $X=[a,b]\subset \R$ being a compact real interval, is a continuous function presenting only a finite number of local maxima and minima points---which are therefore isolated. To avoid ambiguities we specify that $x$ is an isolated minimum point for $f$ if $f(y)>f(x)$ on some open neighborhood of $x$. Similarly, $x$ is an isolated maximum point for $f$ if $f(y)<f(x)$ on some open neighborhood of $x$.
    \end{itemize}
\end{itemize}

Note that if $f$ satisfies (B), then it also satisfies (A1) and (A0). Clearly, (A1) implies (A0).

In this work, our approach is to present and discuss everything under assumptions (B), simplifying the exposition by minimizing notation and technical details, and introducing our contributions in a more accessible setting. However:
\begin{itemize}
\item \Cref{sec:top_rmk} includes definitions and constructions for building merge trees under assumptions (A0);
\item \Cref{prop:invariance}
and \Cref{teo:stability} are stated under assumptions (A0) and (A1), respectively, ensuring they hold in full generality. Notably, adopting (A0) and (A1) instead of (B) does not require significant changes to the proofs.
\end{itemize}
Additionally, we note that the results in \Cref{sec:estimators} and \Cref{sec:approx} have been proven exclusively for functions satisfying (B).

\subsection{Tree Structures, Critical Values, and Topological Changes}
\label{sec:make_merge_trees}

Coherently with \cite{pegoraro2025finitely}, we now define what we mean with {\em tree} and with \emph{merge tree}.

\begin{defi}
A tree structure $T$ is given by a set of vertices $V_T$ and a set of edges $E_T\subset V_T\times V_T$ which form a connected rooted acyclic graph.  We indicate the root of the tree with $r_T$. We say that \(T\) is finite if \(V_T\) is finite. The degree of a vertex of \(T\) is the number of edges which have that vertex as one of the endpoints.
Any vertex with an edge connecting it to the root is its child and the root is its father: this is the first step of a recursion which defines the father and children relationship for all vertices in \(V_T.\)
The vertices with no children are called leaves  or taxa. The father--child relation induces a partial order on $V_T$: for $v,w\in V_T$ we write $v\leq w$ if $w$ lies on the unique path from $v$ to the root. The edges in $E_T$ are identified in the form of ordered couples $(v,w)$ with $v<w$.
The subtree of a vertex $v$ is the tree structure whose set of vertices is $\{x \in V_T\mid  x\leq v\}$.
\end{defi}

\begin{defi}[\citealp{merge_intrins}; \citealp{pegoraro2025finitely}]\label{defi:merge}
A finite tree structure T such that $r_T$ is of degree $1$, coupled with a monotone increasing height function $h_T:V_T\rightarrow \mathbb{R}\cup\{+\infty\}$ with $h_T(r_T)=+\infty$ and $h_T(v)\in\R$ if $v<r_T$, is called merge tree.
\end{defi}

The heuristic idea behind the construction of the merge tree representation of \(f:X\rightarrow \R\) is that, since along the sequence $\{X_t\}_{t\in \mathbb{R}}$ the path-connected components of $X_t$ can only arise, merge with others, or stay the same, it is  natural to represent this merging structure with a tree structure \(T.\)
However, this tree $T$ would not encode the values of the function at which these changes happen, hence we enrich it by defining
a monotone increasing height function $h_T:V_T\rightarrow \mathbb{R}\cup \{+\infty\}$ encoding them. The reader may look at \Cref{fig:func_tree} to have a visual interpretation of the construction. The height function is given by the dotted red lines.

We work under assumptions (B), and so we have that the \emph{critical values} of $f$ in the sense of \Cref{defi:tame}, correspond to the values of $f$ at its isolated maxima and minima points, excluding local maxima, if any at the extremes of $[a,b]$.
We may refer to local maxima and minima associated to critical values, as the critical points of the function.

As also made clear by \Cref{fig:func_tree}, local maxima and minima points of $f$ are in fact where the connectivity of the sublevel sets changes: at local minima we have the birth of new path-connected components, while at local maxima we have the merging of two path-connected components. In particular, local minima points will be associated to the leaves of the merge tree, while local maxima points to the internal vertices; the height of internal vertices and leaves is given by the respective critical values of $f$.
Roughly speaking, the edges then describe the merging pattern of the path-connected components along the sublevel sets, connecting the critical points and values in the graph of $f$, as in \Cref{fig:func_tree}.

We briefly comment on the case of a function $f:[a,b]\rightarrow\R$ having a local maximum in $a$ or in $b$.
Given a local maximum point $x_0\in (a,b)$, at $f(x_0)$ we have the merging of a path connected component \virgolette{on the left} of $x_0$, that is, of the form $(x_0-\varepsilon^-,x_0)$, for some $\varepsilon^->0$, with one \virgolette{on the right} of $x_0$, that is, of the form $(x_0,x_0+\varepsilon^+)$, for some $\varepsilon^+>0$. Clearly this does not happen if $x_0=a$ or $x_0=b$. In that case no path-connected components merge,
and for this reason we ignore such local maxima.

Let $\{t_1,\ldots,t_n\}$ be the critical values of $f$, listed in increasing order. The tree structure \(T\) and the height function \(h_T\) are built along the following rules, in a recursive fashion  starting from an empty set of vertices $V_T$ and an empty set of edges $E_T$. We simultaneously add points and edges to $T$ and define $h_T$ on the newly added vertices.
From now on, we indicate with $\#C$ the cardinality of a finite set $C.$

Rules for the construction of the merge tree \(T:\)
\begin{itemize}
\item[(i)] For the critical value $t_1$ add to $V_T$ a leaf $v_{U_{t_1}}$, with height $t_1$, for every element $U_{t_1}\in\pi_0(X_{t_1})$. These vertices correspond to the global minima points of $f$;
\item[(ii)] for $i>1$, we add a leaf with height $t_{i}$ for each local minimum point of $f$ with value $t_i$. More formally, for every $U_{t_{i}}\in \pi_0(X_{t_{i}})$ such that $U_{t_{i}}\notin \text{Im}(\pi_0(i_{t_{i-1}}^{t_{i}}))$, add to $V_T$ a leaf $v_{U_{t_{i}}}$ with height $t_{i}$;
\item[(iii)] similarly, for $i>1$, we add an internal vertex with height $t_{i}$ for each local maximum point of $f$ with value $t_i$, connecting the path-connected components that merge at the local maximum. That is, if $U_{t_{i}}=\pi_0(i_{t_{i-1}}^{t_{i}})(U_{t_{i-1}})=\pi_0(i_{t_{i-1}}^{t_{i}})(U'_{t_{i-1}})$, with $U_{t_{i-1}}$ and $U'_{t_{i-1}}$ distinct path connected components in $\pi_0(X_{t_{i-1}})$, add a vertex $v_{U_{t_{i}}}$ with height $t_{i}$, and add edges so that the newly added vertex $v_{U_{t_{i}}}$ is connected with both, previously added, vertices:
\[
v = \arg\max \{h_T(v'_{U})\mid  v'_U \in V_T \text{ s.t. }U\subset U_{t_{i-1}}  \}
\]
\[
w = \arg\max \{h_T(w'_{U})\mid  w'_U \in V_T \text{ s.t. }U\subset U'_{t_{i-1}}  \}.
\]
\end{itemize}

The last merging happens at height $t_n$ and, since $X$ is path connected, at height $t_n$ there is only one point $v_U$. Thus we can add a vertex $r_T$ and an edge $(v_U,r_T)$ with $h_T(r_T)=+\infty$ to obtain a merge tree.
Looking at  \Cref{fig:func_tree}, we can appreciate that the merge tree of $f$ is heavily dependent on the shape of $f$, in particular on the displacement of its maxima and minima points.

\subsection{Isomorphism Classes}\label{sec:iso_class}

If we change the parametrization of a function
$f$ by considering $g= f\circ \eta$, with $\eta$ being a homeomorphism, the path-connected components of $f^{-1}(t)$ and $g^{-1}(t)$ will in general be different sets.
Since the vertex sets of the merge trees of $f$ and $g$ are obtained from the path-connected components of sublevel sets, this implies that, to obtain invariance properties, we must consider merge trees up to relabeling of their vertices. In other words, as in \cite{pegoraro2025finitely}, we don't want to distinguish trees if they differ just by the names of their vertices.

\begin{defi}
Two tree structures $T$ and $T'$ are isomorphic if there exists a bijection $\eta:V_T\rightarrow V_{T'}$ inducing a bijection between the edges sets $E_T$ and $E_{T'}$ such that $(v,w)\mapsto (\eta(v),\eta(w))$. Such $\eta$ is an isomorphism of tree structures.
\end{defi}

\begin{defi}\label{defi:iso_class}
Two merge trees $(T,h_T)$ and $(T',h_{T'})$ are isomorphic if $T$ and $T'$ are isomorphic as tree structures and the isomorphism $\eta:V_T\rightarrow V_{T'}$ is such that $h_T = h_{T'} \circ \eta$. Such $\eta$ is an isomorphism of merge trees.
\end{defi}

\subsection{Invariance Property} %

In this section, we state the invariance property anticipated in the introduction.

\begin{prop}[Invariance]
\label{prop:invariance}
The isomorphism class of the merge trees of a function $f:X\rightarrow \mathbb{R}$ satisfying (A0), is invariant under any homeomorphic re-parameterization of $f$.
\end{prop}

\begin{rmk}
As an immediate consequence of  \Cref{prop:invariance} we obtain that, if the functions $f$ and $g$ can be aligned by means of a homeomorphism, that is if $f = g \circ \eta$ being $\eta$ a homeomorphism, then their associated merge trees $T_f$ and $T_g$ are isomorphic.
\end{rmk}

In other words, we can warp, deform, move the domain \(X\) of a function \(f\) by means of any homeomorphism, and this will have no effect on its associated merge tree.
As a consequence, if each element of a sample of functions is represented by its merge tree, one can carry out the statistical analysis without worrying about possible misalignments, that is without first singling out, for each function of the sample, the specific warping function, identified by a homeomorphism, which decouples its phase and amplitude variabilities.

\subsection{The Metric for Merge Trees}

The metric for weighted graphs introduced in \cite{pegoraro2023persistence} and later adapted to merge trees in \cite{pegoraro2025finitely} is based on edit distances \citep{survey_ted, TED}.
Edit distances quantify the minimal cumulative cost of transforming one object into another through a sequence of allowed modifications.
To fit merge trees into this framework, we first equip each tree with a suitable weight function \(w_T\).

\subsubsection{Height and Weight Functions}
\label{sec:merge_and_weight}

Our goal here is to associate each merge tree with a positive weight function, thus turning it into a \emph{positively weighted tree} through an appropriate transformation.
To this end, we complement a merge tree \(T\) with a weight function \(w_T\) derived from its height function \(h_T\).
The function \(w_T\) is defined on \(V_T \setminus \{r_T\}\) and takes values in the space \(\mathbb{R}_{> 0}\).

Following \cite{pegoraro2025finitely}, we employ a \emph{truncation strategy} to handle the edge \((v, r_T)\) extending to infinity.
This approach relies on the following

\begin{itemize}
\item[(C)]\label{assump:K}
\emph{Auxiliary assumption:}
We assume the existence of a universal constant $K\in\R$ bounding above all the functions for which we will adopt a merge tree representation.
\end{itemize}

In \cite{pegoraro2025finitely}, it is demonstrated that all subsequent steps in constructing the specific merge trees discussed in this paper are independent of \(K\). That is, using any \(K' > K\) would yield the same results. This point is further elaborated in the following \Cref{rmk:K}.

\begin{figure}
	\begin{subfigure}[c]{0.49\textwidth}
    	\centering
    	\includegraphics[width = \textwidth]{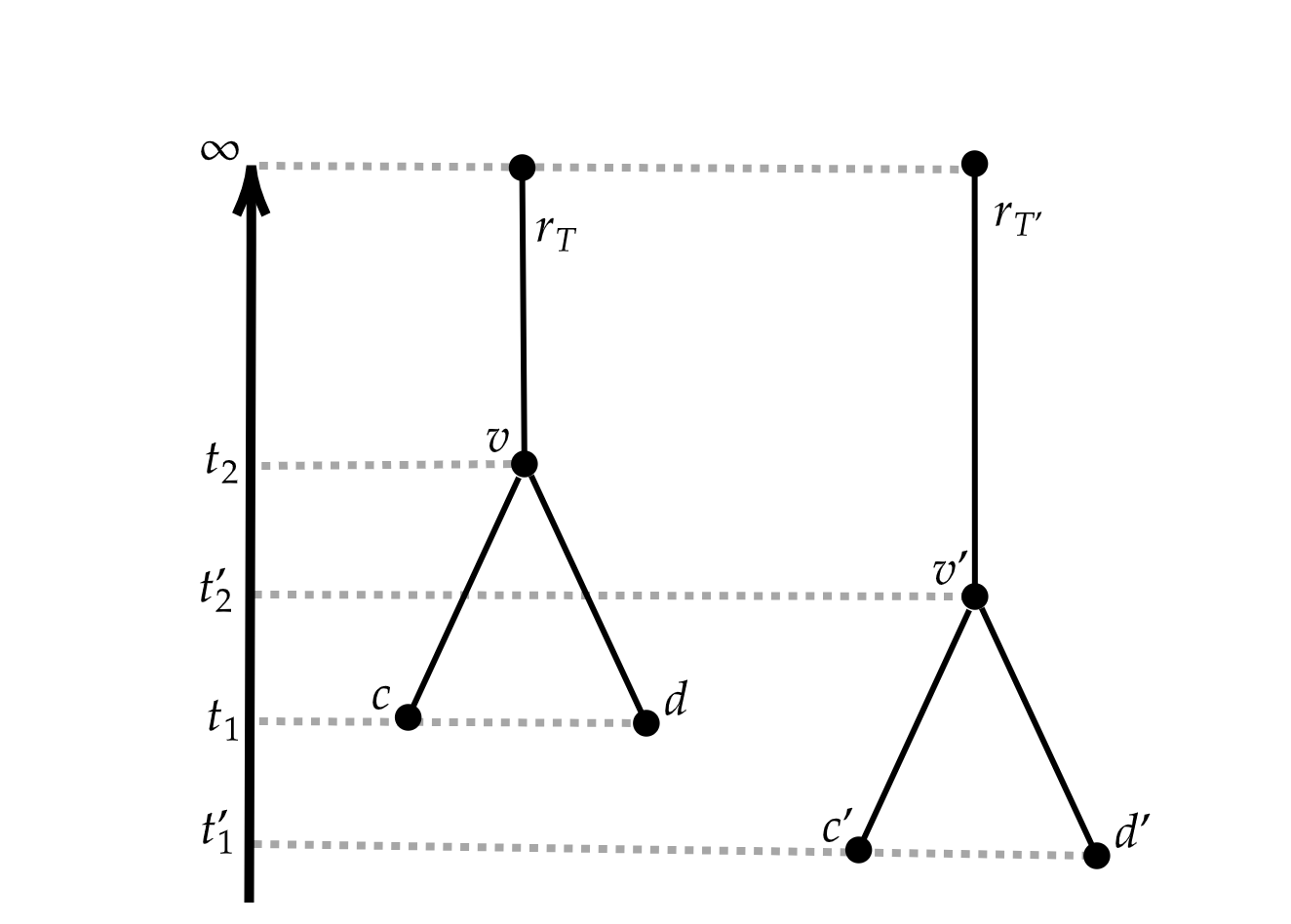}
    	\caption{Two merge trees $T$ and $T'$.}
    	\label{fig:truncate_1}
    \end{subfigure}
   	\begin{subfigure}[c]{0.49\textwidth}
    	\centering
    	\includegraphics[width = \textwidth]{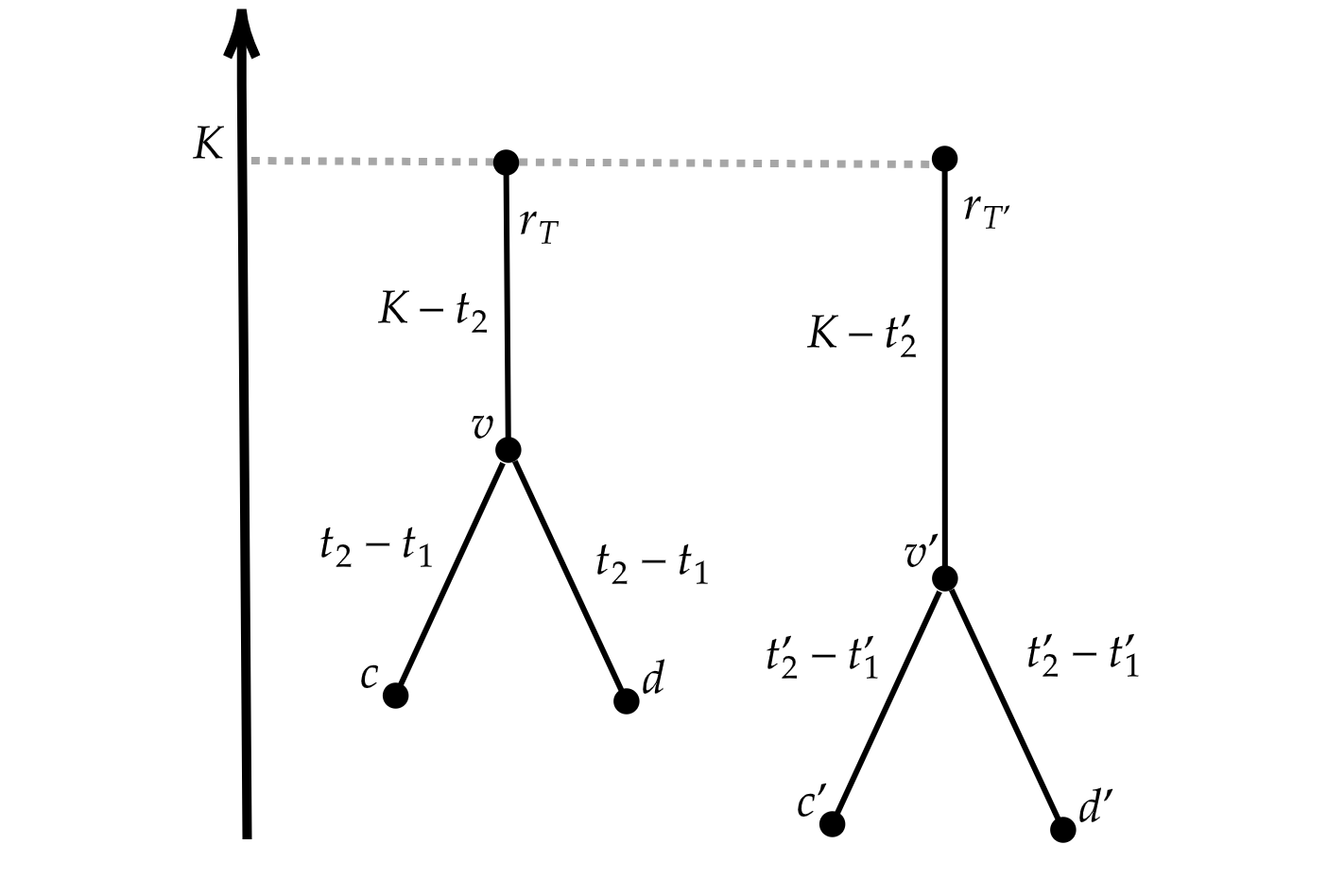}
    	\caption{A graphical representation of the weight functions obtained by truncating the merge trees at height $K$.}
    	\label{fig:truncate_2}
    \end{subfigure}

\caption{A graphical representation of the truncation process described in \Cref{sec:merge_and_weight}.}
\label{fig:truncation}
\end{figure}

Given a merge tree $(T,h_T)$, as a first step we define the function $h'_T:V_T\rightarrow \R$ as $h'_T(v)=h_T(v)$ for all $v<r_T$ and $h'_T(r_T)=K$. Then for every vertex $v \in V_T \setminus \{r_T\},$ we consider the unique edge between \(v\) and its father \(w\) and we define $w_T(v)= h'_T(w)-h'_T(v);$ the weight function \(w_T\) also codes the weight of the edge \((v,w)\), via the rule $w_T((v,w))=w_T(v)$, which identifies the set of edges $E_T$ with vertices in $V_T \setminus\{r_T\}$. Note that, because of (C),
there is a one-to-one correspondence between \(h_T\) and \(w_T.\)  Finally, the monotonicity of \(h_T\) and (C)
guarantee that \(w_T(v) \in \mathbb{R}_{\geq 0},\) for all \(v \in V_T \setminus\{r_T\}.\) See \Cref{fig:truncation} for a visual example.

From now on, the description of a merge tree will always involve both its height function \(h_T\) and the associated weight function \(w_T\).

\subsubsection{Edit Distance}

The space of merge trees can now be endowed with a metric based on the edit distance in \cite{pegoraro2025finitely}, denoted by \(d_E\).

The distance $d_E$ is very different from previously defined edit distances, since it is specifically designed for comparing topological summaries, roughly meaning that all vertices of a merge tree which are topologically irrelevant can be disregarded when computing $d_E.$ To make things more formal, we here introduce the edits, as defined in \cite{pegoraro2025finitely}. They can be used to modify any edge $(v,v')$ of a merge tree, or equivalently its lower vertex $v$:

\begin{itemize}

\item {\em Shrinking} an edge means changing the weight value of the edge with a new positive value.
The inverse of this transformation is the shrinking which restores the original edge weight.

\item {\em Deleting} an edge $(v_1,v_2)$ results in a new tree, with the same vertices apart from $v_1$ (the lower one), and with the father of the deleted vertex which gains all of its children.
With a slight abuse of language, we might also refer to this edit as the deletion of the vertex $v_1$, which indeed means deleting the edge between \(v_1\) and its father.

The inverse of deletion is the {\em insertion} of an edge along with its child vertex.
We can insert an edge at a vertex \(v\) specifying the child of \(v\) and its children (that can be either none or any portion of the children of \(v\)) and the weight of the edge.

\item Lastly, we can eliminate a degree two vertex \(v\), that is a father with exactly one child, connecting the two adjacent edges which arrive and depart from \(v.\) The weight of the resulting edge is the sum of the weights of the joined edges.
This transformation is the {\em ghosting} of the vertex $v$.
Its inverse transformation is called the {\em splitting} of an edge.
\end{itemize}

\begin{figure}
	\centering
	\begin{subfigure}[c]{0.47\textwidth}
    	\centering
    	\includegraphics[width = \textwidth]{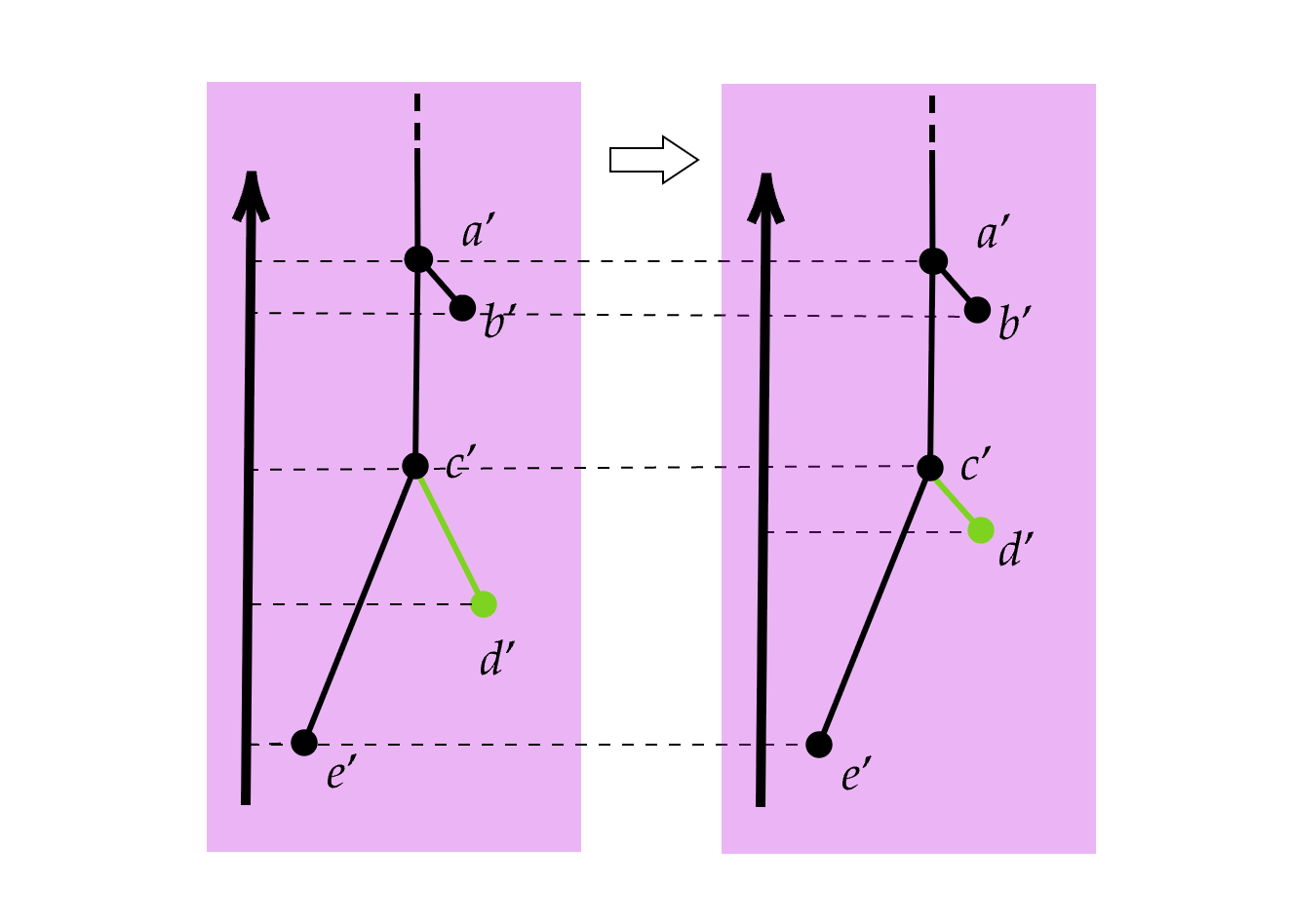}
    	\caption{Shrinking of $d'$.}
    	\label{fig:shrinking_leaf}
    \end{subfigure}
    \begin{subfigure}[c]{0.47\textwidth}
    	\centering
    	\includegraphics[width = \textwidth]{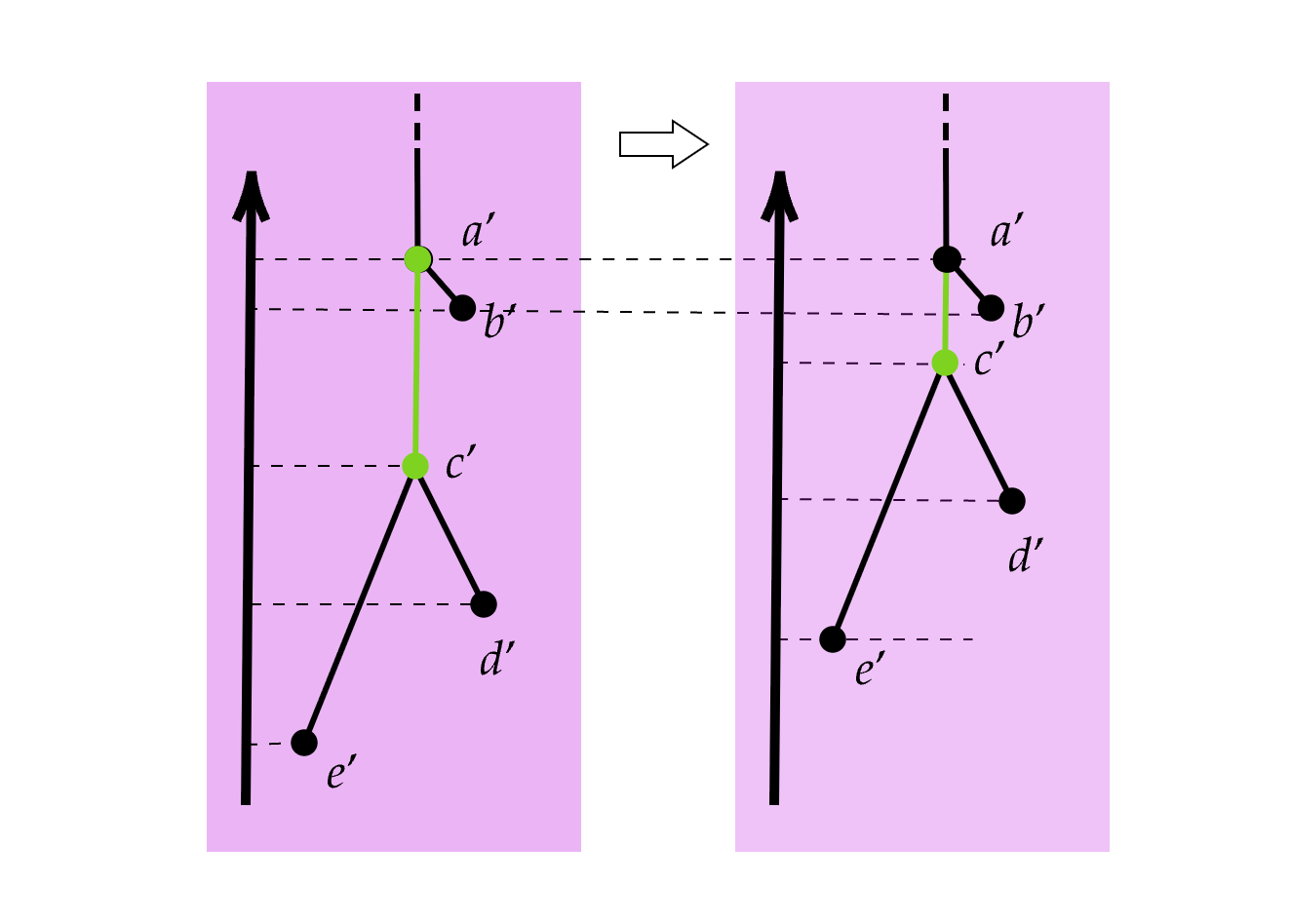}
    	\caption{Shrinking of $c'$.}
    	\label{fig:shrinking_int}
    \end{subfigure}

    \begin{subfigure}[c]{0.47\textwidth}
    	\centering
    	\includegraphics[width = \textwidth]{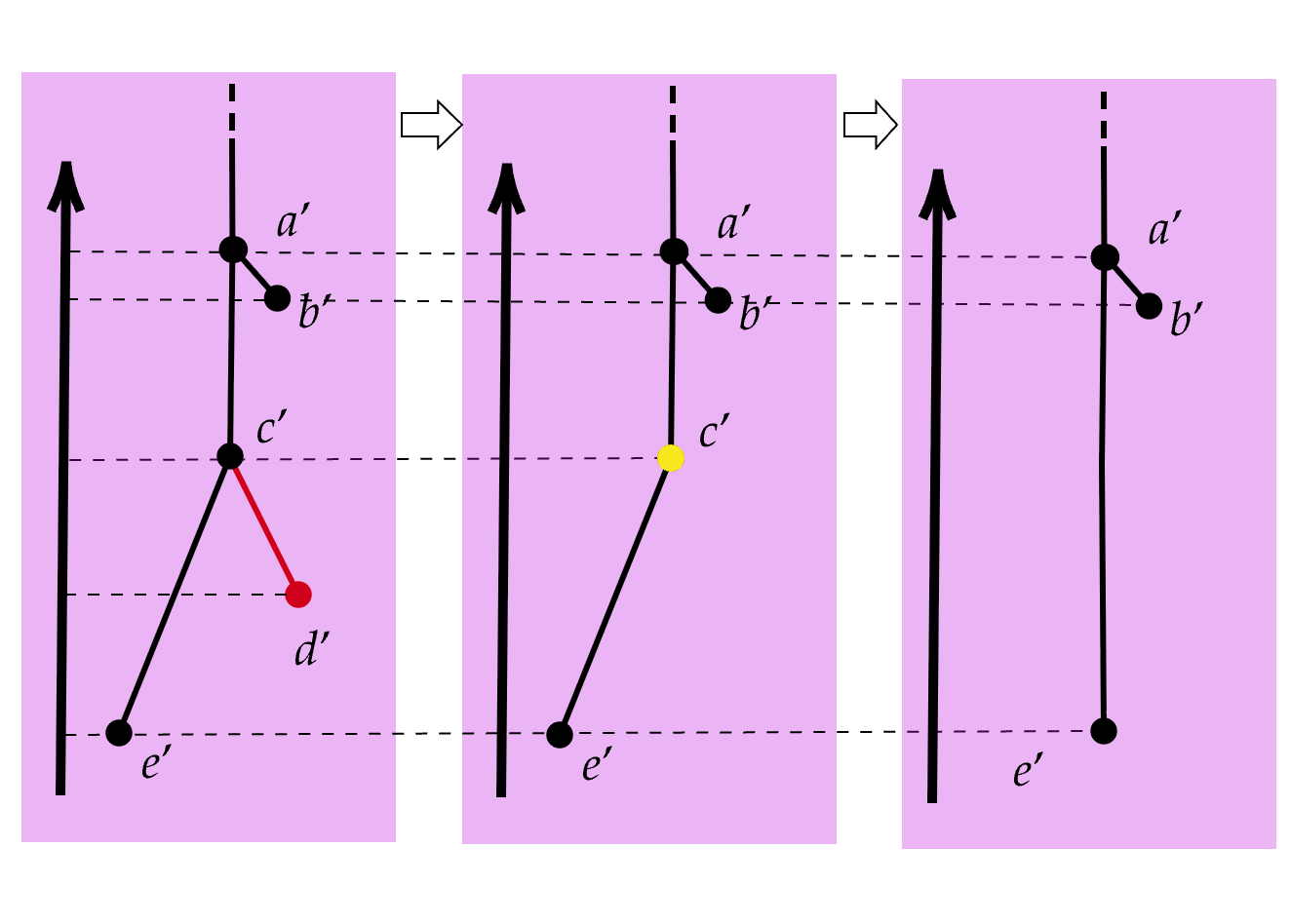}
    	\caption{Deletion of $d'$ and ghosting of $c'$.}
    	\label{fig:deletion_leaf}
    \end{subfigure}
	\begin{subfigure}[c]{0.47\textwidth}
		\centering
		\includegraphics[width = \textwidth]{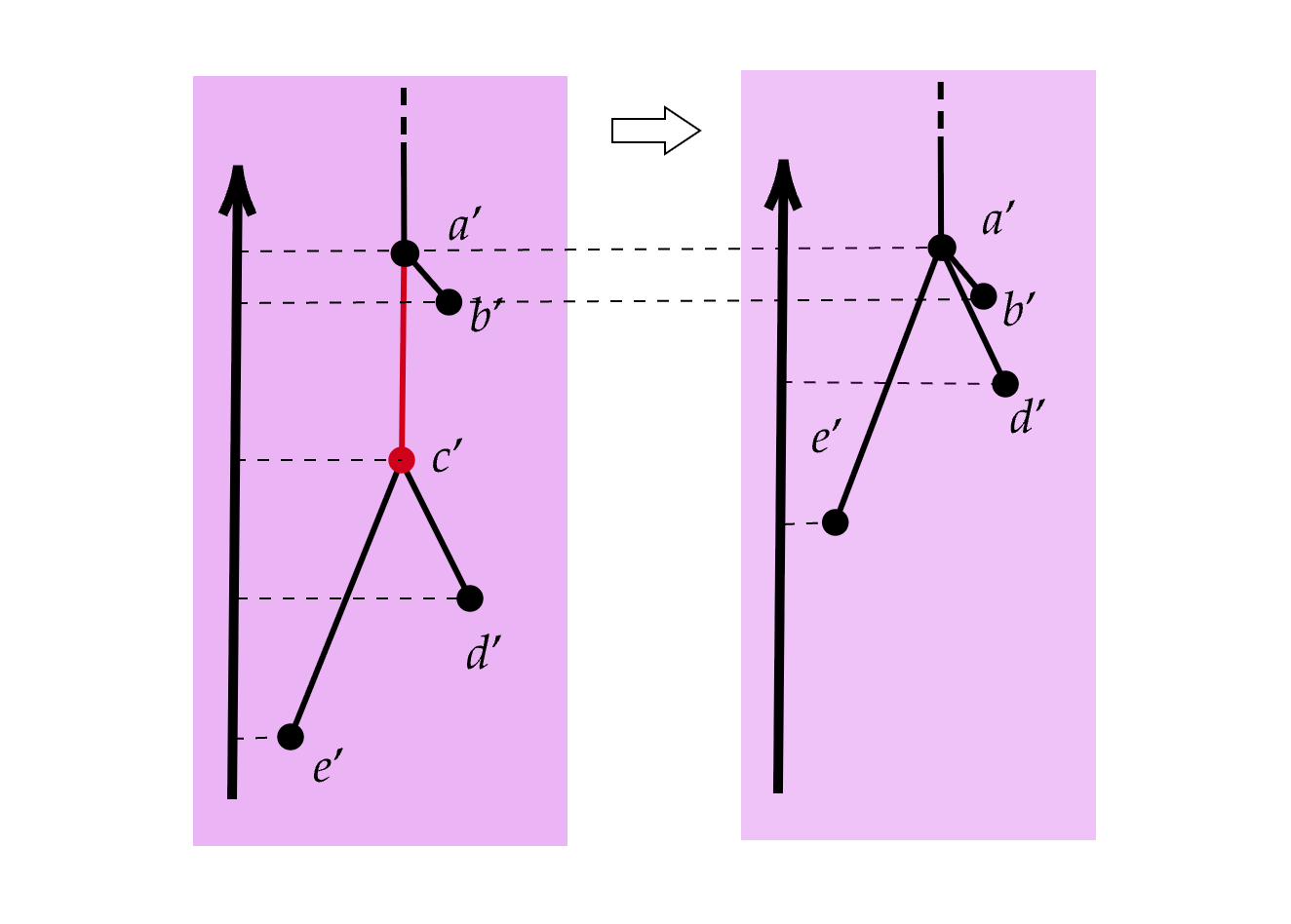}
		\caption{Deletion of $c'$.}
		\label{fig:deletion_int}
	\end{subfigure}
    \hfill
\caption{Visual interpretation of shrinkings, deletions, and ghostings.}
\label{fig:edits}
\end{figure}

\begin{figure}
	\centering
	\begin{subfigure}[c]{0.47\textwidth}
    	\centering
    	\includegraphics[width = \textwidth]{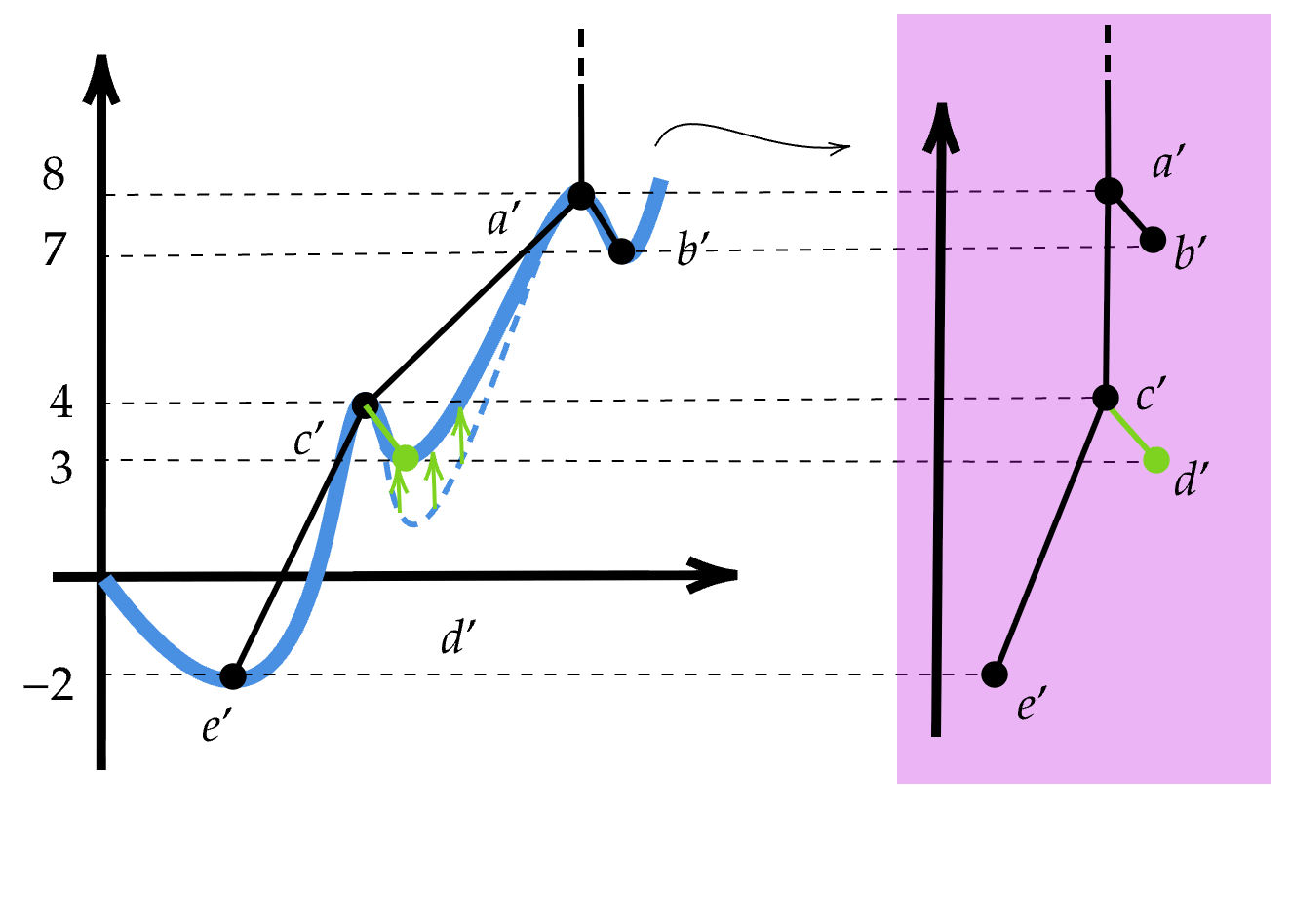}
    	\caption{Qualitative functional interpretation of the edit in \Cref{fig:shrinking_leaf}.}
    	\label{fig:f_shrinking_leaf}
    \end{subfigure}
    \hspace{0.04\textwidth}
    \begin{subfigure}[c]{0.47\textwidth}
    	\centering
    	\includegraphics[width = \textwidth]{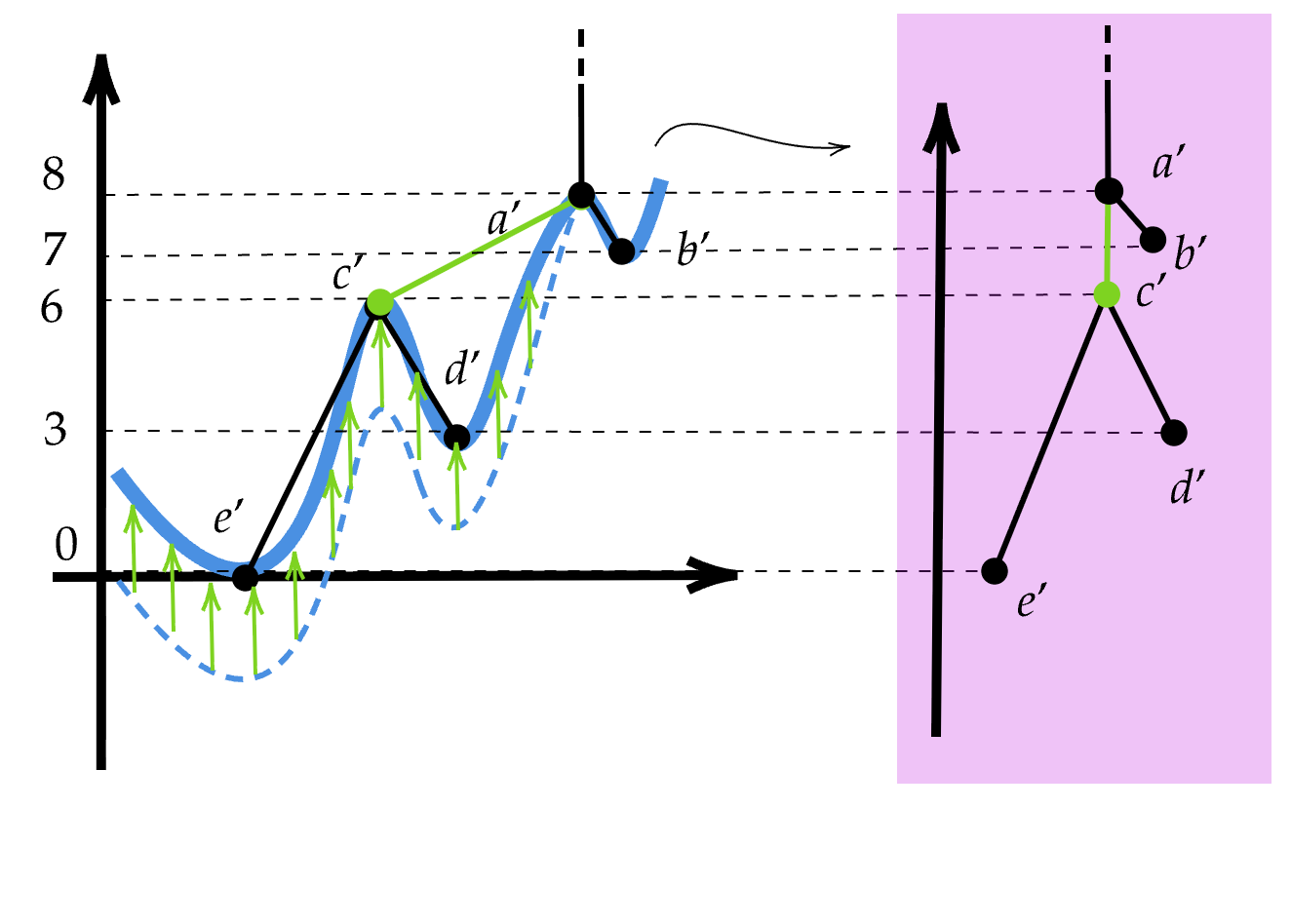}
    	\caption{Qualitative functional interpretation of the edit in \Cref{fig:shrinking_int}.}
    	\label{fig:f_shrinking_int}
    \end{subfigure}

    \medskip

    \begin{subfigure}[c]{0.47\textwidth}
    	\centering
    	\includegraphics[width = \textwidth]{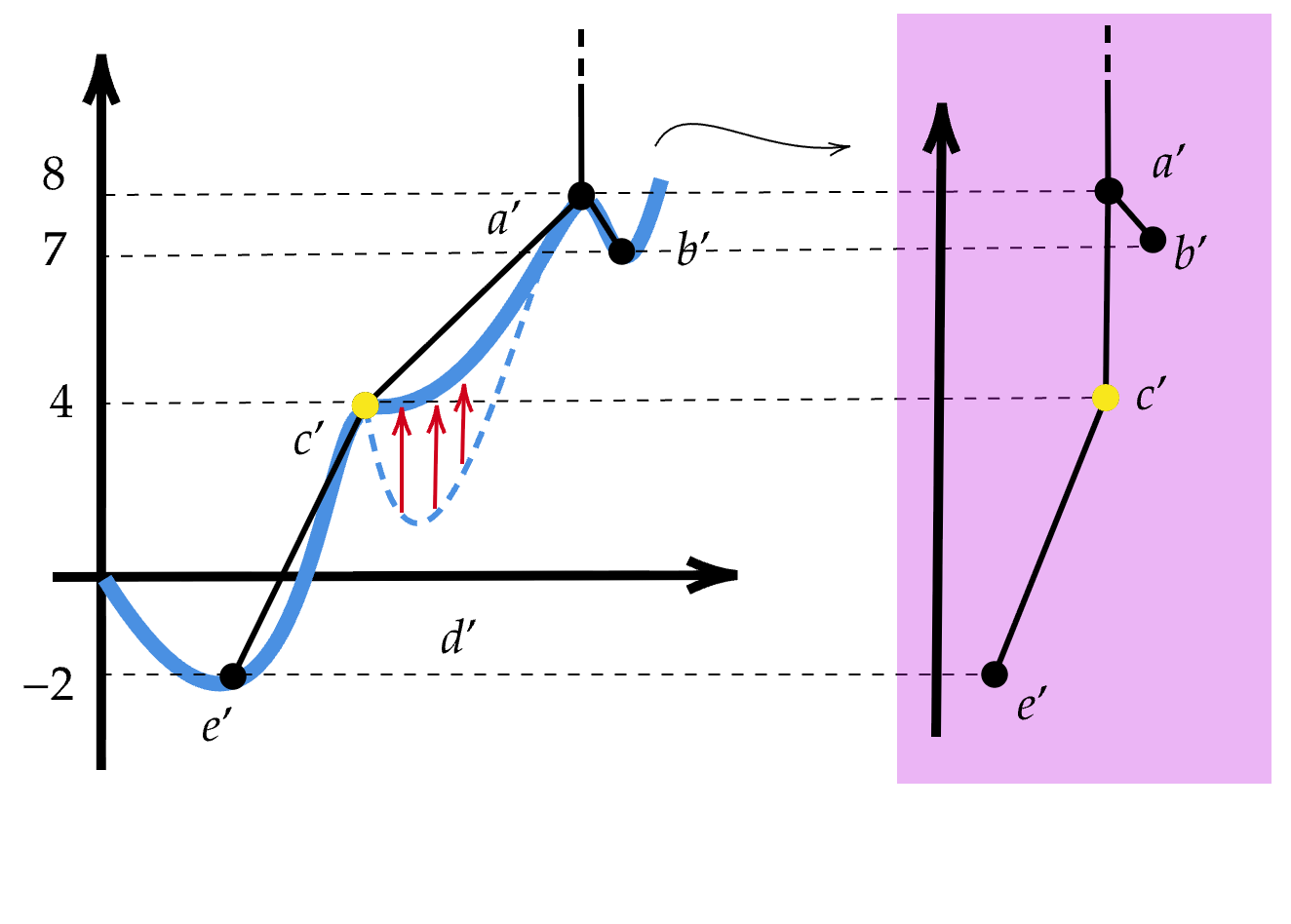}
    	\caption{Qualitative functional interpretation of the edits in \Cref{fig:deletion_leaf}.}
    	\label{fig:f_deletion_leaf}
    \end{subfigure}
    \hspace{0.04\textwidth}
	\begin{subfigure}[c]{0.47\textwidth}
		\centering
		\includegraphics[width = \textwidth]{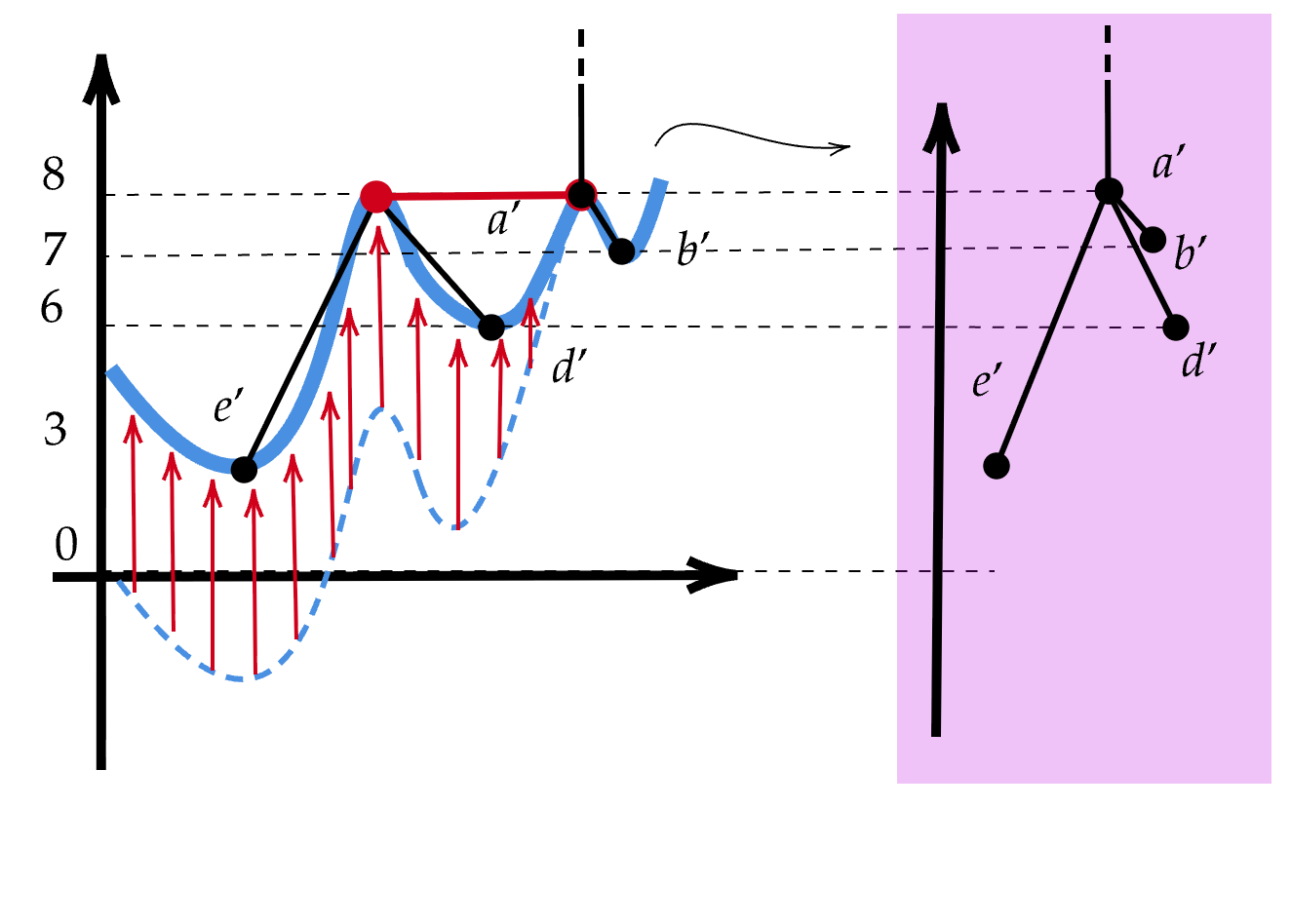}
		\caption{Qualitative functional interpretation of the edit in \Cref{fig:deletion_int}.}
		\label{fig:f_deletion_int}
	\end{subfigure}

\caption{Representation of functions which reflect the edits depicted in \Cref{fig:edits}.}
\label{fig:f_edits}
\end{figure}

The reader may refer to \Cref{fig:edits} for a visual interpretation of these edits, and to \Cref{fig:f_edits} for their interpretation in functional terms: for each edit in \Cref{fig:edits}, we provide functions that reflect the corresponding modifications of the merge trees. Note that the choice of such functions is not unique.

The costs of the edit operations are defined as follows:
\begin{itemize}
\item the cost of shrinking an edge is equal to the absolute value of the difference of the two weights;
\item for any deletion/insertion, the cost is equal to the weight of the edge deleted/inserted;
\item the cost of ghosting or splitting is zero.
\end{itemize}

Given a tree $T$ we can edit it, thus obtaining another tree, on which we can apply a new edit to obtain a third tree and so on.
Any finite composition of edits is called an {\em edit path}.
The cost of an edit path is the sum of the costs of its edit operations.
Putting all the pieces together, we can define the edit distance $d_E$  as:
\[
d_E(T,T')=\inf_{\gamma\in\Gamma(T,T')} cost(\gamma)
\]
where $\Gamma(T,T')$ indicates the set of edit paths which start in $T$ and end in $T'$.

\begin{rmk}
Edit operations are not globally defined as operators mapping merge trees into merge trees. They are defined on the individual tree. Similarly, their inverse is not the inverse in the sense of operators, but it indicates that any time we travel from a tree $T$ to a tree $T'$ by making a sequence of edits, we can also travel the inverse path going from $T'$ to $T$ and restore the original tree.
\end{rmk}

\begin{rmk}\label{rmk:K}
The results in \cite{pegoraro2025finitely} show that
the metric $d_E(T,T')$ does not depend on the value $K$ used in the truncation process and introduced with the Auxiliary Assumption (C).
To be sure, if, after having fixed $K$ to analyze a set of merge trees \({\cal T}\), we add to the set a new merge tree corresponding to a function $f$ which is not bounded above by \(K\), we proceed by fixing a novel $K'$ bounding \(f\) and all the other functions represented in the set \({\cal T}\), and compute $d_E(T_f,T_g)$, for all \(g\in {\cal T},\) after truncating  all merge trees in \({\cal T}\bigcup \{T_f\}\) at height $K'.$ This won't affect  the distances between merge trees in \({\cal T}\) computed before the addition of \(T_f,\) when the truncation constant was $K$, since the metric $d_E$ is the same for such merge trees. In this sense (C)
is unnecessary, since we do not need to fix $K$ uniformly for the entire dataset but only for pairs of data. However for our applications such assumption is never violated, so we can assume it and avoid some formal complications arising from having to fix $K$ for every couple of functions.
\end{rmk}

\begin{rmk}
The null cost of ghosting guarantees that
degree $2$ vertices are completely irrelevant when computing the cost of an edit path.
In \cite{pegoraro2025finitely} it is proved that $d_E$ is a metric on the space
of merge trees, identified up to degree $2$ vertices.
As explained in \cite{pegoraro2025finitely}, the fact that degree $2$ vertices are irrelevant is precisely what makes the metric $d_E$ suitable for comparing merge trees and is fundamental to obtain the results in \Cref{sec:stability}.
\end{rmk}

\section{Persistence Diagrams}
\label{sec:vs_PD}

Persistence diagrams are arguably among the most well-known tools in TDA; for a detailed survey, see, for instance \citep{PH_survey}.
We briefly recall their definition here, as they will be used in the following sections to draw comparisons with merge trees. See \Cref{sec:examples_vs_PD} for a more detailed comparison and examples.

Loosely speaking, a persistence diagram is a collection of points \((c_x, c_y)\) in the first quadrant of \(\mathbb{R}^2\), with \(c_y > c_x\), where \(c_x\) represents the parameter value \(t\) at which a homology class in \(X_t\) first appears (its \emph{birth}), and \(c_y\) is the value of \(t\) at which the same class merges with a previously born class (its \emph{death}).
Homology classes generalize path-connected components to ``holes'' in higher dimensions; path-connected components can be seen as zero-dimensional holes.
For further details, we refer to \cite{hatcher}.

\begin{figure}
	\centering
	\begin{subfigure}[t]{0.32\textwidth}
    	\centering
    	\includegraphics[width = \textwidth]{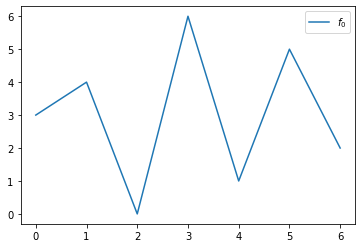}
    \end{subfigure}
	\begin{subfigure}[t]{0.32\textwidth}
    	\centering
    	\includegraphics[width = \textwidth]{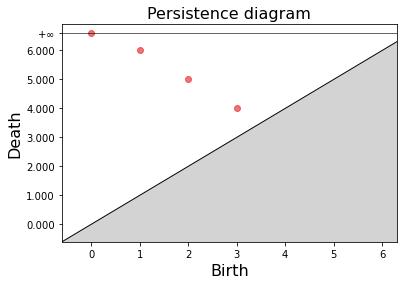}
    \end{subfigure}
	\begin{subfigure}[t]{0.32\textwidth}
    	\centering
    	\includegraphics[width = \textwidth]{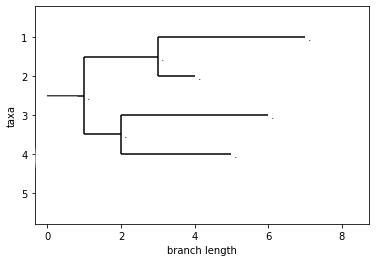}
    \end{subfigure}

\caption{A function (left) with its associated persistence diagram (center) and merge tree (right).  On the PD axes we see the birth and death coordinates of its points. The plot of the merge tree features the length of its branches (given by the weight function - \Cref{sec:merge_and_weight}) on the horizontal axis, and the leaves (taxa) are displaced on the vertical axis. The vertical axis scale is only for visualization purposes. The merge tree is truncated at height $7$ - see \Cref{sec:merge_and_weight}.}
\label{fig:PD_intro}
\end{figure}

In this work, we focus on PDs associated to path-connected components, since we want to compare them with the merge trees introduced in the previous section. Given a function $f:X\rightarrow \mathbb{R},$ we associate to \(f\) the zero-dimensional persistence diagram ($PD(f)$) of the sequence of sublevel sets  $\{X_t\}_{t\in\mathbb{R}}$.
We highlight that, in such representation, in general, there is no information about which path-connected component merges with which; in fact a component represented by the point $(c_x,c_y)$, at height $c_y$ could merge with any of the earlier born and \virgolette{still alive} components.
Of course, this collection of points still depends on the shape of the function and in particular depends on its amplitude and the number of its oscillations (see \Cref{fig:PD_intro}).
Note that, while for merge trees one needs to be careful and consider appropriate isomorphism classes so that the representation does not depend, for instance, on the names chosen for the vertices (that is, the set $V_T$), this issue does not appear with PDs. Topological features are represented as points in the plane, without labels or other kinds of set-dependent information. Thus, two PDs are isomorphic if and only if they are made of the same set of points.
Moreover, since PDs can be completely reconstructed from merge trees, if two functions induce isomorphic merge trees, they also have the same PDs, see \Cref{prop: from_merge_to_PD} in the supplementary. This also implies that the invariance property of merge trees extends to zero dimensional PDs.

\subsection{Metrics for Persistence Diagrams}

The space of PDs can be given a metric structure by means of a family of metrics which derives from Wasserstein distances for bivariate distributions.

Given two diagrams $D_1$ and $D_2$, the expression of such metrics is the following:
\[
W_p(D_1,D_2) = \left( \inf_\gamma  \sum_{x \in D_1} \parallel x-\gamma(x)\parallel^p_\infty \right)^{1/p}
\]
where \(p \geq 1\) and $\gamma$ ranges over the functions partially matching points between diagrams $D_1$ and $D_2$, and matching the remaining points of both diagrams with the line $y=x$ on the plane (for details see \citealp{cohen_PD}). In other words, we measure the distances between the points of the two diagrams, pairing each point of a diagram either with a point on the other diagram, or with a point on $y=x$. Each point can be matched once and only once. The minimal cost of such matching provides the distance.

\section{Stability, Estimators, and Approximations}
\label{sec:stability}

\subsection{Stability}
\label{sec:stab}

As stated in the introduction of the paper, any time we use a data representation -- or we further transform a representation -- it is important to understand and explore the properties of the operators involved. In particular, in this section we want to establish some continuity properties for the operator $f\mapsto T_f$, which maps a function to its merge tree.
Conditional on the topology endowing the functional space where the function \(f\) is embedded, these properties dictate
how the variability between functions is captured by the variability between their merge tree representations.

 \Cref{prop:invariance} implies that the merge tree representation of a function \(f\) is unaffected by a large class of warpings of its domain, which would strongly perturb $f$ if it was embedded, for instance, in an $L_p$ space, with $p\neq \infty$. As an example, if $f:\mathbb{R}\rightarrow \mathbb{R}$ is bounded with compact support, shrinking $f$ by setting $f_n(x)=f(x\cdot \lambda_n)$ with $\lambda_n\rightarrow + \infty$, produces no effect on the merge tree representation of \(f\) since $T_{f_n}=T_f$, while the $p$-norm of $f_n$ goes to zero.

It might therefore be more natural to study the behavior of $f\mapsto T_f$ endowing the space of functions \(f:X \rightarrow \mathbb{R}\) with the topology of uniform convergence, which captures pointwise closeness between functions. This topology, available for any domain $X$, has also the advantage of showing the effect of pointwise noise on merge tree representations. For these reasons, stability results in TDA are often stated via the sup norm of functions.

The main result of this section is the following.

\begin{teo}
\label{teo:stability}
Let $f,g:X\rightarrow \R$ be functions satisfying (A1) and such that $$\text{sup}_{x\in X}\mid f(x)-g(x)\mid \leq \varepsilon.$$
Let $T_f$ and $T_g$ be the merge trees associated to $f$ and $g$ respectively and let $N_f=\# E_{T_f}$ and $N_g = \# E_{T_g}$.

Then, there exists an edit path $e_1\circ\cdots\circ e_{N_f + N_g} \in \Gamma(T_f,T_g)$ which edits each non-root vertex at most once and such that $cost(e_i)<2\cdot \varepsilon,$ for \(i=1,...,N_f+ N_g.\)
\begin{proof}
We now provide a qualitative sketch of the proof of this result.

\begin{itemize}
    \item We construct two maps between the merge trees using the hypothesis $\text{sup}_{x\in X}\mid f(x)-g(x)\mid \leq \varepsilon$. Such maps arise naturally
    using the framework of the interleaving distance between merge trees \citep{merge_interl}, exploiting, at each $t$, the inclusions $f^{-1}((-\infty,t])\subseteq g^{-1}((-\infty,t+\varepsilon])$ and $g^{-1}((-\infty,t])\subseteq f^{-1}((-\infty,t+\varepsilon])$. Concretely, the first inclusion yields a map
$\pi_0\bigl(f^{-1}((-\infty,t])\bigr)\longrightarrow \pi_0\bigl(g^{-1}((-\infty,t+\varepsilon])\bigr)$ (\Cref{sec:path_conn}),
which sends a point of $T_f$ at height $t$ to a point of $T_g$ at height $t+\varepsilon$ (a merge tree parameterizes the path-connected components of sublevel sets across heights, \Cref{sec:make_merge_trees}). The same is done reversing the role of $f$ and $g$ to obtain the second map.
    \item We then fix one tree as the source and the other as the target. Exploiting the interleaving maps and a key technical condition that guarantees an upper bound on the cost of each edit, we identify the leaves of the source tree that we wish to retain and map them into the target tree. We show that any vertex in the source tree that does not lie above one of the retained leaves, and any vertex in the target tree that does not lie above the image of such leaves, can be deleted at cost at most $2\varepsilon$. We therefore remove all such vertices via deletions.
    \item We further simplify the pruned trees obtained after these deletions by removing all degree-two vertices (via ghosting).
    \item
    Next, our aim is to pair edges across the two trees via the interleaving maps and adjust the weight of each edge in the source tree to match the weight of its counterpart in the target tree by means of shrinking edits. To achieve this, we first couple the leaves of the source tree with vertices of the target tree using the interleaving maps, and then delete all internal vertices that obstruct a coherent matching between the least common ancestors of already matched leaves. We prove that the cost of each deletion introduced at this stage is also bounded by $2\varepsilon$.
    \item Finally, we complete the matching between the remaining vertices/edges in a way that respects the tree structures. The resulting pairs correspond to the shrinking edits needed to adjust the edge weights of the source tree to those of the target tree, each with cost at most $2\varepsilon$.
\end{itemize}

\end{proof}
\end{teo}

As a result of \Cref{teo:stability}, we have the following corollary.

\begin{cor}
    In the setting of \Cref{teo:stability}, the following holds:
    \begin{equation}\label{eq:stability}
        d_E(T_f,T_g)\leq 2 (N_f+ N_g) \parallel f-g \parallel_\infty.
    \end{equation}
\end{cor}

We now discuss several consequences of \Cref{teo:stability}, informed by the theoretical analyses and numerical experiments presented in \cite{pegoraro2025finitely}.

\begin{rmk}[Interleaving Distance]
The proof of \Cref{teo:stability} in fact yields a stronger statement than the one explicitly formulated: it shows that
\[
d_E(T_f,T_g)\leq 2 (N_f + N_g)\, d_I(T_f, T_g),
\]
where $d_I$ denotes the interleaving distance. In particular, for merge trees $T_f$ and $T_g$ induced by functions $f$ and $g$, the inequality
\[
d_I(T_f, T_g) \leq \| f - g \|_\infty
\]
\citep{merge_interl} immediately implies the bound in \eqref{eq:stability}.
\end{rmk}

\begin{rmk}[Finite Stability]
\Cref{teo:stability} states that if two functions are pointwise close, then the merge tree associated with one can be transformed into that of the other through at most one edit per vertex, each of small cost.
However, when the functions exhibit a large number of oscillations, the distance between their merge trees may still be substantial.
Indeed, even if $\|f_n - f\|_\infty \xrightarrow{n} 0$, \eqref{eq:stability} does not guarantee that $d_E(T_f, T_{f_n}) \to 0$.
Nonetheless, \Cref{teo:stability} implies that if the cardinalities $\#E_{T_{f_n}}$ remain uniformly bounded, then $d_E(T_f, T_{f_n}) \to 0$.
This property is referred to as \emph{finite stability} in \cite{pegoraro2025finitely}.
Such behavior contrasts with the Lipschitz-type stability enjoyed, for instance, by the bottleneck distance between PDs or the interleaving distance between merge trees.
At the same time, it parallels the stability properties of the $1$-Wasserstein distance between PDs, which has proven more suitable for data analysis, particularly when combined with appropriate smoothing or denoising techniques such as those discussed in \Cref{sec:estimators}.
A more extensive theoretical and empirical discussion of this phenomenon, including comparisons with the interleaving distance under varying noise levels, is provided in \cite{pegoraro2025finitely}.
\end{rmk}

\begin{rmk}[Gromov--Wasserstein]\label{rmk:GH}
Very recently, \cite{li2025flexible} introduced a framework for comparing merge trees via the Gromov--Wasserstein distance and established a stability result that is comparable to ours. In their setting, \(f\) and \(g\) are defined on (the geometric realization of) a finite, connected mesh (or, more generally, a simplicial complex) \(X\) with vertex set \(V\). They encode these functions as merge trees \(G_f\) and \(G_g\), constructed in a slightly different way from our trees. For \(q=1\) and \(p\) the uniform distribution on \(V\), they define
\[
\|f-g\|_{L_1(p)} := \frac{1}{|V|}\sum_{v\in V} |f(v)-g(v)| \leq \|f-g\|_\infty,
\]
which leads to the stability bound
\[
d^{GW}_1(G_f,G_g) \leq \frac{1}{2}\,|V|\,\|f-g\|_\infty.
\]

Recasting our result in their setting, we can bound the number of vertices in the merge trees \(T_f\) and \(T_g\) by \(|V|\), the cardinality of the vertex set. This yields
\[
d_E(T_f,T_g) \leq 4\,|V| \,\|f-g\|_\infty.
\]

We emphasize that, in our framework, \(|V|\) is only a coarse upper bound on the number of edges in the merge trees, whereas in \cite{li2025flexible} both merge trees share the same vertex set \(V\), so the bound depends explicitly on \(X\). In contrast, our formulation produces bounds that are independent of the underlying complex. For example, it is not clear how their bound could be adapted to derive asymptotic results, as we do in later sections.

Furthermore, \cite{pegoraro2025finitely} shows that \(d_E\) is at least as discriminative as the interleaving distance (i.e., \(d_I \le d_E\)). Whether a similar relation holds for the Gromov--Wasserstein distance remains an open question. We conjecture that it does not, since the probability measure on \(V\) attenuates discrepancies as the trees grow, causing differences in subtrees to be downweighted, unlike what happens for \(d_E\) or \(d_I\).
\end{rmk}

\subsection{Merge Tree Approximations}\label{sec:approx}

We now discuss practical aspects of merge trees and of the edit distance $d_E$, with an emphasis on (i) how the merge tree of a given function can be computed or approximated, and (ii) how such approximation behaves under a standard generative model.

When the function $f$ is fully known, computing its merge tree exactly would in principle require locating and classifying all critical points of $f$ (local minima and maxima). For general functions, this is delicate. We therefore adopt a simpler and efficient approximation scheme based on sampling $f$ on a grid and taking its piecewise linear interpolant.

To formalize the resulting error bounds, we assume mild regularity of $f$  and briefly recall the function spaces that will be used. Specifically, we work with Sobolev spaces $H^p([a,b])$ and with functions of bounded variation $\BV([a,b])$. Since our arguments involve only continuously differentiable functions, or functions that are $C^1$ away from finitely many points, we do not need to dwell on the theory of weak derivatives. In this setting, weak derivatives coincide with the classical derivatives wherever they exist, up to the usual Lebesgue a.e.\ identification (see \citealp{quarteroni2008numerical}). If $f$ satisfies assumptions (B), its degree of smoothness determines the Sobolev space to which it belongs. Finally, we recall the definition of bounded variation and refer to \cite{rudin1987real} for further background.

\begin{defi}

    Define $\mathcal{P}([a,b])$ to be the set of all ordered finite sets of distinct points $x_0=a<x_1<\cdots < x_m=b$.

    Given $f:[a,b]\rightarrow \R$, we define $\V_{[a,b]}(f)$ as the total variation of $f$, given by
    \[
    \V_{[a,b]}(f) = \sup_{p\in \mathcal{P}([a,b])} \sum_{i=0}^{\#p-2} \mid  f(x_{i+1}) -f(x_i)\mid .
    \]
    The functions with finite total variation on ${[a,b]}$ are collected in $\BV({[a,b]})$.
\end{defi}

For continuous functions with absolutely integrable (weak) derivative, one has the classical identity
\[
\V_{[a,b]}(f)=\int_a^b |Df(x)|\,dx=\|Df\|_{1},
\]
where $D^k f$ denotes the $k$-th (weak) derivative of $f$.
More generally, if $f$ satisfies assumptions (B), then its total variation can be written as
\[
\V_{[a,b]}(f)=\sum_{i=0}^{n-1}\bigl|f(x_{i+1})-f(x_i)\bigr|,
\]
where $\{x_0,\dots,x_n\}$ is the ordered list of all local minima and maxima of $f$.

Now, let $f:[a,b]\rightarrow \R$ satisfy assumptions (B), and let $\{z_0,\ldots,z_m\}\subset [a,b]$ be a grid such that $z_i<z_{i+1}$. Set
\[
\delta_m := \max_{i=1,\ldots, m} (z_i - z_{i-1}).
\]
Define $f_{\delta_m}^{PL}$ as the piecewise affine function interpolating the points $\{(z_i,f(z_i))\}$. The merge tree of $f_{\delta_m}^{PL}$ provides an approximation of $T_f$, whose convergence is quantified by the following result.

\begin{prop}\label{prop:approx}
Assume $f:{[a,b]}\rightarrow \R$ is Lipschitz with Lipschitz constant $L$. Then:
\begin{enumerate}
    \item $\| f - f_{\delta_m}^{PL} \|_\infty \leq 2L \delta_m$.
\end{enumerate}
If, in addition, $f$ is twice continuously differentiable, then
\begin{enumerate}
    \item[2] $\V_{[a,b]}(f - f^{PL}_{\delta_m}) \leq (b-a)^{1/2} A \delta_m \|D^2 f\|_2$ for some constant $A>0$ independent of $f$;
    \item[3] $d_E(T_f,T_{f^{PL}_{\delta_m}})\leq \Delta_f \delta_m,$
    where $\Delta_f$ is a constant depending only on $f$.
\end{enumerate}
\end{prop}

\begin{rmk}
The proof of \Cref{prop:approx} relies on \Cref{teo:BV}, which is stated and proved in the next section. For clarity and to improve the overall flow of the exposition, we anticipate the statement of the proposition here. In the appendix, however, the proof of \Cref{prop:approx} is presented \emph{after} the proof of \Cref{teo:BV}, so as to make all dependencies explicit and to avoid any possible ambiguity regarding circularity.
\end{rmk}

\Cref{prop:approx} shows that piecewise affine interpolants provide effective merge-tree approximations. Their main practical advantage is that $T_{f_{\delta_m}^{PL}}$ can be computed exactly: all critical points of a piecewise linear function occur at grid points, and their type (minimum/maximum) can be read off directly from the neighboring sampled values. In all experiments and simulations reported in this work, we therefore compute merge trees through this  pipeline.

\subsection{Merge Tree Estimators}
\label{sec:estimators}

Leveraging on \Cref{teo:stability} and on previous works in FDA, we now tackle the problem of estimating merge trees of unknown functions from noisy samples.
In particular, we assume the following standard functional model, which we call (M).

\begin{defi}[Functional Statistical Model (M)]

    Let  $\Xvar$ be a real-valued random variable with density $p$, supported on a compact interval, and assume that $p$ is bounded away from 0 on its support. Let $\Yvar$ be a real-valued random variable such that
    \begin{displaymath}
    \Yvar \mid  \Xvar = x \sim f(x)+\varepsilon \tag*{(M)}
    \end{displaymath}
    where $f$ is an unknown function and $\varepsilon$ is an independent noise variable with zero mean and finite variance.

\end{defi}

To estimate the merge tree of the unknown \(f\), we consider an independent sample $\{(x_i,y_i)\}_{i=1}^n$ from (M); hence, for $i=1,\ldots,n$,
\begin{displaymath}
y_i = f(x_i) + \varepsilon_i.
\end{displaymath}
We will estimate the merge tree of \(f\) using a functional regression estimator that smooths \(f,\) that is, a measurable mapping $\hat{f}_n$ defined on $(\R \times \R)^n$ and transforming any independent sample $\{(x_i,y_i)\}_{i=1}^n$ into a smooth estimate of the function \(f.\)  As a result, additional assumptions may be necessary depending on the selected \(\hat{f_n}\).

\begin{rmk}
As anticipated, the following theoretical developments have been derived under assumptions (B) for the considered function $f$.
We expect that this approach should yield analogous results when extended to the multivariate setting; however, we leave this investigation for future work. \end{rmk}

The pivotal result of this section is the following.

\begin{teo}\label{teo:BV}
    Let $f,g:{[a,b]}\rightarrow \R$ be functions satisfying assumptions (B).
    Let $C_f$ and $C_g$ be the number of local minima of, respectively, $f$ and $g$, then:
\[
d_E(T_f,T_g)\leq 16 \min\{C_f,C_g\}\parallel f-g\parallel_\infty + \mid  V_{[a,b]}(f)-V_{[a,b]}(g)\mid .
\]

\end{teo}

\begin{rmk}\label{rmk:BV}
    Let $f,g \in \BV({[a,b]})$ continuously differentiable functions. We know that $\V_{[a,b]}(f-g) = \int_{[a,b]} \mid  Df(x)-Dg(x)\mid  dx$. Thus $\mid  \V_{[a,b]}(f)-\V_{[a,b]}(g)\mid  \leq \V_{[a,b]}(f-g) \leq (b-a) \parallel Df-Dg\parallel_\infty$. As a consequence, $\parallel Df-Df_n\parallel_\infty\rightarrow 0$ implies $\mid  \V_{[a,b]}(f)-\V_{[a,b]}(f_n)\mid  \rightarrow 0$. In particular, this implies that, if \(\parallel f_n - f\parallel_\infty \rightarrow 0\) and \(\parallel Df_n - Df \parallel_\infty \rightarrow 0\), then \(d_E(T_{f_{n}}, T_f) \rightarrow 0\).
\end{rmk}

\Cref{teo:BV} can be seen as some sort of \virgolette{remainder} result, stating that if the oscillations of the difference between $f$ and $g$ decrease in amplitude fast enough, then we can control the distance between their merge trees. Using \Cref{teo:BV}, we can establish the following result, which provides sufficient conditions for the consistency of estimators of merge trees.

\begin{prop}\label{prop:estimator}
    Let $[a,b]$ be the support of $p$. Given $f:{[a,b]}\rightarrow \R$ continuously differentiable and satisfying assumptions (B), let $\hat{f}_n$ be a functional regression estimator of $f$ mapping any independent sample $\{(x_i,y_i)\}_{i=1}^n$ from (M) into a smooth estimate of \(f\).
    Suppose that, for all \(n,\) the estimator $\hat{f}_n$ satisfies assumptions (B) with probability one. Assume the existence of positive functions $h_n$ and $g_n$ such that, for \(\varepsilon>0\):
    \begin{itemize}
        \item[(i)] $P(\parallel f-\hat{f}_n\parallel_\infty \leq \varepsilon ) \geq 1- h_n(\varepsilon)$;
        \item[(ii)] $P(\parallel Df-D\hat{f}_n\parallel_\infty \leq \varepsilon) \geq 1-g_n(\varepsilon)$.
    \end{itemize}
    Then:
    \[
    P(d_E(T_f,T_{\hat{f}_n})\leq 2\varepsilon) \geq 1- h_n(C_1\varepsilon)- g_n(C_2\varepsilon),
    \]
    for some constants $C_1,C_2>0$.
\end{prop}

Leveraging \Cref{prop:approx}, we now close the loop and use it to construct practical estimators for merge trees. We fix a uniform grid of points \(\{z_0, \ldots, z_m\} \subset [a,b]\), so that $\delta_m = (b-a)/m$. Given a functional regression estimator \(\hat{f}_n\) of \(f: [a,b] \to \mathbb{R}\), we define \(\hat{f}_{n,\delta_m}^{PL}\) as the piecewise affine function interpolating the points \(\{(z_i, \hat{f}_n(z_i))\}\). The merge tree of \(\hat{f}_{n,\delta_m}^{PL}\) serves as our computational estimator.

It is important to note that, in contrast to \Cref{prop:estimator}, proving the convergence of the estimator \(T_{\hat{f}_{n,\delta_m}^{PL}}\) requires stronger assumptions regarding the regularity of \(f\) and the convergence properties of \(\hat{f}_n\).

\begin{prop}\label{prop:comp_est}
        Given $f:{[a,b]}\rightarrow \R$ two times continuously differentiable and satisfying assumptions (B), let $\hat{f}_n$ be a functional regression estimator of $f.$
        Assume that, for all n, the estimator \(\hat{f_n}\) satisfies assumptions (B) with
probability one.  Assume also the existence of positive functions $h_n$, $g_n$ and \(q_n\) such that, for \(\varepsilon>0\):
    \begin{itemize}
        \item[(i)] $P(\parallel f-\hat{f}_n\parallel_\infty \leq \varepsilon ) \geq 1- h_n(\varepsilon)$;
        \item[(ii)] $P(\parallel Df-D\hat{f}_n\parallel_\infty \leq \varepsilon) \geq 1-g_n(\varepsilon)$;
    \item[(iii)] $P(\parallel D^2f-D^2\hat{f}_n\parallel_\infty \leq \varepsilon) \geq 1-q_n(\varepsilon).$
    \end{itemize}
    Let $\hat{f}_{n,\delta_m}^{PL}$ be the piecewise affine interpolant of $\hat{f}_n$ on the
    uniform
    grid \(\{z_0, \ldots, z_m\} \subset [a,b]\).

    Then, there exist constants $C_1,\ldots, C_4,$ depending only on $f$, such that, for $m>C_4/\varepsilon,$
\begin{equation}\label{eq:PL_estimator}
P(d_E(T_{f},T_{\hat{f}_{n,\delta_m}^{PL}})< 4\varepsilon) \geq 1- h_n(C_1\varepsilon) -g_n(C_2\varepsilon) -q_n(C_3m\varepsilon ).
\end{equation}

\end{prop}

\begin{rmk}
In \eqref{eq:PL_estimator}, the dependence on the chosen grid (which is fixed independently of the data) enters through the parameter $m$, which determines the step size $\delta_m$, since the grid points are equally spaced.
\end{rmk}

In the introduction, we highlighted data smoothing as a fundamental step in any functional data analysis pipeline. This applies equally to the topological approach we here propose for analyzing functions, and it operates on different levels:
\begin{itemize}
\item \emph{Functional Representation of Data}:
   to construct a merge tree from a sample \(\{(x_i, y_i)\}_{i=1}^n\)from (M), we must first make assumptions on the regularity of the function $f$ and of its functional estimator $\hat{f}_n$.

\item \emph{Impact of Smoothing on Topological Representation}:
   the topological representation derived from the chosen functional representation of $\hat{f}_n$ inherits part of the variability present in the noisy observations. The roughness of the functional representation directly influences this variability: the rougher the representation, the noisier the merge tree. This noise manifests as small artifacts (e.g., extraneous edges), increasing the size of the merge tree and distorting metrics such as \(d_E\).
   \Cref{prop:estimator} and \Cref{prop:comp_est} demonstrate that carefully selecting the smoothness of the functional representation of the functional regression estimator can mitigate the issue and allow for accurate estimation of merge trees.

\item \emph{Consistency}: finally, the convergence of merge tree estimators is controlled by the convergence of the functional regression estimators. Therefore, the selection of hyperparameters for \(\hat{f}_n\) (such as the bandwidth in kernel estimators) should follow established criteria developed specifically for this purpose.
\end{itemize}

\begin{rmk}
    We conjecture that the proof of \Cref{teo:BV}, and consequently of the other results of \Cref{sec:estimators} and \Cref{sec:approx}, could be adapted to work also for PDs embedded in a $1$-Wasserstein metric space. The key points for proving a result analogous to \Cref{teo:BV} are that matching points of a persistence diagram to the diagonal is equivalent to removing leaves from merge trees, and the similarity between the stability properties of $d_E$ and of the Wasserstein distance (see \citealp{pegoraro2025finitely}). To the best of our knowledge, consistent estimators of persistence diagrams compatible with Wasserstein distances are yet to be considered in the literature.
\end{rmk}

We now summarize in a theorem several results established in \cite{schuster1979contributions}, which demonstrate the existence of a functional estimator that, under appropriate conditions, satisfies the requirements of \Cref{prop:comp_est}.

\begin{teo}[adapted from \cite{schuster1979contributions}]\label{teo:NW}
    Consider the setting of model (M) and assume that:
\begin{enumerate}
    \item[(i)] The density $p$ is three times continuously differentiable;
    \item[(ii)] If $q$ is the joint density of $(\Xvar,\Yvar),$
    \[
    w(x)=\int_\R y q(x,y) dy,
    \]
    is three times continuously differentiable;
    \item[(iii)] The function $f$ is three times continuously differentiable.
\end{enumerate}

Moreover, let \(K\) be a density defined on \(\R\) such that:
\begin{itemize}
    \item[(iv)]
    \[
    \int_\R \mid  x K(x)\mid  dx <\infty;
    \]
    \item[(v)] $D K$, $D^2K$ and $D^3K$ are continuous and of bounded variation on $\R$;
    \item[(vi)] For \(x \in \R,\) the functions $x\phi$ and   $x^2\phi(x)$ are absolutely integrable, with
    \[
    \phi(x) = \int_\R e^{iux} K(u) du,
    \]
    being the characteristic function of $K$.
\end{itemize}

Given an iid sample $\{(x_i,y_i)\}_{i=1}^n$
from (M)
and a sequence of bandwidths $\{a_n\}$ such that \(\lim_{n \rightarrow \infty} a_n=0,\) define
\begin{equation}\label{eq:estimator}
\hat{f}_n(x):=\frac{\sum_{i=1}^n y_i K(\frac{(x-x_i)}{a_n})/(na_n) }{\sum_{i=1}^n K(\frac{(x-x_i)}{a_n})/(na_n)}
\end{equation}

extended with $0$ when the denominator is $0$.

Then, there is a constant \(C\) such that, for $\varepsilon>0$ and $n$ sufficiently large,
\begin{equation*}
    P(\parallel D^jf-D^j\hat{f}_n\parallel_\infty \leq\varepsilon) \geq 1-C/(n a_n^{2j+2}\varepsilon^2),
\end{equation*}
for \(j=0,1,2.\)
\end{teo}

Note that, for all $n$, the estimator \(\hat{f_n}\) built as in \eqref{eq:estimator}, satisfies assumptions (B) with
probability one and that the Gaussian density satisfies the hypotheses (iv)-(vi) of \Cref{teo:NW}.

To conclude this section, we note that the literature on derivative estimation is extensive and diverse \citep{stone1985additive, muller1987bandwidth, delecroix1996nonparametric, chaudhuri1999sizer, zhou2000derivative, gijbels2005data, aneurisk_splines, fan2018local, liu2020smoothed}. It also includes results for multivariate data \citep{akima1984estimating, lu1996multivariate}, which can be utilized when extending these findings to multivariate functions.
For simplicity, in \Cref{teo:NW} and our case study, we focus on works \citep{schuster1979contributions, muller1984smooth, mack1989derivative} that allow the use of the well-known Nadaraya-Watson kernel estimator \citep{nadaraya1964estimating}. This estimator has been widely employed for nonparametric regression and has also supported consistency results in the TDA context \citep{fasy2014confidence, bobrowski2017topological}.
However, the generality of \Cref{prop:comp_est} accommodates alternative estimator choices, offering flexibility in its application.

\section{Case Study}
\label{sec:case_study}

\subsection{The Data}
\label{sec:case_study_sub1}

We now turn to a comparative analysis based on the real-world AneuRisk65 dataset. Originally introduced in \cite{aneurisk_jasa}, this dataset has become a benchmark for evaluating FDA methods designed to classify misaligned functional data, both in supervised and unsupervised settings (see the 2014 \emph{Electronic Journal of Statistics} special issue on phase and amplitude variability).
Here, we reproduce the classification experiment from \cite{aneurisk_jasa} to compare merge trees and PDs as representations of the AneuRisk65 data, and to assess their performance relative to traditional FDA approaches. To strengthen this comparison, we also add a pipeline based on Fisher-Rao alignment \citep{sriva_1}.

The data were collected within the multidisciplinary AneuRisk Project, aimed at understanding how vessel morphology, hemodynamics, and wall biomechanics contribute to cerebral aneurysm formation. The project involved 65 patients examined at Ospedale Niguarda (Milan), producing 3D angiographic images of their internal carotid arteries (ICAs). From these images, 3D reconstructions of the ICAs were obtained, along with \emph{centerline curves} (connecting centers of maximal inscribed spheres) and corresponding \emph{radius functions} \citep{aneurisk_kmeans}.

\begin{figure}[p]
\centering
\captionsetup{font=small,skip=4pt}
\setlength{\tabcolsep}{2pt}

\begin{tabular}{@{}cccc@{}}
\begin{subfigure}[t]{0.23\textwidth}\centering
  \includegraphics[width=\linewidth]{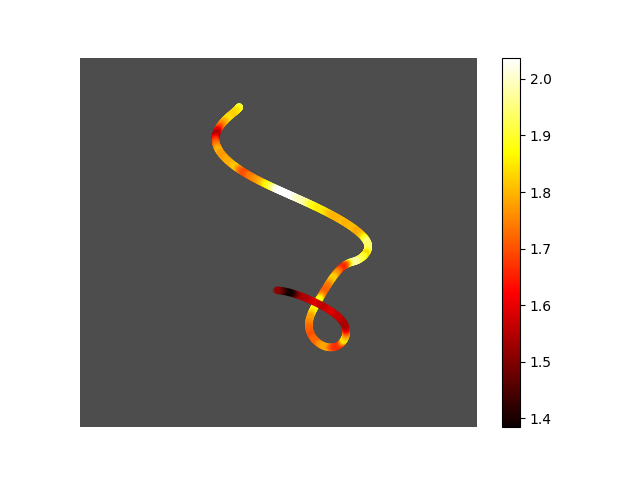}
  \caption{ICA patient 1 (L).}\label{fig:R_1_heat}
\end{subfigure} &
\begin{subfigure}[t]{0.23\textwidth}\centering
  \includegraphics[width=\linewidth]{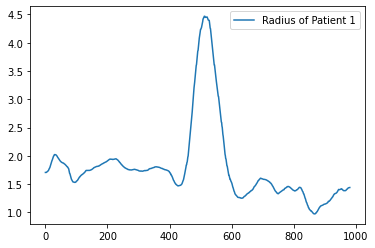}
  \caption{Radius along ICA (L).}\label{fig:R_1_graph}
\end{subfigure} &
\begin{subfigure}[t]{0.23\textwidth}\centering
  \includegraphics[width=\linewidth]{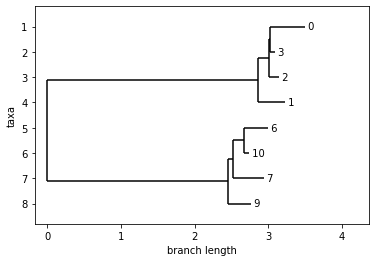}
  \caption{Merge tree (L).}\label{fig:R_1_tree}
\end{subfigure} &
\begin{subfigure}[t]{0.23\textwidth}\centering
  \includegraphics[width=\linewidth]{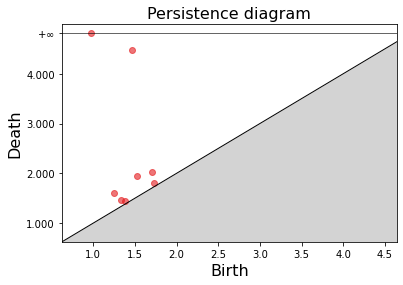}
  \caption{PD (L).}\label{fig:R_1_PD}
\end{subfigure}
\\[-2pt]
\begin{subfigure}[t]{0.23\textwidth}\centering
  \includegraphics[width=\linewidth]{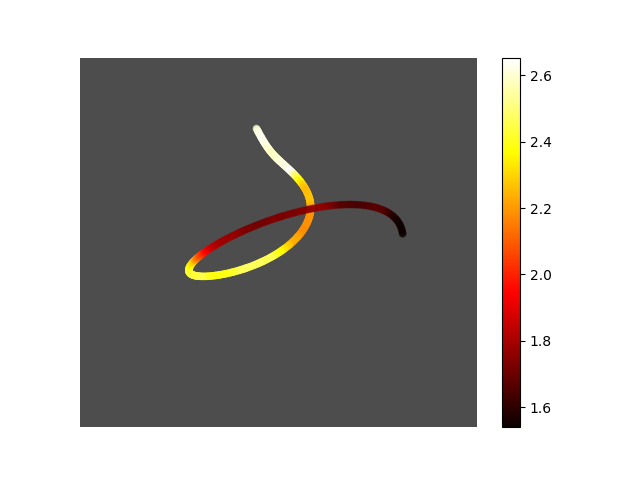}
  \caption{ICA patient 55 (U).}\label{fig:R_55_heat}
\end{subfigure} &
\begin{subfigure}[t]{0.23\textwidth}\centering
  \includegraphics[width=\linewidth]{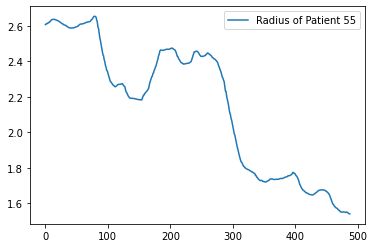}
  \caption{Radius along ICA (U).}\label{fig:R_55_graph}
\end{subfigure} &
\begin{subfigure}[t]{0.23\textwidth}\centering
  \includegraphics[width=\linewidth]{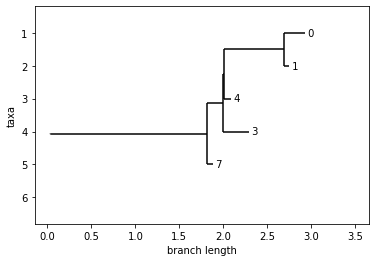}
  \caption{Merge tree (U).}\label{fig:R_55_tree}
\end{subfigure} &
\begin{subfigure}[t]{0.23\textwidth}\centering
  \includegraphics[width=\linewidth]{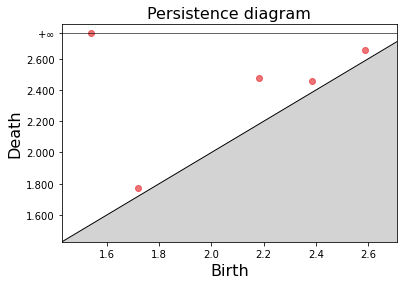}
  \caption{PD (U).}\label{fig:R_55_PD}
\end{subfigure}
\\[-2pt]
\begin{subfigure}[t]{0.23\textwidth}\centering
  \includegraphics[width=\linewidth]{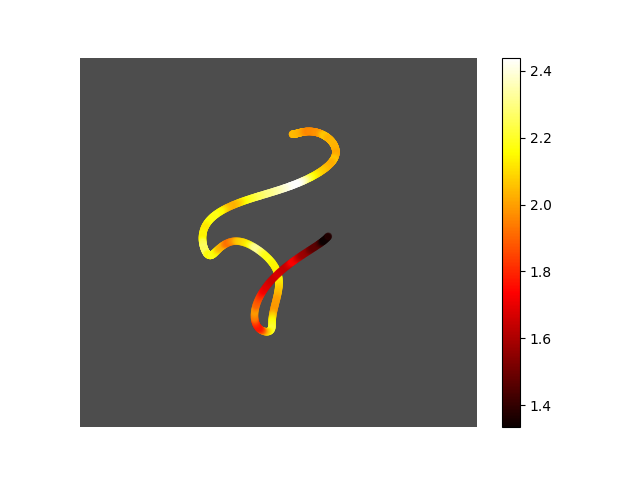}
  \caption{ICA patient 56 (U).}\label{fig:R_56_heat}
\end{subfigure} &
\begin{subfigure}[t]{0.23\textwidth}\centering
  \includegraphics[width=\linewidth]{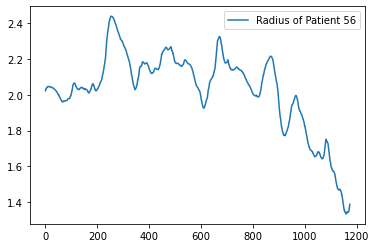}
  \caption{Radius along ICA (U).}\label{fig:R_56_graph}
\end{subfigure} &
\begin{subfigure}[t]{0.23\textwidth}\centering
  \includegraphics[width=\linewidth]{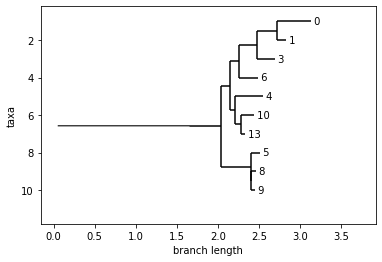}
  \caption{Merge tree (U).}\label{fig:R_56_tree}
\end{subfigure} &
\begin{subfigure}[t]{0.23\textwidth}\centering
  \includegraphics[width=\linewidth]{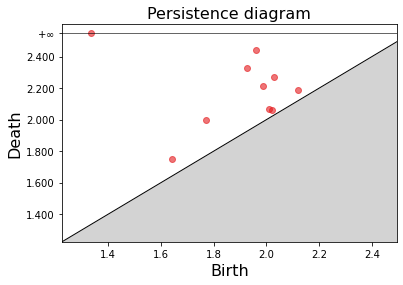}
  \caption{PD (U).}\label{fig:R_56_PD}
\end{subfigure}
\end{tabular}

\caption{Three patients in the AneuRisk65 dataset. From left to right: ICA colored by radius, radius function, associated merge tree, and PD. Patient~1 belongs to the Lower group (L), while patients~55 and~56 belong to the Upper group (U). Merge trees are truncated with $K=\max$ of the radius range in \Cref{fig:R_1_graph}.}
\label{fig:aneu}
\end{figure}

This dataset poses several challenges. The reconstruction process introduces measurement and smoothing errors, leading to noisy, highly oscillatory centerlines \citep{aneurisk_splines}. Moreover, differences in head positioning, vessel geometry, and scanning volume make direct spatial comparison of ICAs impossible. As illustrated in \Cref{fig:aneu}, vessel segments can vary greatly in length and embedding, requiring alignment methods to enable meaningful comparison.

The AneuRisk65 data are therefore a paradigmatic example of the threefold challenge, \emph{smoothing}, \emph{embedding}, and \emph{alignment}, that motivates a number of original works in FDA. In our study, we use the smoothed centerline curvatures obtained via free-knot regression splines \citep{aneurisk_splines}, while the data relative to the radius of the blood vessel, although measured on a very fine grid of points along the centerline, is still in its raw format. Hence the AneuRisk65 data also allow us to compare the behavior of tree representations on smoothed data and on raw data.

\subsection{The Pipeline for Supervised Classification}
The patients in the AneuRisk65 dataset are divided into three groups:
the \emph{Upper} group (U), comprising patients with aneurysms in the Willis circle at or beyond the terminal bifurcation of the ICA;
the \emph{Lower} group (L), with aneurysms located along the ICA before its terminal bifurcation;
and the \emph{None} group (N), consisting of patients without aneurysms.
Our classification goal, as in \cite{aneurisk_jasa}, is to discriminate between subjects in \(L \cup N\) and those in \(U\), using geometric features of the ICA.

We develop two additional classification pipelines, one based on topological representations and one using the Fisher–Rao functional alignment method, closely following the benchmark procedure in \cite{aneurisk_jasa}.
The benchmark pipeline includes data smoothing, alignment, dimensionality reduction via Functional Principal Components Analysis (FPCA) on curvature and radius functions, and subsequent quadratic discriminant analysis (QDA) based on the (concatenated) first two FPCA scores from each function.

In the Fisher–Rao pipeline, we replicate all steps of \cite{aneurisk_jasa} but modify the alignment procedure.
Specifically, (1) all functions are first re-parameterized onto $[0,1]$ by affine transformations, and (2) alignment is refined using the Fisher–Rao metric applied separately to curvature and radius functions, using the \texttt{scikit-fda} \citep{scikit-fda} package.

In the topological pipelines, we omit alignment altogether, leveraging the invariance property of merge trees established in \Cref{prop:invariance}.
This ensures a fair comparison between the functional and topological approaches, with the key difference being the data representation.

Data are smoothed using kernel estimators as in \cite{muller1984smooth, mack1989derivative}, with smoothed curvature functions serving as a control for bandwidth selection.
For the functional pipelines, aligned functions are processed via FPCA and classified using QDA.
For the topological pipelines, we extract merge tree and PD representations from both curvature and radius functions, compute pairwise distance matrices using the edit distance \(d_E\) or the 1-Wasserstein distance respectively, and combine them coherently with the functional approach. Formally, this amounts to compute new matrices of mixed pairwise distances according to the formula:
\begin{equation}\label{eq:d_mixed}
d_\text{mixed}^2 = d_\text{curvature}^2 + d_\text{radius}^2,
\end{equation}
where \(d_\text{curvature}\) and \(d_\text{radius}\) are the pairwise distances of merge tree or PD representations derived from the smoothed curvature and radius functions, respectively.  In fact, concatenating the scores from radius and curvature, as done in \cite{aneurisk_jasa}, amounts to treating the information of radius and curvature with the same weights. Thus, combining the distance matrices of merge trees and PDs with different weights would result in a less principled comparison of the different approaches.

An Euclidean nonlinear embedding (Isomap; \citealp{balasubramanian2002isomap}) is then applied to \(d_{\text{mixed}}\), mapping the data into a finite-dimensional Euclidean space of dimension \(m\).
Finally, a QDA classifier is fitted on the embedded points, as in the original study \cite{aneurisk_jasa}.

The proposed pipelines involve several hyperparameters that require tuning.

The smoothing step depends on the selection of the kernel bandwidth.
In the Fisher–Rao pipeline, an additional hyperparameter (\(\lambda\)) is the penalization term controlling the degree of warping.
Moreover, we vary the number of principal components retained in the FPCA step (\(fpca\_comp\)) to assess its effect on classification performance.
The topological pipelines, in turn, require setting two hyperparameters: the dimension \(m\) of the Euclidean embedding in Isomap, and the number of neighbors (\(neigh\)) used to construct the neighborhood graph.

The bandwidth parameter is selected to optimize the convergence of the merge tree estimators (see \Cref{sec:band}), whereas the remaining hyperparameters are chosen by maximizing the discriminatory power of the QDA, estimated via leave-one-out cross-validation.

\subsection{Classification Results}

\Cref{table:conf_mat} reports the detailed leave-one-out cross-validation results for all classification pipelines.
We further assessed the performance of the topological approaches when using only curvature or only radius data, in order to explore differences between PDs and merge trees, beyond the simulations discussed in \Cref{sec:examples_vs_PD}.

From the first two rows of \Cref{table:conf_mat}, we observe that PDs were more effective than merge trees in extracting discriminant information from the radius data. This behavior is potentially explained by the phenomenon illustrated in \Cref{sec:exampleI}, where the discriminant signal in functions depends more on the number and amplitude of oscillations than on their ordering.
When either curvature or, jointly, curvature and radius are considered (through the mixed distance $d_\text{mixed}$ defined in \eqref{eq:d_mixed}), merge trees and PDs achieve comparable levels of accuracy. Despite their structural differences (see \Cref{sec:examples_vs_PD}), both representations capture similar discriminant information for the classification task. In particular, the two methods misclassified the same eight patients (out of ten total misclassifications).

The last row of \Cref{table:conf_mat} reports the results for the benchmark classifier described in \cite{aneurisk_jasa} and for the Fisher–Rao pipeline. Although the Fisher–Rao metric is among the most sophisticated tools for functional alignment, it yielded the poorest performance overall. These results further support the effectiveness of the topological approach, whose invariance properties appear advantageous in this context.
For reference, the Fisher–Rao pipeline shares only two misclassified patients with either of the topological models, indicating that the variability captured by the topological metrics is of a fundamentally different nature, yet remains interpretable, thanks to the stability results established earlier, thus providing a valuable complement to classical FDA methods.

\begin{table}[H]
\begin{center}
\small
\begin{tabular}{l cc @{\qquad} cc @{\qquad} cc}
\multicolumn{7}{c}{\textbf{Merge Trees}}\\
\multicolumn{7}{c}{Predicted}\\
 & \multicolumn{2}{c}{Curvature} & \multicolumn{2}{c}{Radius} & \multicolumn{2}{c}{Mixed} \\
\cmidrule(lr){2-3}\cmidrule(lr){4-5}\cmidrule(lr){6-7}
True & $\{U\}$ & $\{L,N\}$ & $\{U\}$ & $\{L,N\}$ & $\{U\}$ & $\{L,N\}$ \\
\midrule
$\{U\}$     & 22 & 10 & 21 & 11 & 25 & 7 \\
$\{L,N\}$   &  3 & 30 &  4 & 29 &  3 & 30 \\
\midrule
& \multicolumn{2}{c}{$n=4$, $neigh=10$}
& \multicolumn{2}{c}{$n=7$, $neigh=24$}
& \multicolumn{2}{c}{$n=7$, $neigh=14$} \\
\end{tabular}

\vspace{1.2em}

\begin{tabular}{l cc @{\qquad} cc @{\qquad} cc}
\multicolumn{7}{c}{\textbf{Persistence Diagrams}}\\
\multicolumn{7}{c}{Predicted}\\
 & \multicolumn{2}{c}{Curvature} & \multicolumn{2}{c}{Radius} & \multicolumn{2}{c}{Mixed} \\
\cmidrule(lr){2-3}\cmidrule(lr){4-5}\cmidrule(lr){6-7}
True & $\{U\}$ & $\{L,N\}$ & $\{U\}$ & $\{L,N\}$ & $\{U\}$ & $\{L,N\}$ \\
\midrule
$\{U\}$     & 23 & 9 & 23 & 9 & 24 & 8 \\
$\{L,N\}$   &  4 &29 &  2 &31 &  2 &31 \\
\midrule
& \multicolumn{2}{c}{$n=6$, $neigh=10$}
& \multicolumn{2}{c}{$n=2$, $neigh=19$}
& \multicolumn{2}{c}{$n=2$, $neigh=22$} \\
\end{tabular}
\end{center}

\vspace{1.5em}

\begin{center}
\begin{tabular}{@{}c @{\hspace{4em}} c@{}}
\textbf{Benchmark} & \textbf{Fisher--Rao} \\[6pt]
\begin{tabular}{c|cc}
    & \multicolumn{2}{c}{Predicted} \\
    True & $\{U\}$ & $\{L,N\}$ \\
    \hline
    $\{U\}$ & 26 & 6 \\
    $\{L,N\}$ & 6 & 27 \\
\end{tabular}
&
\begin{tabular}{c|cc}
    & \multicolumn{2}{c}{Predicted} \\
    True & $\{U\}$ & $\{L,N\}$ \\
    \hline
    $\{U\}$ & 25 & 7 \\
    $\{L,N\}$ & 8 & 25 \\
\end{tabular} \\
 & \textit{fpca\_comp = 7, $\lambda$ = 0.1}
\end{tabular}
\end{center}

    \caption{
    Leave-one-out confusion matrices for the classification pipelines described in \Cref{sec:case_study}, with the optimal parameters.
    }
	\label{table:conf_mat}
\end{table}

\section{Discussion}\label{sec:discussion_FDA}

We believe that methods from TDA can substantially enrich the FDA toolbox, particularly in settings where nontrivial smoothing and alignment are required.
In this work, we focused on two topological representations of functions: PDs, the classical TDA benchmark, and merge trees, relying on the recent metric framework of \cite{pegoraro2025finitely} and deriving new theoretical results tailored to functional data.

Using the AneuRisk65 dataset as a case study, we revisited a well-known FDA problem characterized by challenging smoothing and alignment issues.
By adopting representations invariant to homeomorphic transformations of the abscissa, we reduced alignment dependence and achieved robust classification performance that matched or exceeded traditional FDA approaches based on FPCA and QDA of aligned functions.
Both TDA pipelines also outperformed the Fisher–Rao method, one of the most sophisticated tools for functional alignment, underscoring the effectiveness of invariance-based representations.
The importance of such invariance is well recognized in fields like shape analysis, where alignment procedures are often far more complex or even infeasible, making topological approaches particularly advantageous.
Our findings indicate that these advantages extend even to contexts where alignment methods are readily available and classical approaches are well developed, a perspective further supported by recent work combining the two paradigms \cite{kim2025peak}.

A key challenge lies in the intrinsic complexity of performing statistical learning in the non-Euclidean spaces of PDs and merge trees.
This difficulty has spurred research, for instance, on linearization techniques for PDs and on the definition of Fréchet means and barycenters for merge trees.
Similarly, future work should aim to extend merge tree–based statistical frameworks to multivariate functional data, where alignment procedures are often computationally prohibitive.
Developing robust statistical tools for populations of trees, such as principal component analysis and regression, remains an open problem that could be addressed through dedicated linearization methods for merge trees.
The existence of Fréchet means in merge tree spaces (see earlier versions of \citealp{pegoraro2024finitely}) and more applied approaches proposed in \cite{pont2021wasserstein, wetzels2023merge} offer promising directions, though issues of non-uniqueness, and scalability still need to be addressed through further theoretical and computational work.

Nonetheless, careful smoothing remains essential for meaningful tree estimation, and topological representations are not universally suitable.
Merge trees and PDs encode the order and amplitude of extrema but ignore precise abscissa information; when spatial or temporal localization is crucial, such alignment-invariant representations may be inappropriate.
As a general principle, analysts should follow Occam’s razor, selecting the simplest representation that retains the information relevant to the scientific question.

\section{Outline of the Supplementary Material}\label{sec:supplementary-material}

The Supplementary Material contains the following sections. \Cref{sec:top_rmk} provides an extended discussion on constructing merge trees in more general settings than those considered in the main text. In \Cref{sec:examples_vs_PD}, we present a detailed comparison between merge trees and PDs, with \emph{in silico} examples to further illustrate the distinctions and similarities between the two topological summaries. \Cref{sec:band} delves deeper into the smoothing applied to functions in the case study. \Cref{sec:comp} contains some runtime simulations. \Cref{sec:proofs_FDA} collects the proofs of the paper's results.

\section*{Acknowledgments}

We thank Steve Marron who initiated us to algebraic topology for the statistical analysis of functional data during seminal discussions while one of us (P.S.) was visiting him at UNC. We thank the AneuRisk Project for providing the data analyzed in the case study.
We also deeply thank the editor and the reviewers for their very helpful and constructive comments which contributed to a significant improvement of the original manuscript. P.S. acknowledges the support by MUR, Italy, grant Dipartimento di Eccellenza 2023--2027.

\section*{Disclosure Statement}
No potential conflict of interest was reported by the authors.

\section*{Funding}
This article was supported by Ministero dell'Universit\`a e della Ricerca, grant Dipartimento di Eccellenza 2023--2027.

\section*{Data Availability Statement}
The data that support the findings of this study are public and can be accessed via the AneuRisk Project website \url{https://statistics.mox.polimi.it/aneurisk}.

\bibliography{references}

\clearpage
\FloatBarrier
\begin{center}
{\Large\bf Supplementary Material}
\end{center}
\addcontentsline{toc}{section}{Supplementary Material}

\setcounter{section}{0}
\setcounter{equation}{0}
\setcounter{figure}{0}
\setcounter{table}{0}
\setcounter{teo}{0}
\setcounter{prop}{0}
\setcounter{defi}{0}
\setcounter{rmk}{0}
\setcounter{lem}{0}
\setcounter{assumption}{0}
\renewcommand{\thesection}{S\arabic{section}}
\renewcommand{\theequation}{S\arabic{equation}}
\renewcommand{\thefigure}{S\arabic{figure}}
\renewcommand{\theteo}{S\arabic{teo}}
\renewcommand{\theprop}{S\arabic{prop}}
\renewcommand{\thedefi}{S\arabic{defi}}
\renewcommand{\thermk}{S\arabic{rmk}}
\renewcommand{\thelem}{S\arabic{lem}}
\renewcommand{\theassumption}{S\arabic{assumption}}

\section{Topological Remark}
\label{sec:top_rmk}

In this section, we show how to represent a real valued function $f:X\rightarrow \mathbb{R}$ by means of a merge tree, under conditions (A0), introducing also the formal definition of \emph{tameness}. To achieve this, we rely on some key assumptions, which are well established in the literature for producing \emph{constructible} objects \citep{de2016categorified, patel2018generalized, curry2021decorated}.
We then restate \Cref{defi:tame} in a more detailed form to clarify its components and interpretation.

\begin{assumption}[Tameness]\label{ass:construct}
Given a family of topological spaces $\{X_t\}_{t\in \mathbb{R}}$ with $X_t\subset X_{t'}$, $t\leq t'$, we assume the existence of a finite collection of real numbers $\{t_1<t_2<\ldots<t_n\}$, called critical set, such that, given $t<t',$
if $t,t'\in (t_i,t_{i+1})$ or $t,t'>t_n$, then $\pi_0(i_t^{t'})$ is bijective. The values $t_i$ are called critical values. As in \cite{pegoraro2024finitelyfunc} we always consider a minimal set of critical values, that is, the smallest possible set of critical values. With this condition, for any critical value $t_i$ there is some constant $C>0$ such that for all $\varepsilon\in (0,C) $, $\pi_0(i_{t_i-\varepsilon}^{t_i+\varepsilon})$ is not bijective.
Moreover, we assume that
for every \(t\in \mathbb{R},\) $\pi_0(X_t)$ is finite. A filtration $\{X_t\}_{t\in \mathbb{R}}$ satisfying the assumptions above is called \emph{tame}; likewise, a function \(f:X\rightarrow \mathbb{R}\) is said to be \emph{tame} if its sublevel set filtration $\{X_t\}_{t\in \mathbb{R}}$ is tame \citep{chazal2016structure}.
\end{assumption}

Together with the tameness of \(f\), (A0) also assumes that $X$ is path-connected. Moreover, we make an extra and simplifying \emph{regularity} assumption - not needed for the general construction of a merge tree - which implies a strong property of the critical values of the sublevel set filtration $\{X_t\}_{t\in \mathbb{R}}$ of \(f.\) This assumption can in fact be weakened, at the cost of some nontrivial topological details (for more details see also \cite{pegoraro2024finitelyfunc}).

Let  $t_j$ be a critical value of the sublevel set filtration $\{X_t\}_{t\in \mathbb{R}}$ of \(f.\) Let $\varepsilon>0$ be such that $t_j-\varepsilon >t_{j-1}$ and $t_j+\varepsilon <t_{j+1}$. The properties of $\pi_0$ imply that $\pi_0(i_{t_j-\varepsilon}^{t_j+\varepsilon})= \pi_0(i_{t_j-\varepsilon}^{t_j}) \circ \pi_0(i_{t_j}^{t_j+\varepsilon})$.
Due to the minimality condition stated in Assumption \ref{ass:construct} we know that $\pi_0(i_{t_j-\varepsilon}^{t_j+\varepsilon})$ is not bijective.

\begin{assumption}[Regularity]\label{assunzione}
We assume that the sublevel set filtration \(\{X_t\}_{t\in \mathbb{R}}\)  of \(f\) is \emph{regular}: that is, for every critical value $t_j$ of \(\{X_t\}_{t\in \mathbb{R}},\) there is a $C>0$ such that, for all  $\varepsilon\in (0,C)$, the map $\pi_0(i_{t_j}^{t_j+\varepsilon})$ is bijective.
When $\pi_0(i_{t_j}^{t_j+\varepsilon})$ is bijective, we say that the topological changes happen \emph{at} the critical value $t_j$, as opposed to \emph{across} $t_j$. Hence we are assuming that all the topological changes of \(\{X_t\}\) happen at the critical values.
\end{assumption}

From the topological point of view, what lies behind the requirement that the topological changes happen at critical values is the following. Let $U_{t_j}\in \pi_0(X_{t_j})$ and $U_t=\pi_0(i_{t_j}^t)(U_{t_j})$ for $t\in (t_j,t_{j+1})$. By construction, $U_{t_j} \subset U=\bigcap_{t\in (t_j,t_{j+1})} U_t$ and $f(p)=t_j$ for all $p\in U$. Which means $U\subset X_{t_j}$.
If $U$ is path-connected then $U_{t_j}=U$ and we cannot have another path-connected component $U'_{t_j}\in\pi_0(X_{t_j})$ such that $U'_{t_j}\subset U$.
All of this implies that, for $t\in (t_j,t_{j+1}),$ $\pi_0(i_{t_j}^t)^{-1}(U_t)=\{U_{t_j}\}$ - i.e. $\pi_0(i_{t_j}^t)$ is injective at $U_{t_j}$ and $U_t\in \pi_0(i_{t_j}^t)(\pi_0(X_{t_j}))$. So, if for every path-connected component $U_{t_j}\in \pi_0(X_{t_j})$ the set $U=\bigcap_{t\in (t_j,t_{j+1})} \pi_0(i_{t_j}^t)(U_{t_j})$ is non-empty and path-connected, then $\pi_0(i_{t_j}^t)$ is bijective. However, in general, $U$ need not be non-empty and path-connected!

  Two notable cases where the topological changes happen at critical values are that of a continuous function \(f\) defined on a connected compact subset of \(\mathbb{R}\) and that of an \(f\) defined on a finite graph.

Indeed, let $f:X\rightarrow \mathbb{R}$  be continuous and
$X\subset \mathbb{R}$ be a compact interval.
Then, for all \(t \in  \mathbb{R},\) $X_t$ is closed, since $f$ is continuous, and its path-connected components are compact intervals of the form $[a,b]$. Each path-connected component $U_{t_j}$ is thus a convex set and any intersection of the form
 $U = \bigcap_{t\in (t_j,t_{j+1})} \pi_0(i_{t_j}^t)(U_{t_j})$ is non-empty and convex; that is, $U$ is non-empty and path-connected. Thus, for continuous functions $f:X\rightarrow \mathbb{R}$, with
$X\subset \mathbb{R}$ being a compact interval, the topological changes always happen at critical values.

Consider now the discrete setting of a finite graph
 $X=(V,E)$, with vertices $V$ and edges $E$, such that the sublevel set filtration is well defined (i.e. for any edge  $e_{ij}=(x_i,x_j)\in E$ connecting two vertices $x_i$ and $x_j$, we have $f(e)\geq\max \{f(x_i),f(x_j)\}$). Let $t_1<t_2<\ldots<t_n$ be the image of $f$.
Then, $X_t = X_{t_j}$ for all $t\in [t_j,t_{j+1})$.
This implies that all topological changes happen at critical values.

As a consequence, all the functions considered in the present work, those illustrated in the examples or those pertaining to the case study described in \Cref{sec:case_study}, do satisfy our Assumption \ref{assunzione}, and the same is true for all numerical implementations.

However, Assumption \ref{assunzione} is not always satisfied, as we can see in the upcoming examples.

\textbf{Example I}
Consider the following
 sequences of topological spaces $\{A_t\}_{t\in [0,\infty)}$ and $\{B_t\}_{t\in [0,\infty)}$. For \(t> 0,\) let $A_t = (-t,t)\bigcup (1-t,1+t)$ and $B_t = [-t,t]\bigcup [1-t,1+t]$. Moreover, let $A_0=B_0=\{0,1\}$. $\{A_t\}$ and $\{B_t\}$ share the same set of critical values, namely $\{0,1/2\}$ and they only differ by the number of path-connected components at the critical value $1/2$: $\#\pi_0(A_{1/2})=2$,
 while
$\#\pi_0(B_{1/2})=1$. In $\{A_t\}$ changes happen across the critical values - $\pi_0(A_{1/3})\cong \pi_0(A_{1/2})$ and $\pi_0(A_{1/2})\ncong \pi_0(A_{1})$, while in $\{B_t\}$ changes happen at the critical values - $\pi_0(B_{1/3})\ncong \pi_0(B_{1/2})$ and $B_{1/2}\cong B_{1}$.

\textbf{Example II}
Consider the following
 sequence of topological spaces $\{A_t\}_{t\in [0,\infty)}$. For \(t> 0,\) let $A_t=\{(-\infty, +\infty)\times [-1/t,1/t]\}-\{(0,0)\}$ and $A_0 =  \{(-\infty,0)\cup (0,+\infty)\}\times \{0\}.$ Then $\bigcap_{t>0} A_t=A_0:$ however $A_0$ has two path-connected components while, for $t>0,$ the set $A_t$ is path-connected.

\textbf{Example III}
Let $\gamma:(0,1]\rightarrow \mathbb{R}^2$ be defined by \(\gamma(t)=(t,\sin(1/t)).\) Let $T$ be the closure in the plane of the set \(S=\{(t,\gamma(t)) \in \mathbb{R}^2: t \in (0,1]\}\), which is given by $T=S\cup \{0\}\times[-1,1]$. The set $S$ is usually referred to as the topologist's sine curve and the set $T$ as the closed topologist's sine curve.
Let $S_n$ be:
\[
S_n = \{(t,s)\mid  t \in [1/n,1]\text{ and }s \in [\sin(1/t)-1/n,\sin(1/t)+1/n]\}
\]
That is, $S_n$ is a $1/n$-thickening along the $y$-axis (second component of $\mathbb{R}^2$), of the graph of $\gamma$ restricted to the interval
$[1/n,1]$. Thus, if we add to $S_n$ the rectangle $R_n = [-1/n,1/n]\times [-1-1/n,1+1/n]$ we obtain a set $T_n=S_n \bigcup R_n$ such that:
\begin{itemize}
\item $T=\bigcap_{n\in\mathbb{N}} T_n$
\item $T_n$ is compact and path-connected. It is in fact homeomorphic to $R_n \bigcup [1/n,1]\times [-1/n,1/n]$ and thus homeomorphic to a closed disk in the plane.
\end{itemize}
As $n \rightarrow \infty$ we obtain a family of compact "disks" whose intersection is $T$ which is not path-connected.

\textbf{Example IV}
Set
$A_t = [1/t,+\infty) = f^{-1}((-\infty,t])$ with $f:\mathbb{R}_{>0}\rightarrow \mathbb{R}$ defined by $f(x)=1/x$. Clearly $A_t$ is closed and path-connected, but $\bigcap_{t> 0} A_t$ is empty.

\smallskip

In all the examples above we see different situations in which at some critical point $t$
we have a very \virgolette{unstable} topological scenario, which changes at $t+\varepsilon$ for any small $\varepsilon>0$:
\begin{itemize}
\item in Example I the balls centered in $0$ and $1$ contained in $A_t$ touch right after $t=1/2$; in fact their closures (giving $B_{1/2}$) at $t=1/2$ would intersect;
\item in Example II the horizontal stripe given by $A_t$ suddenly disconnects at $t=0$ because it is no longer thick enough to get around the hole in $(0,0)$;
\item we find a very similar situation also in Example III, where every thickening of $T$ would allow for a bridge between its two path-connected components;
\item lastly, in Example IV, a path-connected component is born with a minimum \virgolette{lying} at $+\infty$, thus producing an empty level set at $t=0$.
\end{itemize}

The general point of view which we assume to build merge trees from irregular tame filtrations, which is formalized in \cite{pegoraro2024finitelyfunc}, is that we deem to be negligible the topological differences between $\{A_t\}$ and $\{B_t\}$ in Example I, as those two filtrations have the same path-connected components but for one point, $t=1/2$, which we may consider as a measure zero subset of the parameter space indexing the filtration (i.e. $\mathbb{R}$).
Thus, for the examples above - and in fact for all tame filtrations of path-connected topological spaces - we propose to build the associated merge tree as if all topological changes happen at critical points: if there is a critical point $t_j$ such that $\pi_0(i_{t_j}^{t_j+\varepsilon})$ is not bijective, instead of looking at the merging information contained in
$\pi_0(i_{t_{j-1}}^{t_j})$ - as we do in
\Cref{sec:make_merge_trees} - one should look at
$\pi_0(i_{t_{j-1}}^{t_j+\varepsilon})$, but still recording the topological changes with a vertex at height $t_j$. In this way, for instance, the merge trees associated to $\{A_t\}$ and $\{B_t\}$ in Example I would be the same, in Example II we would have a single leaf at height $0$ - despite $A_0$ having two path-connected components -, the same for Example III (upon replacing $n$ with $1/\varepsilon$), and, lastly, Example IV would feature a leaf at height $0,$ despite $A_0$ being empty.

Having sorted out all these technicalities, we can finally build a merge tree from a tame and regular filtration of topological spaces $\{X_t\}_{t\in \R}$.

We build the merge tree $(T,h_T)$ from $\{\pi_0(X_t)\}_{t\in \R}$ with the following rules, in a recursive fashion,  starting from an empty set of vertices $V_T$ and an empty set of edges $E_T$. We simultaneously add points and edges to $T$ and define $h_T$ on the newly added vertices. Let $\{t_i\}_{i=1}^n$ be the critical
set of $\{\pi_0(X_t)\}_{t\in \R}$
and let $\pi_0(X_{t}):=a_{t}:=\{a^{t}_1,\ldots,a^{t}_{n_t}\}$. Call $\psi_t^{t'}:=\pi_0(X_{t\leq t'})$.
Considering the critical values in increasing order:

\begin{itemize}
\item for the critical value $t_1$ add to $V_T$ a leaf $a_{t_1}^k$, with height $t_1$, for every element $a_{t_1}^k\in a_{t_1}$;
\item for $i>1$, for every $a_{t_{i}}^k\in a_{t_i}$ such that  $a_{t_{i}}^k\notin \text{Im}(\psi_{t_{i-1}}^{t_{i}})$, add to $V_T$ a leaf $a_{t_i}^k$ with height $t_{i}$;
\item for $i>1$, if $a_{t_i}^k=\psi_{t_{i-1}}^{t_{i}}(a_{t_{i-1}}^s)=\psi_{t_{i-1}}^{t_{i}}(a_{t_{i-1}}^r)$, with $a_{t_{i-1}}^s$ and $a_{t_{i-1}}^r$ distinct basis elements in $a_{t_{i-1}}$, add a vertex $a_{t_i}^k$ with height $t_{i}$, and add edges so that the previously added vertices
\[
v = \arg\max \{h_T(v')\mid  v' \in V_T \text{ s.t. }\psi_{t_{i-1}}^{t_{v'}}(v')=a_{t_{i-1}}^s  \}
\]
and
\[
w = \arg\max \{h_T(w')\mid  w' \in V_T \text{ s.t. }\psi_{t_{i-1}}^{t_{w'}}(w')=a_{t_{i-1}}^r  \}.
\]
 connect with the newly added vertex $a_{t_i}^k$. Non-binary mergings
\[
a_{t_i}^k
=
\psi_{t_{i-1}}^{t_i}(a_{t_{i-1}}^{s_1})
=
\cdots
=
\psi_{t_{i-1}}^{t_i}(a_{t_{i-1}}^{s_\ell}),
\qquad \ell\geq 2,
\]
are handled analogously: one adds a single vertex \(a_{t_i}^k\) at height \(t_i\)
and connects it to the previously added vertices corresponding to
\(a_{t_{i-1}}^{s_1},\ldots,a_{t_{i-1}}^{s_\ell}\).
\end{itemize}

The last merging happens at height $t_n$ and, by construction, at height $t_n$ there is only one point, which is the root of the tree structure.

These rules define a tree structure with a monotone increasing height function $h_T$. In fact, edges are induced by maps $\psi_t^{t'}$ with $t<t'$ and thus no cycles are generated, and the function $h_T$ must be increasing. Moreover, $\psi_t^{t_n}(a_i^t)=a_1^{t_n}$ for every $i$ and $t<t_n$ and thus the graph is path-connected.

\section{Merge Trees vs Persistence Diagrams}
\label{sec:examples_vs_PD}

In this section, we present a detailed comparison between persistence diagrams and merge trees, drawing on existing literature, theoretical results, and practical examples.

\begin{prop} \label{prop: from_merge_to_PD}
If $f:X\rightarrow \mathbb{R}$ satisfies (A0), the associated zero dimensional $PD(f)$ can be obtained from any tree in the isomorphism class of the associated merge tree $T_f.$
\end{prop}

\begin{figure}[H]
\centering
\setlength{\tabcolsep}{2pt} %
\renewcommand{\arraystretch}{0.6} %
\begin{tabular}{ccc}
\includegraphics[width=.32\textwidth]{Immagini_paper/f_0.png} &
\includegraphics[width=.32\textwidth]{Immagini_paper/PD_0.png} &
\includegraphics[width=.32\textwidth]{Immagini_paper/T_0.png} \\
\includegraphics[width=.32\textwidth]{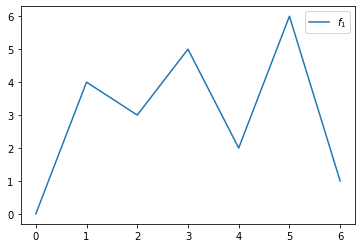} &
\includegraphics[width=.32\textwidth]{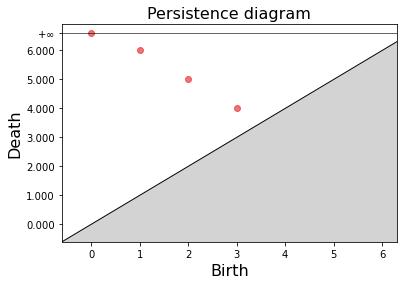} &
\includegraphics[width=.32\textwidth]{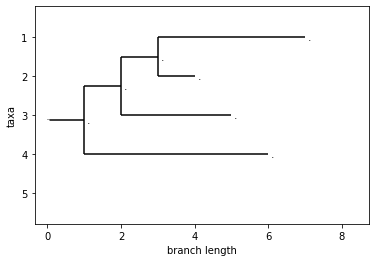} \\
\includegraphics[width=.32\textwidth]{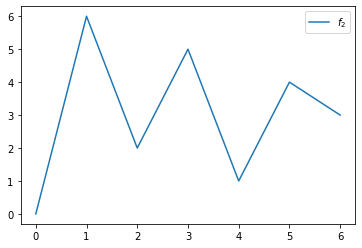} &
\includegraphics[width=.32\textwidth]{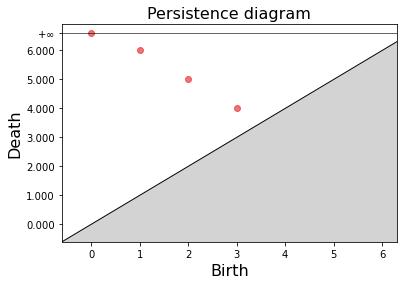} &
\includegraphics[width=.32\textwidth]{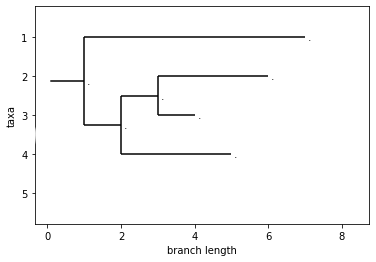} \\
\includegraphics[width=.32\textwidth]{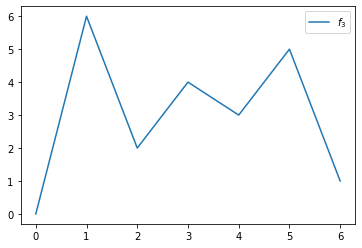} &
\includegraphics[width=.32\textwidth]{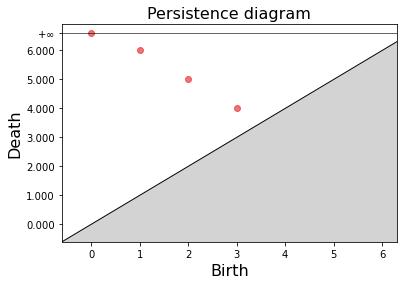} &
\includegraphics[width=.32\textwidth]{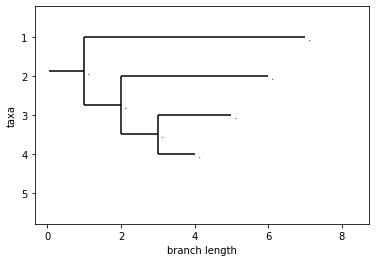} \\
\end{tabular}

\vspace{-2mm}
\caption{A visual example highlighting differences between PDs and merge trees. We consider four functions all associated to the same $0$-dimensional $PD$ but to different merge trees. The first column shows the functions, the second the associated $PD$, and the third the merge tree. All merge trees are truncated at height $7$; see \Cref{sec:merge_and_weight}.}
\label{fig:PD}
\end{figure}

A persistence diagram and a merge tree are not equivalent representations of a function.
Indeed, persistence diagrams do not record information about the merging components: as already mentioned, the death of a path-connected component could be caused by its merging with any other alive component at death-time.
This implies that, for a given persistence diagram PD, there might be more than one merge tree associated to the diagram: the birth and death of the path-connected components of each merge tree coincide with those of the PD, but the merging structure is different from merge tree to merge tree \citep{kanari2020trees, curry2021trees, pegoraro2025finitely, curry2021decorated, mergegrams, smith2022families, dlotko2025combinatorial}. In particular, the works
of \cite{kanari2020trees} and of \cite{curry2021trees} formally address this point, by providing explicit formulas for the \emph{tree realization number}: the number of trees associated to the same persistence diagram. This number can be very high, being $n!$ if $n$ is the number of points in the diagram, for certain configurations of the persistence diagram.
This is the case, for instance, with hierarchical clustering dendrograms with $n$ leaves: all leaves are born at height $0$, and so, at the first merging point, each of the $n$ leaves can merge with any of the $n-1$ remaining ones. At the following merging step we have $n-1$ clusters and each one of them can merge with the other $n-2$ and so on.
Thus, depending on the structure of the persistence diagram representing a function \(f\), the associated merge tree could contain much more information regarding \(f;\) from a different perspective, merge trees can discriminate between functions which are indistinguishable for persistent diagrams.
To see some easy examples of how merge trees capture also the local merging structure of the components which persistent diagrams cannot distinguish, see \Cref{fig:PD}.

In the following subsections, we present examples designed to emphasize the differences between persistence diagrams and merge trees.
Specifically, \Cref{sec:exampleI} provides further intuition on the distinction between functions that can be differentiated by merge trees but share the same persistence diagram. \Cref{sec:exampleII} and \Cref{sec:exampleIII} focus on qualitatively exploring the types of variability between functions that are better captured by persistence diagrams and merge trees. These are analyzed with the $1$-Wasserstein metric and the edit distance, respectively.

\subsection{Example I}
\label{sec:exampleI}

In this first example we produce a set of functions which are all described by the same persistence diagram but are distinguished by merge trees.

We exploit the fact that, for continuous functions of a single real variable, the merging structure of path-connected components (and consequently the tree structure \( T \)) is determined by the distribution of local minima relative to the positions of local maxima.

To illustrate this, we construct a simple toy example: we define a set of functions, each characterized by a single very high peak and several smaller peaks of equal height. The distinguishing feature of these functions lies in the varying arrangements of the smaller peaks relative to the highest one.

For $i=0,\ldots,9$, let $g_i:[0,11]\rightarrow \mathbb{R}$ be such that $g_i \equiv 0$ on
$[0,11]- [i+1/3,i+2/3]$ while, on
$[i+1/3,i+2/3]$, \(g_i\) is the linear interpolation of $(i+1/3,0)$, $(i+1/2,1)$ and $(i+2/3,0)$. Then, for $i=0,\ldots,9$, define $G_i$  as $G_i\equiv 0$ on
$[0,11]- [i+2/3,i+1]$ while, on $[i+2/3,i],$  $G_i$ is the
linear interpolation of $(i+2/3,0)$, $(i+3/4,5)$ and $(i+1,0)$.

\begin{figure}[H]
\centering
\setlength{\tabcolsep}{2pt}
\renewcommand{\arraystretch}{0.7}

\begin{tabular}{@{}ccc@{}}
\includegraphics[width=.32\textwidth]{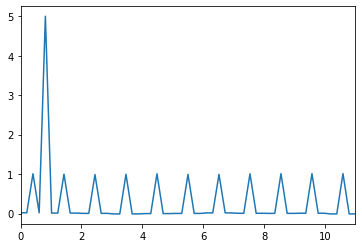} &
\includegraphics[width=.32\textwidth]{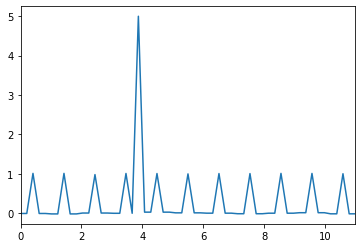} &
\includegraphics[width=.32\textwidth]{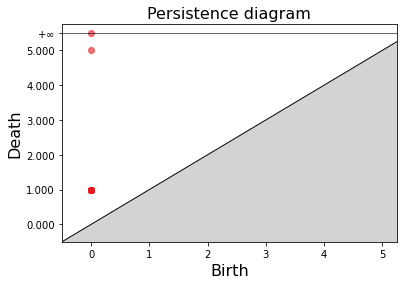} \\
\small (a) $f_0$ &
\small (b) $f_3$ &
\small (c) PD
\end{tabular}

\vspace{1mm}

\begin{tabular}{@{}ccc@{}}
\includegraphics[width=.32\textwidth]{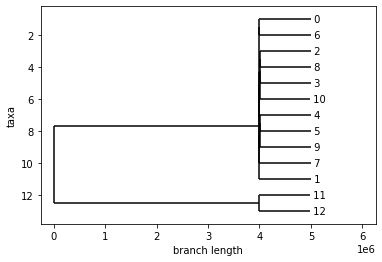} &
\includegraphics[width=.32\textwidth]{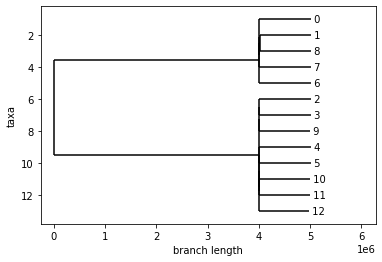} &
\includegraphics[width=.32\textwidth]{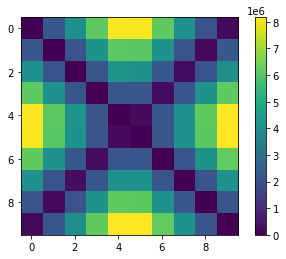} \\
\small (d) $(T_{f_0},h_{f_0})$ &
\small (e) $(T_{f_3},h_{f_3})$ &
\small (f) distance matrix
\end{tabular}

\vspace{1mm}

\includegraphics[width=.50\textwidth]{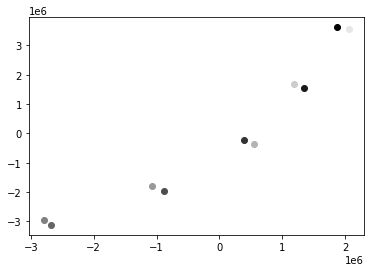}
\\
\small (g) MDS embedding

\vspace{-1mm}
\caption{Plots for the simulated scenario in \Cref{sec:exampleI}.
(a)--(b) Example functions ($f_0$ and $f_3$): the smaller peaks shift relative to the highest one.
(d)--(e) Corresponding merge trees: structural changes reflect the different peak disposition.
(c) Persistence diagram shared by the dataset; $(0,1)$ has multiplicity equal to \#local minima minus $1$.
(f) Pairwise merge-tree distance matrix.
(g) MDS embedding in $\mathbb{R}^2$ of (f); grayscale encodes the ordering (white $\to$ black).}
\label{fig:ex_I}
\end{figure}

Then, for $i=0,\ldots,9$, $f_i$ is obtained as follows:
\[
f_i = G_i+ \sum_{j=0}^{10} g_j.
\]
The functions $f_0$ and $f_3$ are displayed in \Cref{fig:ex_I}. Note that the first and the last peak of every function, by construction, are always small peaks.
The key point is that for every path-connected component we are not changing any of the corresponding critical values and thus the associated persistence diagram is always the same (see again \Cref{fig:ex_I}).

The shortest edit path between two merge trees $T_{f_i}$ and $T_{f_{i+1}}$  is given by the deletion of one leaf so that distribution of leaves is the same for the two trees. The more the distributions of peaks in the two trees are different, the more leaves need to be deleted in both trees to find the shortest path between them. Thus, if we fix the first line of the matrix in \Cref{fig:ex_I}, we see that going left-to-right the distance at first gradually increases. It is also evident that, from a certain point on, the distance decreases to the point of reaching almost zero. This is because
the first function (the one in which the highest peak is the second peak) and the last function (the one in which the highest peak is the second-to-last peak) can be obtained one from the other via a $y$-axis symmetry and a translation -- i.e.
 $x \mapsto -x$ (reflection on the $y$-axis) and $x \mapsto v+x$, with $v\in\mathbb{R}$ fixed (translation) --, these transformations being homeomorphisms of the abscissa.  Similarly, the second function is equal, up to homeomorphic alignment, to the third-to-last one, etc.; thus, because  of \Cref{prop:invariance}, their merge trees are the same. To sum up the situation depicted in the first row of the distance matrix in \Cref{fig:ex_I}, first we get (left-to-right) farther away from the first merge tree, and then we return closer to it. This intuition is confirmed by looking at the multidimensional scaling (MDS) embedding in $\mathbb{R}^2$ of the pairwise distance matrix (see \Cref{fig:ex_I} - note that the shades of gray reflect, from white to black, the ordering of the merge trees). The discrepancies between the couple of points are caused by numerical errors.

\subsection{Example II}
\label{sec:exampleII}

In this second example we want to produce a situation in which the variability between functional data is better captured by PDs than by merge trees.
Accordingly, we generate two clusters of functional data such that membership of a function in one cluster or the other should depend on the amplitude of its oscillations and not on the merging structure of its path-connected components. We then look at the matrices of pairwise distances between functions, comparing merge tree and persistence diagram representations in terms of their goodness in identifying the clustering structure.

\begin{figure}
	\centering
	\begin{subfigure}[c]{0.31\textwidth}
		\centering
		\includegraphics[width = \textwidth]{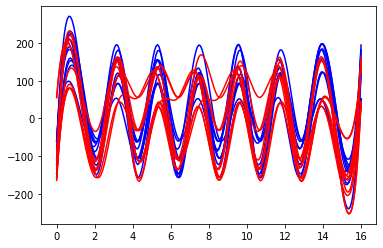}
		\caption{Data being plotted with different colours depending on the cluster.}
		\label{fig:sim_1_dati}
	\end{subfigure}
	\begin{subfigure}[c]{0.31\textwidth}
    	\centering
    	\includegraphics[width = \textwidth]{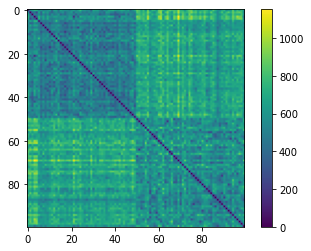}
    	\caption{Merge trees pairwise distances.}
    	\label{fig:t_sim_1}
    \end{subfigure}
    \begin{subfigure}[c]{0.31\textwidth}
    	\centering
    	\includegraphics[width = \textwidth]{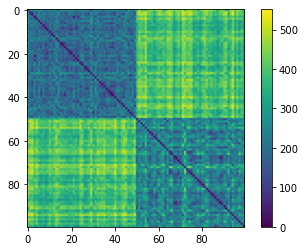}
    	\caption{$PDs$ pairwise distances.}
    	\label{fig:pd_sim_1}
    \end{subfigure}
\caption{Example II. In the first row we can see a few data from the two clusters. In the second row we see the matrices of pairwise distance extracted with trees and $PD$s. The data are ordered according to their cluster. It is clear how $PD$s perform much better in separating the two clusters.}
\label{fig:sim_1}
\end{figure}

To generate each cluster of functions, we draw, for each cluster, an independent sample of 16 critical points, 8 maxima and 8 minima, from two univariate Gaussian distributions with means equal to $+100$ for maxima and to $-100$ for minima, respectively. The standard deviations of the two Gaussian distributions are the same and they are set equal to $50.$
To generate a function inside a cluster, we draw a random permutation of 8 elements and we reorder, according to this permutation, both the set of maxima and the set of minima associated to the cluster. Then,  we take a regular grid of 16 nodes on the abscissa axis: on the ordinate axis we associate to the first point on the grid the first minimum, to the second the first maximum, to the third the second minimum and so on. To obtain a function we interpolate such points with a cubic spline. We thus generate $50$ functions in each cluster.
The key point is that, within the same cluster, the critical points are the same but for their order,  while the two clusters correspond to two different sets of critical points.

In this example, we expect that the clustering structure carried by the amplitude of the functions will be shadowed by the differences in the merging order, when adopting the merge tree representation; persistence diagrams should perform much better because they are less sensitive to peak reordering.
This is in fact confirmed by inspecting the distance matrices in  \Cref{fig:t_sim_1} and \Cref{fig:pd_sim_1}.

\subsection{Example III}
\label{sec:exampleIII}

Here we reverse the state of affairs and we set the feature for discriminating between clusters to be the merging structure of the functions. Hence, we generate two clusters of functions: the members of each cluster have the same merging structure which is however different between clusters.

\begin{figure}
	\begin{subfigure}[c]{\textwidth}
    	\centering
    	\includegraphics[width =0.45\textwidth]{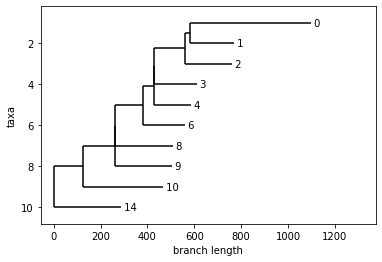}
    	\includegraphics[width =0.45\textwidth]{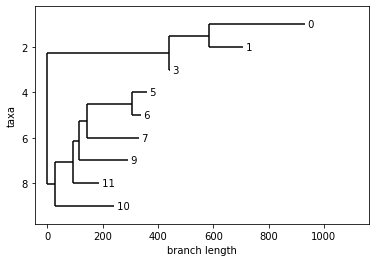}
    	\caption{Tree structures of the two clusters: left the first and right the second.}
    	\label{fig:struct_sim_2}
    \end{subfigure}

	\centering
	\begin{subfigure}[c]{0.31\textwidth}
		\centering
		\includegraphics[width = \textwidth]{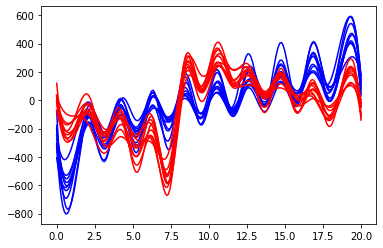}
		\caption{Simulated data.}
		\label{fig:sim_2_dati}
	\end{subfigure}
	\begin{subfigure}[c]{0.31\textwidth}
    	\centering
    	\includegraphics[width = \textwidth]{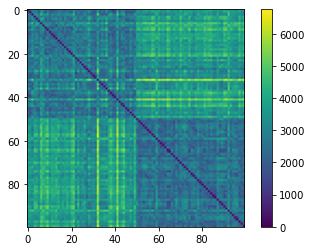}
    	\caption{Merge trees pairwise distances.}
    	\label{fig:t_sim_2}
    \end{subfigure}
    \begin{subfigure}[c]{0.31\textwidth}
    	\centering
    	\includegraphics[width = \textwidth]{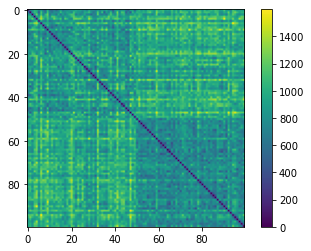}
    	\caption{PD pairwise distances.}
    	\label{fig:pd_sim_2}
    \end{subfigure}
\caption{Example III. In the first row we can find the tree structures associated to the two clusters. In the second row, leftmost plot, we can see a few data from the two clusters. In the central and rightmost column of the second row we see the matrices of pairwise distances between merge tree representations and $PD$s, respectively. The data are ordered according to their cluster. It is clear how in this example merge trees are more suitable to separate the two clusters.}
\label{fig:sim_2}
\end{figure}

To generate the two clusters of 50 functions each, we first draw an independent sample of 10 critical values, $10$ maxima and $10$ minima, shared between the clusters. Such samples are drawn from Gaussian distributions with means $100$ and $-100$ respectively and standard deviation $200$.
Given a regular grid of $20$ nodes on the abscissa axis, on the ordinate axis we associate to the first point of the grid a maximum, to the second a minimum, and so on, as in Example I. To generate every member of one cluster or the other, we add to the ordinate of each maximum or minimum critical point a random noise generated by a Gaussian with mean 0 and standard deviation $100$.
Then we reorder such points following a cluster-specific order. And, lastly, we interpolate with a cubic spline.
We remark that the ordering of the maxima and that of the minima now becomes essential. For the two clusters, these orderings are fixed but different and they are set as follows ($0$ indicates the smallest value and $9$ being the largest value):
\begin{itemize}
\item first cluster: maxima are ordered along the sequence $(0, 1, 2, 3, 4, 5, 6, 7, 8, 9)$, minima along the sequence $(0, 1, 2, 3, 4, 5, 6, 7, 8, 9)$;
\item second cluster: maxima are ordered along the sequence $(3, 2, 1, 0, 8, 9, 7, 6, 4, 5)$, minima along the sequence $(3, 2, 1, 0, 8, 9, 7, 6, 4, 5)$.
\end{itemize}
Such different orderings provide non-isomorphic tree structures for the merge trees associated to the functions of the two clusters, as we can see in  \Cref{fig:struct_sim_2}, while keeping a similar structure in terms of persistence diagrams.

In this example, we expect $PDs$ to be unable to recognise the clustering structure among the data; indeed, the only discriminant feature available to $PDs$ is the different height of critical points, but this bears little information about the clusters.

We can visually observe this by comparing  \Cref{fig:t_sim_2} with  \Cref{fig:pd_sim_2}.

\section{Kernel and Bandwidth Selection for the Case Study}\label{sec:band}

In this section, we take a closer look at the smoothing carried out on the  curvature and radius functions involved in the case study in \Cref{sec:case_study}.

To ensure the validity of the consistency results presented in \cite{mack1989derivative}, we utilize the following kernel function:
\[
K(x) = \frac{35}{32} \cdot (1 - 3x^2 + 3x^4 - x^6),
\]
which is supported on the interval \([-1, 1]\). As proven in \cite{muller1984smooth}, $K$ fits into the desired framework, belonging to the therein defined space of functions $\mathcal{M}_{0,2}$. The functional estimator, then, is:
\[
\hat{f}^h_n(x):=\frac{\sum_{i=1}^n y_i K(\frac{x-x_i}{h})}{\sum_{i=1}^n K(\frac{x-x_i}{h})},
\]
with $n$ being the cardinality of the sample $\{(x_i,y_i)\}_{i=1}^n$, which we assume iid from model (M), and $h>0$ being a fixed bandwidth we need to select.

Following \Cref{prop:comp_est}, the bandwidth \( h \) should be chosen to enhance the convergence of merge tree estimators, which depends on the convergence of \(\hat{f}_n^h\), \(D\hat{f}_n^h\), and \(D^2\hat{f}_n^h\). For the selected kernel \( K \), we employ Equation (3.6) from \cite{mack1989derivative}, setting \( k = 4 \) and \( \nu = 2 \). This yields the following optimal local bandwidth for derivative convergence:
\begin{equation}\label{eq:band}
h_x = \left( \frac{4! \mathbb{E}(\Yvar^2 \mid  \Xvar = x) \int_{-1}^1 D^2 K(t)^2 \, dt}{D^4 f(x)^2 \left( \int_{-1}^1 D^2 K(t) t^4 \, dt \right)} \right)^{1/9} \cdot n^{-1/9}.
\end{equation}

\begin{figure}
	\centering
    	\includegraphics[width = 0.47\textwidth]{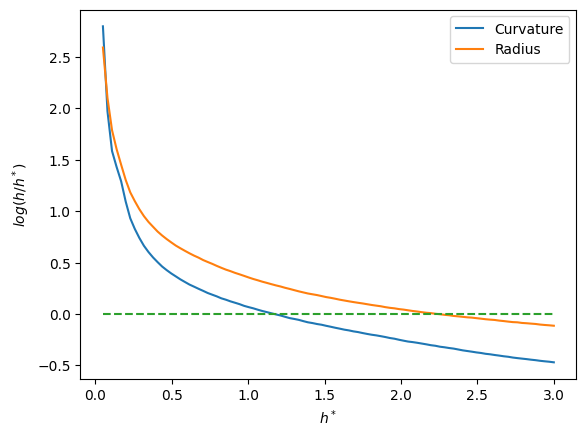}
    	\caption{Plot of $h^*$ versus $\log(h/h^*)$, for both curvature and radius - with the log-scale being used to emphasize the behaviour of the curves for higher values of $h^*$. The green dotted line represents the threshold after which the bandwidth $h^*$ oversmooths the data. The values $h_c$ and $h_r$ are chosen at the intersection of the dotted line with the respective curves. The curves clearly show that the curvature data is much smoother compared to radius, as the ratio $h/h^*$ decreases much faster.}
    	\label{fig:bandwidth}
\end{figure}

Using this local bandwidth requires estimating two additional quantities: \(\mathbb{E}(\Yvar^2 \mid  \Xvar = x)\) and \(D^4 f(x)^2\).
In line with \cite{mack1989derivative}, we estimate \(\mathbb{E}(\Yvar^2 \mid  \Xvar = x)\) using another kernel regression, where the bandwidth is selected via the Generalized Cross Validation (GCV) Score \citep{kauermann2004generalized}. While GCV is known to be susceptible to overfitting, it provides a practical approach in this context.
For \(D^4 f(x)^2\), we estimate it using \(D^4 \hat{f}_n^{h^*}(x)^2\), which involves another bandwidth \(h^*\). Although \cite{mack1989derivative} does not offer specific guidelines for choosing \(h^*\), we treat it as an additional hyperparameter and optimize it to refine the estimate of \(h_x\).
The following iterative procedure was applied to both curvature and radius data (independently):
\begin{enumerate}
\item \emph{Initialization:} fix \( h^* \in (0,3].\)
\item  \emph{Local Bandwidth Estimation}:
   For the \(k\)-th patient’s data \(\{(x_i, y_i)\}_{i=1}^n\), compute the local bandwidth \( h_{x_j} \) by substituting \( D^4 \hat{f}_n^{h^*}(x_j)^2 \) into \Cref{eq:band}.
\item \emph{Patient-Specific Bandwidth}:
   Compute \(\hat{h}_k^*\) as the average of \( h_{x_j} \) across all points \( x_j \).
\item \emph{Global Bandwidth}:
   Calculate \( h \) as the average of \(\hat{h}_k^*\) over all patients.
\item \emph{Bandwidth Comparison}:
   Compare \( h \) and \( h^* \),
   \begin{itemize}
   \item If \( h^* \gg h \), the initial guess \( h^* \) is oversmoothing the data.
   \item Conversely, if \( h^* \ll h \), the initial \( h^* \) is undersmoothing the data.
   \end{itemize}
   This relationship is illustrated in \Cref{fig:bandwidth}, where the ratio \( h/h^* \) decreases as \( h^* \) increases, and vice versa.
\item \emph{Final bandwidth selection}:
Repeat the procedure (1)-(5) with a different initialization, taken from a grid, until we find $h^*$ such that \( h / h^* = 1 \).
\end{enumerate}

This process yielded \( h_r = 2.25 \) for the radius data and \( h_c = 1.15 \) for the curvature data. The results align with expectations, as the radius functions are significantly more oscillatory than the curvature functions, which had already undergone a smoothing step.

To validate these choices, we focused on the curvature data and evaluated the pointwise differences between the function before and after the proposed additional smoothing. These differences were normalized by the range of the raw data.
The normalized pointwise errors presented a median of $0.0017$ and an IQR of $0.0044$.
The results demonstrate that the additional smoothing is consistent with the earlier work in \cite{aneurisk_splines}.

\section{Computational runtimes}\label{sec:comp}

We now present a simulation study illustrating the computational runtimes of the implementation used in this work.
The data were generated as follows.

For each fixed \( i \), we sample two point clouds
\( C_i^k = \{(x_j^k, y_j^k) \mid j = 1, \ldots, n_i\} \subset \mathbb{R}^2 \),
with \( k = 1,2 \), according to:
\[
\begin{aligned}
x_j^k &\sim^{iid} \mathcal{N}(5, \sigma_{x,k}), \quad j = 1,\ldots,n_i, \\
y_j^k &\sim^{iid} \mathcal{N}(5, \sigma_{y,k}), \quad j = 1,\ldots,n_i, \\
\sigma_{x,k} &\sim |\mathcal{N}(3,1)|, \\
\sigma_{y,k} &\sim |\mathcal{N}(3,1)|,
\end{aligned}
\]
where \( \mathcal{N}(\mu, \sigma) \) denotes a Gaussian distribution.

From each point cloud \( C_i^k \), we construct the single-linkage hierarchical clustering
dendrogram \( T_{C_i^k} \) (equivalently, the merge tree associated with the
Vietoris--Rips filtration of \( C_i^k \)) and compute the pairwise distance
\( d_E(T_{C_i^1}, T_{C_i^2}) \).
The parameter \( n_i \) determines the number of leaves in the resulting trees
and is fixed prior to sampling the point clouds \( C_i^k \).
Since each \( T_{C_i^k} \) is a binary tree, the corresponding merge tree contains
\( 2n_i - 1 \) edges.

\begin{figure}[H]
	\centering
	\begin{subfigure}[c]{0.45\textwidth}
		\centering
		\includegraphics[width = \textwidth]{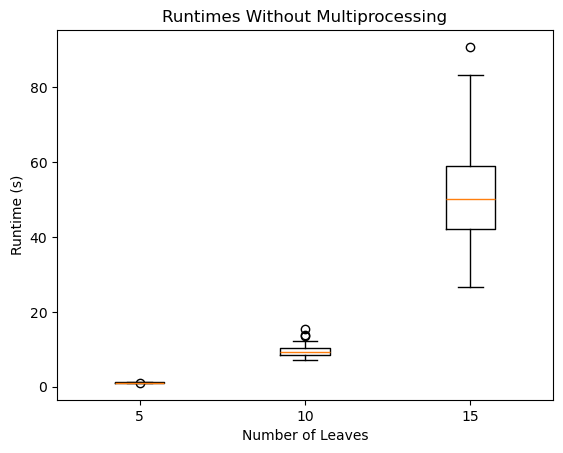}
	\end{subfigure}
	\begin{subfigure}[c]{0.45\textwidth}
		\centering
		\includegraphics[width = \textwidth]{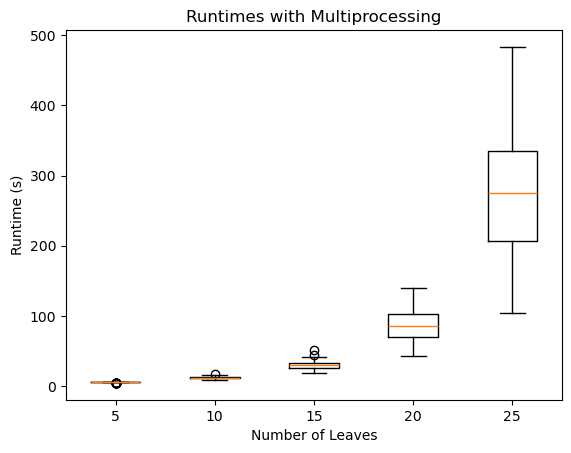}
	\end{subfigure}

	\caption{Boxplots of the runtimes of $d_E$, as a function of the number of leaves in each merge tree, without (left) and with (right) parallelization of the Binary Optimization Problems. The medians of the boxplots of the left figure, from left to right, are: 1.13s, 9.17s, 50.24s.  The medians of the boxplots of the right figure, from left to right, are: 6.04s,
12.40s, 30.18s, 86.28s, 275.67s.}
	\label{fig:runtimes}
\end{figure}

The parameter \( n_i \) is varied over the set \( \{5, 10, 15, 20, 25\} \) for the parallelized experiment and \( \{5, 10, 15\} \) otherwise;
for each value, we independently sample 100 pairs of point clouds.
The distributions of the resulting runtimes are summarized by the boxplots shown in
\Cref{fig:runtimes}.
For reference, the number of leaves in the merge trees considered in the
\textit{AneuRisk65} case study ranged from 5 to 14.

The computation of \( d_E \) follows the algorithm introduced in
\cite{pegoraro2023persistence}, which is based on binary linear programming (BLP).
This method entails solving \( O(N \cdot M) \) BLP problems, each involving
\( O(N \log N \cdot M \log M) \) variables and
\( O(\log_2 N + \log_2 M) \) constraints.
\Cref{fig:runtimes} reports runtimes both with and without solving in parallel some of the BLP problems.

For reference, the classical edit distance between unlabeled trees can be obtained
by solving \( O(N \cdot M) \) BLP problems with \( O(N \cdot M) \) variables
and the same number of constraints \cite{TED}.
Furthermore, as shown in \cite{pegoraro2025finitely}, one can incorporate heuristics
from \cite{wetzels2025accelerating} to accelerate computations or rely on \cite{wetzels2022deformation} to derive a polynomial-time (albeit unstable) upper bound.

We conclude the simulation study with the following disclaimer.
Given the scope of the present work, our objective was not to optimize the code
or the computational environment.
For instance, all computations were carried out in native \texttt{Python} using a
non-commercial license of the Gurobi~12 solver, without access to ample dedicated RAM.
As a result, the solver consistently wrote intermediate data to disk when parallelizing
across multiple BLP problems.
Furthermore, the implementation was designed to accommodate more general uses of the
distance introduced in \cite{pegoraro2023persistence}, which was also employed in the
simulations of \cite{pegoraro2024finitelyfunc}.
Consequently, the reported runtimes should not be interpreted as indicative of the
potential efficiency of a fully optimized implementation.
For reference, see state-of-the-art implementations of metrics with comparable
computational complexity, such as \cite{TED}.

\section{Proofs}\label{sec:proofs_FDA}

\subsection{\underline{\textit{Proof of}  \Cref{prop:invariance}.}}

Let $f:X\rightarrow \mathbb{R}$ be a bounded function defined on a path-connected topological space $X$ and let $\varphi:Y\rightarrow X$ be a homeomorphism.
We prove that the merge trees
associated to the function $f$ and $f'=f\circ \varphi$ are isomorphic.

We know:
\[
Y_t=(f')^{-1}((-\infty,t])
=\{y\in Y\mid f'(y)\le t\}
=\{y\in Y\mid f(\varphi(y))\le t\}
=\varphi^{-1}(X_t).
\]
This means that $y\in Y_t$ if and only if $\varphi(y)\in X_t$, and so $Y_t=\varphi^{-1}(X_t)$.
Since the restriction of a homeomorphism is still a homeomorphism, we can take its inverse, and by the composition properties of $\pi_0$, we obtain
$\pi_0(X_t)\cong \pi_0(Y_t)$.
Given $t'<t$, we have the following commutative diagram:

\[
\begin{tikzcd}
X_{t'}\ar[d,"\varphi"]\ar[r]&X_t\ar[d,"\varphi"]\\
Y_{t'}\ar[r]&Y_t
\end{tikzcd}
\]
and passing to path-connected components:
\[
\begin{tikzcd}
\pi_0(X_{t'})\ar[d,"\pi_0(\varphi)"]\ar[r," "]&\pi_0(X_t)\ar[d,"\pi_0(\varphi)"]\\
\pi_0(Y_{t'})\ar[r," "]&\pi_0(Y_t).
\end{tikzcd}
\]
The diagram gives the isomorphism of merge trees.

\hfill $\blacksquare$

\subsection{\underline{\textit{Proof of}  \Cref{prop: from_merge_to_PD}.}
}

Each leaf in $(T,h_f)$ corresponds to a point in $PD(f)$.
The $x$ coordinate of each point is given by its height, which can be retrieved through $h_f$. Consider $v\in L_T$ and let $\gamma_v$ be the ordered set $\{v'\in V_T \mid  v'\geq v\}$ i.e. the path from $v$ to $r_T$. The $y$ coordinate of the points associated to $v$ is the minimal height at which $\gamma_v$ intersects another path $\gamma_l$, with $l$ being a leaf with height less than $v$.
\hfill$\blacksquare$

\subsection{\underline{\textit{Proof of}  \Cref{teo:stability}.}}

To avoid dealing with unpleasant technicalities we work under the hypothesis
that for any merge tree \((T,h_T)\), \(h_T\) is an injective function. We call
this assumption (G). It is used only to avoid extra bookkeeping when several
incomparable vertices have the same height, so that the choice of representatives
in the discrete vertex sets is unambiguous. Comparable vertices already have
distinct heights.

First we define the \emph{least common ancestor} (LCA) of a set of vertices in a merge tree.

\begin{defi}
Given a merge tree $(T,h_T)$ and set of vertices $A=\{a_1,\ldots,a_n\}\subset V_T$, define $\LCA(a_1,\ldots,a_n)=\min\bigcap^n_{i=1} \{v\in V_T  \mid  v\geq a_i\}$.
\end{defi}

Consider now two functions $f,g:X\rightarrow \R$ satisfying assumptions (A1), and such that $\text{sup}_{x\in X}\mid  f(x)-g(x)\mid  \leq \varepsilon$.  Let $(F,h_f)$ and $(G,h_g)$ be their associated merge trees.
For \(t \in \mathbb{R},\) we set $X^f_t=f^{-1}((-\infty,t])$.
Since $\mid  f(x)-g(x)\mid  \leq \varepsilon$ we have $X^f_t\subset X^g_{t+\varepsilon}$ and of course $X^g_t\subset X^f_{t+\varepsilon}$.
Set:
\begin{align*}
    &f_{t}^{t+\varepsilon}:= \pi_0(X^f_t\hookrightarrow X^f_{t+\varepsilon}),\\
    &g_{t}^{t+\varepsilon}:= \pi_0(X^g_t \hookrightarrow X^g_{t+\varepsilon}),\\
    &\alpha_{t}^{t+\varepsilon}:= \pi_0(X^f_t \hookrightarrow X^g_{t+\varepsilon}),\\
    &\beta_{t}^{t+\varepsilon}:= \pi_0(X^g_t \hookrightarrow X^f_{t+\varepsilon}).
\end{align*}

And call $F_t:=\pi_0(X_t^f)$ and $G_t:=\pi_0(X_t^g)$. With these pieces of notation we write following zig-zag diagram:
\[
\begin{tikzcd}
F_t\ar[r,"f_t^{t+\varepsilon}"]\ar[dr]&F_{t+\varepsilon}\ar[r,"..."]&F_{t'}\ar[dr]\ar[r,"f_{t'}^{t'+\varepsilon}"]&F_{t'+\varepsilon}\\
G_t\ar[ur]\ar[r,"g_t^{t+\varepsilon}"]&G_{t+\varepsilon}\ar[r,"..."]&G_{t'}\ar[ur]\ar[r,"g_{t'}^{t'+\varepsilon}"]&G_{t'+\varepsilon}.
\end{tikzcd}
\]
Note that the diagonal maps are $\alpha: F_t\rightarrow G_{t+\varepsilon}$ and $\beta:G_t\rightarrow F_{t+\varepsilon}$. Lastly, if $a_t'=f_{t}^{t'}(a_t)$, we say that $a_t<a_{t'}$.

If we collect together the path-connected components $\{F_t\}_{t\in\mathbb{R}}$ and $\{G_t\}_{t\in\mathbb{R}}$ taking the disjoint unions $\mathbf{F}:=\coprod_{t\in\mathbb{R}} F_t$ and $\mathbf{G}:=\coprod_{t\in\mathbb{R}} G_t$ we can define $\alpha:\mathbf{F}\rightarrow \mathbf{G}$
 and $\beta:\mathbf{G}\rightarrow \mathbf{F}$, so that
given $a_t \in F_t$, $\alpha(a_t):=\alpha_{t}^{t+\varepsilon}(a_t)$.
Given $a_{t'}\in \mathbf{F}$, we also set $h_f(a_{t'})=t'$. The same for $h_g$.

We point out that the vertices of the merge trees $F$ and $G$ are contained in some $F_t$ or $G_t$, respectively, and thus we have $V_F\hookrightarrow \mathbf{F}$ and $V_G\hookrightarrow \mathbf{G}$. Given $v\in V_F$, we will often write $v\in \mathbf{F}$  without explicitly mentioning the inclusion map, and thus, for instance, take $\alpha(v)$. Note that the partial order we defined for $\mathbf{F}$ and $\mathbf{G}$ is compatible with the one of the vertices of the merge trees.

In a more technical language, $\mathbf{F}$ and $\mathbf{G}$ are the display posets of the two persistence sets $\pi_0(X^f)$ and $\pi_0(X^g)$
\citep{curry2021decorated}, but we want to avoid introducing such more technical definition.
The work of \cite{merge} shows that $\alpha$ and $\beta$ give an $\varepsilon$-interleaving of merge trees (see \cite{merge}), which, by \cite{interl_approx}, implies that $\alpha$ satisfies:

\begin{itemize}
\item $h_g(\alpha(a_{t}))=h_f(a_{t})+\varepsilon=t+\varepsilon$ for all $a_{t}\in\mathbf{F}$
\item if $\alpha(a_t)<\alpha(a_{t'})$ then there is $a_{t''}$ such that $a_t<a_{t''}$, $a_{t'}\leq a_{t''}$ and $t''-t' <\varepsilon$.
\item if $b_{t'}\in\mathbf{G}-\alpha(\mathbf{F})$, then, given $b_t = \min \{b_{t''}>b_{t'}  \mid  b_{t''}\in\alpha(\mathbf{F})\}$, we have $t-t'\leq 2\varepsilon$.
\end{itemize}

A map which satisfies these properties is called $\varepsilon$-good \citep{interl_approx}.

To bridge between the continuous nature of $\mathbf{F}$ and
$\mathbf{G}$ and the discrete $(F,h_f)$ and $(G,h_g)$, we define the following maps:
\[
L_f:\mathbf{F}\rightarrow V_{F}
\]
$L_f(a_t)=\max\{v\in V_{F} \mid  v\leq a_t\}$ and similarly for $L_g$. Leveraging these definitions we set $\phi:= L_g\circ\alpha$
and
$\psi:= L_f\circ\beta$.

Finally we start building an edit path between $(F,h_f)$
and $(G,h_g)$. To do so we progressively add couples to an empty set $M$: couples of the form $(v,\mathfrak{D})$ mean that the vertex $v\in V_F$ is deleted, while $(\mathfrak{D},w)$ means that $w\in V_G$
is deleted, $(v,\mathfrak{G})$ means that the vertex $v\in V_F$ is ghosted, $(\mathfrak{G},w)$ means that $w\in V_G$
is ghosted and $(v,w)$ means that we shrink $(v,father(v))$ so that the weight of
$(v,father(v))$ becomes equal to the one of $(w,father(w))$. The set $M$ is a \emph{mapping}, as defined in \cite{pegoraro2023persistence}.

By working simultaneously on $F$ and $G$, we collect all the \virgolette{edits} in $M\subset \left( V_F\cup \{\mathfrak{D},\mathfrak{G}\}\right) \times \left( V_G \cup \{\mathfrak{D},\mathfrak{G}\}\right)$ and then, in the last part of the proof, we use $M$ to induce an edit path between $F$ and $G$.
We will call $\pi_F:M\rightarrow V_F\cup \{\mathfrak{D},\mathfrak{G}\}$
the projection on the first factor of the cartesian product, and $\pi_G$ the projection on the second.

\begin{center}
\textbf{Step 1: Leaves of $F$}
\end{center}

In this step we take care of the leaves of the merge tree $F$.

\smallskip

\textbf{Step 1.1: Selecting the Coupled Leaves}

\smallskip

Consider the following non-empty set of leaves:
\begin{equation}
\mathcal{L}_F=\{v\in L_F\mid  \nexists v'\in L_F \text{ such that }\alpha(v)< \alpha(v')\}
\end{equation}
We give a name to the condition used to assert if for a vertex $v\in L_F$, we have $v\in \mathcal{L}_F$:
\begin{itemize}
\item[$(a)$] $\nexists v'\in L_F \text{ such that }\alpha(v)< \alpha(v')$
\end{itemize}
so that we can more easily refer to it during the proof.
Observe that we never have $\alpha(v)=\alpha(v')$ thanks to condition $(G)$.

The set $\mathcal{L}_F$ is the set of leaves which will be coupled by $M$: we add to $M$ all the couples of the form $(v,\phi(v))$ with $v\in \mathcal{L}_F$ and add $(v,\mathfrak{D})$ for all $v\in L_F-\mathcal{L}_F$.

\begin{lem}\label{prop:L_T}
Given $v,v'\in \mathcal{L}_F$, then $\phi(v)\geq \phi(v')$ if and only if $v=v'$. Moreover, for every $v'\in L_F$ for which $(a)$ does not hold, there is $v\in\mathcal{L}_F$ such that $\alpha(v)<\alpha(v')$.

\begin{proof}
The first part of the proof reduces to observing that $\phi(v)\leq \phi(v')$ if and only if $\alpha(v)\leq \alpha(v')$.

Now consider $v'\in L_F$ such that $(a)$ does not hold.
We know there is $v_0$ such that $\alpha(v_0)<\alpha(v')$. If $v_0\in\mathcal{L}_F$ we are done, otherwise there is $v_1$ such that $\alpha(v_1)<\alpha(v_0)<\alpha(v')$. Note that $f(v_1)<f(v_0)$. Thus we can carry on this procedure until we find $v_i\in \mathcal{L}_F$. Note that $\arg\min_{a\in V_F} f(a) \in \mathcal{L}_F$, thus, in a finite number of steps we are done.
\end{proof}
\end{lem}

\smallskip
\textbf{Step 1.2: Height Bounds on Couples}
\smallskip

The following proposition represents an upper bound for the cost of some edits induced by the couples $(v,\phi(v))$ added to $M$.

\begin{lem}\label{prop:cost_leaves}
Given $v\in \mathcal{L}_F$, then $\mid  h_f(v)-h_g(\phi(v))\mid  \leq \varepsilon$.

\begin{proof}
Suppose the thesis does not hold.
Since $h_g(\phi(v))\leq h_f(v)+\varepsilon$, contradicting the thesis means that we have $v\in\mathcal{L}_F$ such that:
\begin{itemize}
\item[$(b)$] $h_f(v)-h_g(\phi(v))>\varepsilon $.
\end{itemize}
Let $w=\phi(v)$.
If $(b)$ holds, then $h_g(father(w))-h_g(w)>h_g(\alpha(v))-h_g(w)> 2\varepsilon$.
Let $v'=\psi(w)\leq \beta(w)$. Note that $h_f(v')<h_f(v)$. We have $\phi(v')\leq \alpha(v')\leq  \alpha(\beta(w))$. But since $h_g(father(w))-h_g(w)>2\varepsilon$, we also have
$\alpha(v')\leq \alpha(v)$ with $v'\neq v$ which is absurd by \Cref{prop:L_T}.
\end{proof}
\end{lem}

\smallskip
\textbf{Step 1.3: Cost Bound on Deletions}
\smallskip

We add to $M$ the couples $(x,\mathfrak{D})$ for every $x \notin \{v'>v\mid  v \in \mathcal{L}_F\}$.

\begin{lem}\label{prop:cost_deletions}
For any $x\notin \{v'\in V_F\mid  \exists v<v', v \in \mathcal{L}_F \}$, we have
$w_F((x,father(x)))\leq 2\varepsilon$.

\begin{proof}
We observe that, if $x\notin \{v'\in V_F\mid  \exists v<v', v \in \mathcal{L}_F \}$ then there
is $v\in \mathcal{L}_F$ such that $\alpha(v)<\alpha(x)$. By property (P2) of $\alpha$, since $h_f(v)<h_f(x)$, we have that $(x,\mathfrak{D})$ has cost at most $2\varepsilon$.
\end{proof}
\end{lem}

\begin{center}
\textbf{Step 2: Leaves and deletions of $G$}
\end{center}

Similarly we add to $M$ the couples $(\mathfrak{D},y)$ for every $y  \notin \{w'\in V_G\mid  \exists w<w', w \in \pi_G(M)\cap V_G\}$.

\begin{lem}\label{prop:G_leaves}
For any $y\in V_G$ such that $y\notin \{w'\in V_G\mid  \exists w<w', w \in \pi_G(M)\cap V_G\}$, we have $w_G((y,father(y)))\leq 2\varepsilon$.

\begin{proof}
Consider $\beta(y)$.
Let $v\leq \beta(y)$ leaf.
We have $\alpha(\beta(y))\geq \alpha(v)$ and $\alpha(\beta(y))\geq y$.
If $L_g(\alpha(\beta(y)))\neq y$ we are done since $h_g(\alpha(\beta(y)))= h_g(y)+2\varepsilon$.
Suppose instead $L_g(\alpha(\beta(y)))= y$.
We know that there is always $v_0\in \mathcal{L}_F$ such that
$\alpha(v_0)\leq \alpha(v)$ (either $v\notin \mathcal{L}_F$ or $v=v_0$).
Suppose $\alpha(v_0)\geq y$.
Since $\alpha(\beta(y))\geq \alpha(v) \geq \alpha(v_0)$, then $L_g(\alpha(v_0))= y$ and so $y\in \pi_G(M)$. Which is absurd.

Thus, we are left with $\alpha(v_0)\leq y$.
But this is absurd as well as it implies again $y\in \{w'\in V_G\mid  \exists w<w', w \in \pi_G(M)\cap V_G\}$.
\end{proof}
\end{lem}

\begin{center}
\textbf{Step 3: Internal Vertices}
\end{center}

Thanks to \Cref{prop:cost_deletions} and \Cref{prop:G_leaves} we can delete all $x\notin \{v'\in V_F\mid  \exists v<v', v \in \mathcal{L}_F \}$ and $y\in V_G$ such that $y\notin \{w'\in V_G\mid  \exists w<w', w \in \pi_G(M)\cap V_G\}$, each with cost at most $2\varepsilon$. Note that these deletions do not change the heights of any non deleted vertex.

Call $F_1$ and $G_1$ the two trees obtained after such deletions and after the ghosting of all the degree $2$ vertices arising - and consequently adding $(v,\mathfrak{G})$ or $(\mathfrak{G},w)$ to $M$ for all the ghosted vertices respectively in $F$ or $G$.
Let $\mathbf{F}_1$ and $\mathbf{G}_1$ be the subsets of, respectively, $\mathbf{F}$ and $\mathbf{G}$ induced by $F_1$ and $G_1$.

By construction
$\alpha_{\mid  \mathbf{F}_1}:\mathbf{F}_1\rightarrow \mathbf{G}_1$. Similarly
$\beta_{\mid  \mathbf{G}_1}:\mathbf{G}_1\rightarrow \mathbf{F}_1$, thanks to condition (a) and the properties of $\alpha$. Moreover $\alpha(\mathbf{F})=\alpha_{\mid  \mathbf{F}_1}(\mathbf{F}_1)$.
Thus $\alpha_{\mid  \mathbf{F}_1}$
is still $\varepsilon$-good.
Lastly, note that the leaves of $\mathbf{F}_1$ and $\mathbf{G}_1$ are in bijection via (the restriction of) $\phi$ and $\psi$.

In what follows, with an abuse of notation, we avoid explicitly writing the restriction of the maps $\alpha$ and $\beta$, implying that these are always considered as defined on the \virgolette{pruned} trees $F_1$ and $G_1$.

\smallskip
\textbf{Step 3.1: Auxiliary Results on Internal vertices}
\smallskip

In the upcoming lemmas, the roles of $\mathbf{F}_1$ and $\mathbf{G}_1$ can be exchanged .

\begin{lem}\label{prop:LCA_1}
Consider $x\in V_{F_1}-L_{F_1}$ and set $A=\{v\in V_{F_1}\mid  v<x\}\bigcap L_{F_1}$. Take $y= \LCA(\phi(A))$. Then $\mid  h_f(x)- h_g(y)\mid  \leq \varepsilon$.

\begin{proof}
For every $a\in A$ we know $\phi(a)<\phi(x)$ and thus $y\leq \phi(x)$. Thus $h_g(y)\leq h_f(x)+\varepsilon$. Suppose then $h_g(y)<h_f(x)-\varepsilon$. We know, $\beta(y)\geq x$, as $\beta(y)\geq\beta(\phi(a))\geq a$. Which is absurd as $h_f(\beta(y))=h_g(y)+\varepsilon<h_f(x)$.
\end{proof}
\end{lem}

\begin{lem}\label{prop:LCA_2}
Consider $x\in V_{F_1}-L_{F_1}$ and set $A=\{v\in V_{F_1}\mid  v<x\}\bigcap L_{F_1}$. Take $y= \LCA(\phi(A))$ and $B=\{w\in V_{G_1}\mid  w<y\}\bigcap L_{G_1}$.
Suppose there is $w\in B-\phi(A)$. Since the leaves of $\mathbf{F}_1$ and $\mathbf{G}_1$ are in bijection, there is
$v$ such that  $w=\phi(v)$. Take $x'=\LCA(A\cup \{v\})$.

We have $h_f(x')-h_f(x)\leq 2\varepsilon$.

\begin{proof}
Suppose by contradiction that \(h_f(x')-h_f(x)>2\varepsilon\). Since \(w=\phi(v)\in B- \phi(A)\), we have \(w<y\), and by \Cref{prop:LCA_1} we know \(y\le \phi(x)\). Hence \(\phi(v)=w<\phi(x)\le \alpha(x)\). Applying \(\beta\) we obtain \(v\le \beta(w)\le \beta(\alpha(x))\).
Moreover, for every \(a\in A\), since \(a\le x\), we have \(\phi(a)\le \phi(x)\le \alpha(x)\), and therefore \(a\le \beta(\alpha(x))\). Thus \(\beta(\alpha(x))\) is an upper bound for \(A\cup\{v\}\), which implies \(\beta(\alpha(x))\ge x'=\LCA(A\cup\{v\})\).

On the other hand, \(h_f(\beta(\alpha(x)))=h_f(x)+2\varepsilon<h_f(x')\), which is impossible because \(x'\le \beta(\alpha(x))\). Therefore \(h_f(x')-h_f(x)\le 2\varepsilon\).
\end{proof}
\end{lem}

\begin{lem}\label{prop:weights}
Consider $e=(x,father(x))\in E_{F_1}$
 and $A=\{v\in V_{F_1}\mid  v<x\}\bigcap L_{F_1}$. Set $y= \LCA(\phi(A))$ and suppose $\phi(A)=B=\{w\in V_{G_1}\mid  w<y\}\bigcap L_{G_1}$; set
$e'=(y,father(y))\in E_{G_1}$.
Then $\mid  w_{F_1}(e)-w_{G_1}(e')\mid  \leq 2\varepsilon$.

\begin{proof}
We know by \Cref{prop:LCA_1} that $\mid  h_f(x)- h_g(y)\mid  \leq \varepsilon$. Thus we can focus on $x'=father(x)$ and $y'=father(y)$. Let  $A'=\{v\in V_{F_1}\mid  v<x'\}\bigcap L_{F_1}$,  $B'=\{w\in V_{G_1}\mid  w<y'\}\bigcap L_{G_1}$, $w=\LCA(\phi(A'))$ and $v=\LCA(\psi(B'))$.
By \Cref{prop:LCA_1}, again
$\mid  h_f(x')- h_g(w)\mid  \leq \varepsilon$,
and
$\mid  h_f(v)- h_g(y')\mid  \leq \varepsilon$.

We have $A\subset A' \implies \phi(A)\subset \phi(A') \implies B\subset \phi(A')$, and similarly $A\subset\psi(B')$.
Which entails $x'\leq v$ and $y'\leq w$.
Since $x''= \LCA(\psi(\{w'\in V_{G_1}\mid  w'<w\}))\geq v$, by \Cref{prop:LCA_2} we have
$h_f(v)-h_f(x')\leq 2\varepsilon$ and, similarly,
$h_g(w)-h_g(y')\leq 2\varepsilon$.
Putting together $\mid  h_f(x')- h_g(w)\mid  \leq \varepsilon$, $y'<w$ and $h_g(w)-h_g(y')\leq 2\varepsilon$ and the analogous inequalities for $y'$ we obtain $\mid  h_f(x')-h_g(y')\mid  \leq \varepsilon$.

Thus $\mid  h_f(x')-h_f(x)-(h_g(y')-h_g(y))\mid  \leq 2 \varepsilon$.

\end{proof}
\end{lem}

\begin{lem}\label{lem:intern_deletions}
    Consider $y\in V_{G_1}$ such that for every $x\in V_{F_1}$, having set $A=\{v\in V_{F_1}\mid  v<x\}\bigcap L_{F_1}$, we have $y\neq \LCA(\phi(A))$. Then $\mid  h_g(father(y))-h_g(y)\mid  \leq 2\varepsilon$.

    \begin{proof}
        Set $B=\{w\in V_{G_1}\mid  w<y\}\bigcap L_{G_1}$, $x= \LCA(\psi(B))$ and $A=\{v\in V_{F_1}\mid  v<x\}\bigcap L_{F_1}$. By hypothesis we know that $y < \LCA(\phi(A))$ i.e. there is $w\in V_{G_1}$, $w\nleq y$, such that $\LCA(\phi(A)) \geq \LCA(\{y,w\})$. In particular, $father(y)\leq \LCA(\phi(A))$. By \Cref{prop:LCA_1} we know that $\mid  h_g(\LCA(\phi(A)))-h_g(y)\mid  \leq 2\varepsilon$, which implies the thesis.
    \end{proof}
\end{lem}

\smallskip
\textbf{Step 3.2: Internal Vertices}
\smallskip

We add to $M$ all the couples $(x,y)$ such that all of the following hold:
\begin{itemize}
    \item $x\in V_{F_1}- L_{F_1}$;
    \item $y=\LCA(\phi(A))$, with $A=\{v\in V_{F_1}\mid  v<x\}\bigcap L_{F_1}$;
    \item $\phi(A)=B$, with $B=\{w\in V_{G_1}\mid  w<y\}\bigcap L_{G_1}.$
\end{itemize}

Suppose now that $(x,y),(x',y)\in M$. We would have $\phi(A)=\phi(A')$, with $A=\{v\in V_{F_1}\mid  v<x\}\bigcap L_{F_1}$ and $A'=\{v\in V_{F_1}\mid  v<x'\}\bigcap L_{F_1}$. But, since $\phi$ is bijective, we have $A=A'$ and so $x=x'$.
For each couple $(x,y)$, we will shrink  $(x,father(x))$ onto $(y,father(y))$.

Now consider $v=\LCA(L_{F_1})$. By construction we have $\{v\}=child(r_{F_1})$.
Similarly, set $w=\LCA(L_{G_1})$, knowing
$\{w\}=child(r_{G_1})$.
Since $\phi$ is a bijection between the leaves, we know that $(v,w)\in M$. Due to the truncation operator which needs to be applied on $F$ and $G$ to compute $d_E$, we know that the cost of shrinking
$(v,r_{F_1})$ onto $(w,r_{G_1})$ is just $\mid  h_f(v)-h_g(w)\mid  $. By \Cref{prop:LCA_1} the cost of this shrinking is at most $\varepsilon$.
For all the other shrinkings, we can apply  \Cref{prop:weights}, obtaining that their cost is less than $2\varepsilon$.

We delete all internal vertices in $F_1$ and $G_1$ which are not coupled. For each such $v$ in $F_1$ we add $(v,\mathfrak{D})$ and For each such $w$ in $G_1$ we add $(\mathfrak{D},w)$. By \Cref{prop:LCA_2} we know that the deletions on $F_1$ cost less than $2\varepsilon$. By
\Cref{lem:intern_deletions} we have the analogous result for the deletions on $G_1$. Note that all these deletions cannot create any degree two vertices, as the number of children of the father of the deleted vertex cannot decrease after the deletion. Since deleting vertices does not change the cost of any shrinkings to be done afterwards, unless some ghostings are applied, we can freely exchange the order of these deletions and of the aforementioned shrinkings between $F_1$ and $G_1$.

\begin{center}
\textbf{Step 4: Inducing the Edit Path}
\end{center}

To conclude the proof we sum up everything and induce an order on edits operations according to the couples contained in $M$, so that the costs of the edits match the ones described along the previous subsections of the proof.

First we apply all the deletions on $F$ described in Step 1.3,
with the cost of every edit being at most $2\varepsilon$. Then we ghost all degree $2$ vertices. By construction we obtain, from $F$, the tree $F_1$. At this point we delete internal vertices of $F_1$ according to the procedure described in Step 3.2, obtaining $F_2$.
Then we shrink all the edges of $F_2$, according to Step 3.2, obtaining $G_2$.
Then we insert all the edges needed to obtain $G_1$ from $G_2$, which are associated to the couples $(\mathfrak{D},y)$ mentioned in Step 3.2.
Then we go on with the splittings induced by $(\mathfrak{G},w)\in M$, which are needed to
subsequently insert the edges which take us from $G_1$ to $G$, as explained in
Step 2. In the respective paragraphs it is shown
that all the edits we employed have cost less than $2\varepsilon$.

This concludes the proof of \Cref{teo:stability}.

\hfill$\blacksquare$

\subsection{\underline{\textit{Proof of}  \Cref{teo:BV}.}}

We build an edit path \(\gamma_g\) between \(T_g\) and \(T_f\) as in
\Cref{teo:stability}. We then define \(T'\) as the merge tree obtained from
\(T_g\) by applying only the deletions contained in \(\gamma_g\) which delete
leaves of \(T_g\).

At this point, the number of leaves of \(T'\) is not larger than the number of
leaves of \(T_f\). Thus, if \(C_f\) denotes the number of leaves of \(T_f\), we
have
$\#E_{T_f},\#E_{T'}\leq 2C_f.
$

By the construction in \Cref{teo:stability}, the remaining edits needed to
transform \(T'\) into \(T_f\) have cost bounded by
\(2\|f-g\|_\infty\). Therefore
\begin{equation}\label{eq:deletion_ineq}
d_E(T_f,T')\leq 4C_f\|f-g\|_\infty .
\end{equation}

We now bound the cost of transforming \(T_g\) into \(T'\). This transformation
only deletes some leaves of \(T_g\). Recall that, for a function defined on an
interval, each leaf of the merge tree is associated with a local minimum, while
the father of such a leaf is associated with one of the adjacent local maxima.

Let
$x_1<\cdots<x_{2n+1}
$
be the ordered local extrema of \(g\). Without loss of generality, we assume
that \(x_{2i+1}\) are local minima and \(x_{2i}\) are local maxima. As already
mentioned, local maxima at the extremes of \([a,b]\) are ignored when building
merge trees. Let \(W\) be the set of indices \(i\) such that the leaf of \(T_g\)
associated with the local minimum \(x_{2i+1}\) is deleted when passing from
\(T_g\) to \(T'\). For every \(i\in W\), let
\(\widetilde{x}_{2i+1}\) be the local maximum associated with the father of
that leaf. Since \(\widetilde{x}_{2i+1}\) is the local maximum at which the
component born at \(x_{2i+1}\) merges, we have
$g(\widetilde{x}_{2i+1})\geq g(x_{2i+1}).
$

Set
\[
A:=
\sum_{i\in W}
\bigl(g(\widetilde{x}_{2i+1})-g(x_{2i+1})\bigr).
\]
Then
\[
d_E(T',T_g)\leq A .
\]

It remains to estimate \(A\). Let
$z_1<\cdots<z_k
$
be the ordered local extrema of \(f\). Again, we assume that \(z_{2i+1}\) are
local minima and \(z_{2i}\) are local maxima; in particular, the sequence starts
and ends with local minima. Hence \(k\leq 2C_f\). Since \(f\) is monotone on
each interval \([z_i,z_{i+1}]\), we have
\[
\V_{[a,b]}(f)
=
\sum_{i=1}^{k-1}(-1)^{i+1}\bigl(f(z_{i+1})-f(z_i)\bigr).
\]
Define
\[
B:=
\sum_{i=1}^{k-1}
\left|g(z_{i+1})-g(z_i)\right|.
\]
Then \(B\leq \V_{[a,b]}(g)\), and
\[
\left|B-\V_{[a,b]}(f)\right|
\leq
\sum_{i=1}^{k-1}
\left|
\left|g(z_{i+1})-g(z_i)\right|
-
\left|f(z_{i+1})-f(z_i)\right|
\right|
\leq
2k\|f-g\|_\infty
\leq
4C_f\|f-g\|_\infty .
\]

Let
$\{y_1,\ldots,y_m\}
$
be the ordered partition of \([a,b]\) obtained by taking the union of
\(\{z_1,\ldots,z_k\}\) and \(\{x_1,\ldots,x_{2n+1}\}\). Set
\[
C:=
\sum_{r=1}^{m-1}
\left|g(y_{r+1})-g(y_r)\right|.
\]
Thus \(C\) is the variation of \(g\) along the refined partition, while \(B\)
is the variation of \(g\) along the coarser partition given by the extrema of
\(f\). We claim that
\begin{equation}\label{eq:A_excess_variation}
A\leq C-B+8C_f\|f-g\|_\infty .
\end{equation}

To prove the claim, set \(\varepsilon:=\|f-g\|_\infty\). Fix
\(i\in\{1,\ldots,k-1\}\), and write the points of the refined partition lying
in \([z_i,z_{i+1}]\) as
\[
z_i=y_{\alpha_i}<y_{\alpha_i+1}<\cdots<y_{\beta_i}=z_{i+1}.
\]
Define
\[
\Delta_i:=
\sum_{r=\alpha_i}^{\beta_i-1}
\left|g(y_{r+1})-g(y_r)\right|
-
\left|g(z_{i+1})-g(z_i)\right|.
\]
Then, summing over \(i\),
\[
C-B=\sum_{i=1}^{k-1}\Delta_i .
\]

We split the set \(W\) of deleted leaves into an interior part and a boundary
part. For \(i=1,\ldots,k-1\), set
\[
W_i^{\mathrm{int}}
:=
\left\{
j\in W \ \middle|\
x_{2j+1}\in (z_i,z_{i+1})
\ \text{and}\
\widetilde{x}_{2j+1}\in (z_i,z_{i+1})
\right\}.
\]
Thus \(W_i^{\mathrm{int}}\) consists of those deleted leaves whose associated
local minimum and associated father are both strictly inside the same interval
\((z_i,z_{i+1})\). Define
\[
D_i^{\mathrm{int}}
:=
\sum_{j\in W_i^{\mathrm{int}}}
\bigl(g(\widetilde{x}_{2j+1})-g(x_{2j+1})\bigr).
\]
We claim that
$D_i^{\mathrm{int}}\leq \Delta_i .
$

Indeed, set \(v_r:=g(y_r)\). The quantity
\[
\sum_{r=\alpha_i}^{\beta_i-1}|v_{r+1}-v_r|
\]
is the total variation of the finite sequence
$(v_{\alpha_i},v_{\alpha_i+1},\ldots,v_{\beta_i}),
$
whereas \(|v_{\beta_i}-v_{\alpha_i}|\) is the total variation of the sequence
obtained by retaining only the two endpoint values.

Each \(j\in W_i^{\mathrm{int}}\) corresponds to a leaf-edge of the merge tree of
\(g\), associated with the branch born at the local minimum \(x_{2j+1}\) and
dying at the local maximum \(\widetilde{x}_{2j+1}\), with both points strictly
contained in \((z_i,z_{i+1})\). At the level of the ordered finite sequence
$y_{\alpha_i}<\cdots<y_{\beta_i},
$
removing such a branch removes an interior valley whose height is
$g(\widetilde{x}_{2j+1})-g(x_{2j+1}).
$

Since the corresponding leaves are distinct, these removed valleys are disjoint
in the ordered sequence of extrema of \(g\). Hence, after removing all valleys
indexed by \(W_i^{\mathrm{int}}\), the total variation of the resulting finite
sequence is at most
\[
\sum_{r=\alpha_i}^{\beta_i-1}|v_{r+1}-v_r|
-
D_i^{\mathrm{int}}.
\]
Moreover, because all these removals are strictly internal to
\((z_i,z_{i+1})\), the first and last values of the sequence remain
\(v_{\alpha_i}=g(z_i)\) and \(v_{\beta_i}=g(z_{i+1})\). The total variation of
any finite sequence is bounded from below by the absolute difference between
its first and last values. Therefore
\[
\sum_{r=\alpha_i}^{\beta_i-1}|v_{r+1}-v_r|
-
D_i^{\mathrm{int}}
\geq
|v_{\beta_i}-v_{\alpha_i}|
=
|g(z_{i+1})-g(z_i)|.
\]
Equivalently,
\[
D_i^{\mathrm{int}}
\leq
\sum_{r=\alpha_i}^{\beta_i-1}
|g(y_{r+1})-g(y_r)|
-
|g(z_{i+1})-g(z_i)|
=
\Delta_i .
\]

Now set
\[
W^{\partial}
:=
W\setminus \bigcup_{i=1}^{k-1}W_i^{\mathrm{int}}.
\]
These are precisely the deleted leaves whose associated local minimum and
associated father are not both strictly contained in the same interval
\((z_i,z_{i+1})\). Since each deleted leaf in \(W^\partial\) is one of the
deletions appearing in the edit path \(\gamma_g\), and since the edit path
\(\gamma_g\) is the one constructed in \Cref{teo:stability}, each such deletion
has cost bounded by \(2\varepsilon\). Hence
\[
\sum_{j\in W^\partial}
\bigl(g(\widetilde{x}_{2j+1})-g(x_{2j+1})\bigr)
\leq
2\varepsilon\,\# W^\partial .
\]

It remains to bound \(\#W^\partial\). If \(j\in W^\partial\), then the adjacent
extrema \(x_{2j+1}\) and \(\widetilde{x}_{2j+1}\) are not both strictly contained
in any single interval \((z_i,z_{i+1})\). Therefore the closed interval with
endpoints \(x_{2j+1}\) and \(\widetilde{x}_{2j+1}\) contains at least one point
of the critical sequence \(\{z_1,\ldots,z_k\}\). Distinct indices
\(j\in W^\partial\) correspond to distinct adjacent extremum intervals of \(g\).
The interiors of such intervals are pairwise disjoint, and each point
\(z_\ell\) can belong to the closure of at most two of them. Hence
$\#W^\partial\leq 2k.
$

Since \(k\leq 2C_f\), we get
$\#W^\partial\leq 4C_f.
$

Consequently,
\[
\sum_{j\in W^\partial}
\bigl(g(\widetilde{x}_{2j+1})-g(x_{2j+1})\bigr)
\leq
2\varepsilon\,\#W^\partial
\leq
8C_f\varepsilon .
\]

Combining the interior and boundary estimates gives
\[
A
=
\sum_{i=1}^{k-1}D_i^{\mathrm{int}}
+
\sum_{j\in W^\partial}
\bigl(g(\widetilde{x}_{2j+1})-g(x_{2j+1})\bigr)
\leq
\sum_{i=1}^{k-1}\Delta_i
+
8C_f\varepsilon.
\]
Since \(\sum_{i=1}^{k-1}\Delta_i=C-B\), we conclude that
\[
A\leq C-B+8C_f\varepsilon .
\]
Since \(C\leq V_{[a,b]}(g)\), this gives
\[
A\leq V_{[a,b]}(g)-B+8C_f\varepsilon .
\]
Moreover,
\[
V_{[a,b]}(g)-B
\leq
\left|V_{[a,b]}(g)-V_{[a,b]}(f)\right|
+
\left|V_{[a,b]}(f)-B\right|.
\]
Using the bound on \(\left|B-\V_{[a,b]}(f)\right|\) proved above, we get
\[
A
\leq
\left|V_{[a,b]}(g)-V_{[a,b]}(f)\right|
+
12C_f\|f-g\|_\infty .
\]
Therefore
\[
d_E(T',T_g)
\leq
\left|V_{[a,b]}(g)-V_{[a,b]}(f)\right|
+
12C_f\|f-g\|_\infty .
\]

Combining this inequality with \eqref{eq:deletion_ineq}, and using the triangle
inequality, gives
\[
d_E(T_f,T_g)
\leq
\left|V_{[a,b]}(g)-V_{[a,b]}(f)\right|
+
16C_f\|f-g\|_\infty .
\]
Repeating the same argument with the roles of \(f\) and \(g\) exchanged gives
the same estimate with \(C_g\) in place of \(C_f\). Hence
\[
d_E(T_f,T_g)
\leq
\left|V_{[a,b]}(g)-V_{[a,b]}(f)\right|
+
16\min\{C_f,C_g\}\|f-g\|_\infty .
\]

\hfill \(\blacksquare\)

\subsection{\underline{\textit{Proof of}  \Cref{prop:estimator}.}}

    By \Cref{teo:BV} and \Cref{rmk:BV} we know:
    \[
    d_E(T_f,T_{\hat{f}_n}) \leq 16C_f \parallel f-\hat{f}_n\parallel_\infty + (b-a)\parallel Df-D\hat{f}_n\parallel_\infty.
    \]
    With $C_f$ being the number of local minima points in $f$.
    Thus via the union bound:
    \begin{align*}
     & P\bigl(d_E(T_f,T_{\hat f_n})\le 2\varepsilon\bigr)
\ge
1-h_n\!\left(\frac{\varepsilon}{16
C_f}\right)-g_n\!\left(\frac{\varepsilon}{b-a}\right),
    \end{align*}
    which is our thesis.

\hfill $\blacksquare$

\subsection{\underline{\textit{Proof of}  \Cref{prop:approx}.}}

    \begin{enumerate}
        \item On every interval $[z_i,z_{i+1}]$ we have $\mid  f(z_i)-f(x)\mid  \leq L \cdot \mid  z_i- x\mid  $ and
        $\mid  f_\delta^{PL}(z_i)-f_\delta^{PL}(x)\mid  \leq L_i \cdot \mid  z_i- x\mid  $ with $L_i = \mid  f_\delta^{PL}(z_{i+1})-f_\delta^{PL}(z_i)\mid  /(z_{i+1}-z_i)$. At the same time, we have that $\mid  f_\delta^{PL}(z_{i+1})-f_\delta^{PL}(z_i)\mid  /(z_{i+1}-z_i) = \mid  f(z_{i+1})-f(z_i)\mid  /(z_{i+1}-z_i) \leq L$. Thus, for $x\in (z_i,z_{i+1})$ we have:
        \[
        \mid  f(x)-f_\delta^{PL}(x) \mid  \leq \mid  f(x)-f(z_i) \mid  + \mid  f_\delta^{PL}(x)-f_\delta^{PL}(z_i) \mid  \leq 2L (z_{i+1}-z_i).
        \]
        \item To prove this point, we make use of the following result, which we adapt from Theorem 3.4.1 in \cite{quarteroni2008numerical}.

    \begin{teo}[adapted from \cite{quarteroni2008numerical}]\label{teo:PL}
        Fix $\{z_0=a,\ldots,z_m=b\}\subset X=[a,b]$, increasing sequence of numbers and let $\delta = \max_{i=0,\ldots,m-1} z_{i+1}-z_i$.
        There exist a constant $A>0$ such that for every $f\in H^2({[a,b]})$, its piecewise linear interpolant $f_\delta^{PL}$ on $\{z_i\}_{i=0,\ldots,m}$, satisfies for every $i=0,\ldots,m-1$:
        \begin{equation}
        \int_{z_i}^{z_{i+1}}\mid  f-f_\delta^{PL}\mid ^2dx\leq A^2 \delta^4 \int_{z_i}^{z_{i+1}} \mid  D^2 f\mid ^2 dx,
        \end{equation}
        and
        \begin{equation}
        \int_{z_i}^{z_{i+1}}\mid  D f -D f_\delta^{PL}\mid ^2dx\leq A^2 \delta^2 \int_{z_i}^{z_{i+1}} \mid  D^2 f\mid ^2 dx.
        \end{equation}
    \end{teo}

    By \Cref{teo:PL}, we have that:
    \[
    \parallel Df-Df^{PL}_\delta\parallel_2^2 \leq A^2 \delta^2 \parallel D^2f\parallel_2^2.
    \]
    By Holder inequality we then obtain:
    \[
    \parallel Df-Df^{PL}_\delta\parallel_1 \leq (b-a)^{1/2}  A \delta \parallel D^2f\parallel_2.
    \]
    Thus $\V(f-f^{PL}_\delta) \leq (b-a)^{1/2} A \delta \parallel D^2f\parallel_2.$
    \item Putting the pieces together and using \Cref{teo:BV}:
\[
d_E(T_f,T_{f^{PL}_\delta})\leq 32 C_f  L \delta + (b-a)^{1/2} A \delta \parallel D^2f\parallel_2.
\]
    This concludes the proof.
    \end{enumerate}

\hfill $\blacksquare$

\subsection{\underline{\textit{Proof of}  \Cref{prop:comp_est}.}}

Using \Cref{prop:approx} we have:
\[
\|f-\hat f^{PL}_{n,\delta_m}\|_\infty
\le
\|f-\hat f_n\|_\infty
+2\frac{b-a}{m}\|D\hat f_n-Df\|_\infty
+2\frac{b-a}{m}\|Df\|_\infty.
\]
Similarly:
\begin{align*}
&V(f-\hat{f}_{n,\delta_m}^{PL})\leq V(f-\hat{f}_{n}) + (b-a)^{3/2} \frac{A}{m} (\parallel D^2\hat{f}_n\parallel_2) \leq \\
&V(f-\hat{f}_{n}) + (b-a)^{3/2} \frac{A}{m} (\parallel D^2\hat{f}_n- D^2f\parallel_2 + \parallel  D^2f\parallel_2) \leq \\
&\parallel Df-D\hat{f}_{n}\parallel_\infty (b-a) + (b-a)^{2} \frac{A}{m} ( \parallel D^2\hat{f}_n- D^2f\parallel_\infty + \parallel  D^2f\parallel_\infty) \leq \\
&C' \cdot \parallel Df-D\hat{f}_{n}\parallel_\infty + C'' \frac{\parallel D^2\hat{f}_n- D^2f\parallel_\infty}{m} + \frac{C'''}{m},
\end{align*}
for some constants $C',C'',C'''>0$, and with $A$ being the constant as in \Cref{teo:PL}.
Thus, for some constants $C_1,\ldots, C_4$, we have:
\[
d_E(T_{f},T_{\hat{f}_{n,\delta_m}^{PL}}) \leq C_1^\prime \parallel f-\hat{f}_{n}\parallel_\infty + C_2^\prime \parallel Df-D\hat{f}_{n}\parallel_\infty + C_3^\prime \frac{\parallel D^2\hat{f}_n- D^2f\parallel_\infty}{m} + \frac{C_4}{m}.
\]

The positive constants, in particular, are:
\begin{itemize}
    \item $C_1^{-1}:=C_1^\prime= 16C_f$;
    \item $C_2^{-1}:=C_2^\prime=(b-a)(1+32C_f)\ge (b-a)+\frac{32C_f(b-a)}{m}$;
    \item $C_3^{-1}:=C_3^\prime= (b-a)^{2}A$;
    \item $C_4=(b-a)(32 C_f \parallel Df \parallel_\infty + (b-a)A\parallel D^2f \parallel_\infty)$;
\end{itemize}

where $C_f$ is the number of local minima of $f$.

To conclude, if $m>C_4/\varepsilon$, via the union bound
 we have:
\[
P\bigl(d_E(T_{f},T_{\hat{f}_{n,m}^{PL}})\le 4\varepsilon\bigr)
\ge
1-h_n\!\left(\varepsilon C_1\right)
-g_n\!\left(\varepsilon C_2\right)
-q_n\!\left(m\varepsilon C_3\right),
\]
ending the proof.

\hfill $\blacksquare$

\end{document}